\documentclass[twocolumn,prd,aps,superscriptaddress,preprintnumbers,tightenlines,showpacs,nofootinbib,eqsecnum,amsfonts,amsmath]{revtex4}
\pdfoutput=1

\usepackage{color}
\usepackage{calc}
\usepackage{amsmath,amssymb,graphicx}
\usepackage{tensor}
\usepackage{bm}
\usepackage{times}
\usepackage[varg]{txfonts}
\usepackage{float}
\usepackage{dcolumn}

\usepackage[nolist,nohyperlinks]{acronym}
\usepackage{xspace}

\usepackage[abs]{overpic}
\usepackage{pict2e}

\usepackage[caption=false]{subfig} % prevent loading of caption package: http://tex.stackexchange.com/questions/22388/subfigures-with-revtex

\usepackage[utf8]{inputenc}

\usepackage{hyperref} 
\usepackage{braket}
\usepackage{pgffor}
\allowdisplaybreaks
\usepackage{float}
\usepackage{listings}
\usepackage{tensor}
\usepackage{ulem}
\usepackage[makeroom]{cancel}

\DeclareMathAlphabet{\mathcalstd}{OMS}{cmsy}{m}{n}
\DeclareMathAlphabet{\mathpzc}{OT1}{pzc}{m}{it}

\DeclareUnicodeCharacter{2009}{ }

\newcommand{\UIB}{Departament de F\'isica, Universitat de les Illes Balears and Institut d'Estudis Espacials de Catalunya, 
Crta. Valldemossa km 7.5, E-07122 Palma, Spain}

\newcommand{\UoB}{School of Physics and Astronomy and Institute for Gravitational Wave Astronomy, University of Birmingham, Edgbaston, Birmingham, B15 9TT, United Kingdom}

\acrodef{PN}{post-Newtonian}
\acrodef{EOB}{effective-one-body}
\acrodef{NR}{numerical relativity}
\acrodef{GW}{gravitational-wave}
\acrodef{BBH}{binary black hole}
\acrodef{BH}{black hole}
\acrodef{BNS}{binary neutron star}
\acrodef{NSBH}{neutron star-black hole}
\acrodef{SNR}{signal-to-noise ratio}
\acrodef{aLIGO}{Advanced LIGO}
\acrodef{AdV}{Advanced Virgo}

\newcommand{\phB}{\textsc{IMRPhenomB}\xspace}
\newcommand{\phC}{\textsc{IMRPhenomC}\xspace}
\newcommand{\phD}{\textsc{IMRPhenomD}\xspace}
\newcommand{\phX}{\textsc{IMRPhenomXAS}\xspace}
\newcommand{\phP}{\textsc{IMRPhenomP}\xspace}
\newcommand{\phHM}{\textsc{IMRPhenomHM}\xspace}
\newcommand{\phXHM}{\textsc{IMRPhenomXHM}\xspace}
\newcommand{\surro}{\textsc{NRHybSur3dq8}\xspace}
\newcommand{\vfourom}{\textsc{SEOBNRv4\_ROM}\xspace}
\newcommand{\pvtwotidal}{\textsc{IMRPhenomPv2\_NRTidal}\xspace}
\newcommand{\pvthree}{\textsc{IMRPhenomPv3}\xspace}

\begin{document}

%\preprint{LIGO-PXXX}
%%%%%%%%%%%%%%%%%%%%%%%%%%%%%% Title page %%%%%%%%%%%%%%%%%%%%%%%%%%%%%

\title{IMRPhenomXHM: A multi-mode frequency-domain model for the gravitational wave signal from non-precessing black-hole binaries}

\author{Cecilio Garc{\'i}a-Quir{\'o}s}
\affiliation{\UIB}

\author{Marta Colleoni}
\affiliation{\UIB}

\author{Sascha Husa}
\affiliation{\UIB}
%\affiliation{\ICTS}

\author{H{\'e}ctor Estell{\'e}s}
\affiliation{\UIB}

\author{Geraint Pratten}
\affiliation{\UoB}
\affiliation{\UIB}

\author{Antoni Ramos-Buades}
\affiliation{\UIB}

\author{Maite Mateu-Lucena}
\affiliation{\UIB}

\author{Rafel Jaume}
\affiliation{\UIB}

%%%%%%%%%%%%%%%%%%% Abstract %%%%%%%%%%%%%%%%%%%%%%%%%%%%%%
\begin{abstract}
We present the \phXHM  frequency domain phenomenological waveform model for the inspiral, merger and ringdown of quasi-circular non-precessing black hole binaries.
The model extends the \phX waveform model \cite{phenX}, which describes the dominant quadrupole modes
$\ell = |m| = 2$, to the harmonics $(\ell, |m|)=(2,1), (3,3), (3,2), (4,4)$, and includes mode mixing effects for the $(3,2)$ spherical harmonic.
\phXHM is calibrated against hybrid waveforms, which match an inspiral phase described by the effective-one-body model and post-Newtonian amplitudes for the subdominant harmonics to numerical relativity waveforms and numerical solutions to the perturbative Teukolsky equation for large mass ratios up to 1000.
A computationally efficient implementation of the model is available as part of the LSC Algorithm Library Suite \cite{lalsuite}.

\end{abstract}


%\today

\maketitle

%%%%%%%%%%%%%%%%%%%%%%%%%%%%%%%%%%%%%%%%%%%%%%%%%%%%%%%
\section{Introduction}\label{sec:introduction}
%%%%%%%%%%%%%%%%%%%%%%%%%%%%%%%%%%%%%%%%%%%%%%%%%%%%%%%

Frequency domain phenomenological waveform models for compact binary coalescence, such as \cite{Khan:2015jqa,Hannam:2013oca,Bohe:PPv2,London:2017bcn} have become a standard tool for gravitational wave data analysis \cite{TheLIGOScientific:2016wfe,LIGOScientific:2018mvr}. These models describe the amplitude and phase of spherical harmonic modes in terms of piecewise closed form expressions. The low computational cost to evaluate these models makes them particularly valuable for applications in Bayesian inference \cite{Veitch:2014wba,Ashton:2018jfp}, which typically requires millions of waveform evaluations to accurately determine the posterior distribution of the source properties measured in observations, such as the mass, arrival time, or sky location.

Until recently the modelling of the gravitational wave signal from such systems, and consequently gravitational wave data analysis, have  focused on the dominant $\ell=|m|=2$ harmonics. For high masses or high mass ratios this leads however to a significant loss of detection rate \cite{Varma:2014hm,Varma:2017hm,BustilloHusa:2016}, systematic bias in the source parameters (see e.g.~\cite{Littenberg:2012uj,Varma:2014hm,Bustillo:2015qty,Chatziioannou:2019dsz,Kalaghatgi:2019log,Shaik:2019dym}), and implies a degeneracy between distance and inclination of the binary system.
As the sensitivity of gravitational wave detectors increases, accurate and computationally efficient waveform models that include subdominant harmonics are required in order to not limit the scientific scope of gravitational wave astronomy.

Recently both time domain and frequency domain inspiral-merger-ringdown (IMR) models have been extended to sub-dominant spherical harmonics, i.e.~modes other than the $(2,\pm2)$ modes: In the time domain this has been done in the context of the effective-one-body (EOB) approach \cite{Cotesta:2018fcv}, however EOB models are computationally expensive and usually a reduced order model (ROM) is constructed to accelerate evaluation \cite{Puerrer2014,Purrer:2015tud} (see however \cite{Nagar:2018gnk} for an analytical method to accelerate the inspiral). 
Furthermore, the \surro surrogate model \cite{Varma:2018mmi} has been directly built from hybrid waveforms, but is restricted to mass ratios up to eight.
For a precessing surrogate model, calibrated to numerical relativity waveforms, see \cite{Varma:2019csw}.
Fast frequency domain models have previously been developed for the non-spinning sub-space \cite{Mehta:2017jpq,Mehta:2019wxm}, and for spinning black holes through an approximate map from the $(2,2)$ harmonic to general harmonics as described in \cite{London:2017bcn}, which presented the \phHM model, which is  publicly available as part of the LIGO Algorithm Library Suite (\texttt{LALSuite}) \cite{lalsuite}.
This approximate map is based on the approximate scaling behaviour of the subdominant harmonics with respect to the $(2,2)$ mode, the \phHM model is thus only calibrated to numerical data for the $(2,2)$ mode. This information from numerical waveforms enters through the \phD model, which is calibrated to numerical relativity (NR) waveforms up to mass ratio $q=18$. 

Here we present the first frequency domain model for the inspiral, merger and ringdown of spinning black hole binaries, which calibrates subdominant harmonics to a set of numerical waveforms for spinning black holes, instead of using an approximate map as \phHM.
For the $(2,\pm2)$ modes the model is identical to \phX \cite{phenX}, which presents a thorough update of the \phD model, extends it to extreme mass ratios, drops the approximation of reducing the two spin parameters of the black holes to effective spin parameters, and replaces ad-hoc fitting procedures by the hierarchical method presented in \cite{Jimenez-Forteza:2016oae}.

Our modelling approach largely follows our work on  \phX, with some adaptions to the phenomenology of subdominant modes, as first summarized in Sec.~\ref{sec:prelim}. As for \phX we construct closed form expressions for the amplitude and phase of each spherical harmonic mode in three frequency regimes, which correspond to the inspiral, ringdown, and an intermediate regime.
In the inspiral and ringdown the model can be based on the perturbative frameworks of post-Newtonian theory \cite{Blanchet2014}
and black hole perturbation theory \cite{Kokkotas1999}. The intermediate regime, which models the highly dynamical and strong field transition between the physics of the inspiral and ringdown still eludes a perturbative treatment.
An essential goal of frequency domain phenomenological waveform models is computational efficiency. To this end, an accompanying paper \cite{our_mb} presents techniques to further accelerate the model evaluation following \cite{Vinciguerra:2017ngf}.

We model the complete observable signal, from the inspiral phase to the merger and ringdown to the remnant Kerr black hole, but we restrict our work to the quasi-circular (i.e.~non-eccentric) and non-precessing part of the parameter space of astrophysical black hole binaries in general relativity, which is 3D and given by mass ratio $q = m_1/m_2 \geq 1$ and the dimensionless spin components $\chi_i $ of the two black holes which are orthogonal to the preserved orbital plane,
\begin{equation}
\chi_i = \frac{\vec S_i \cdot \vec L}{m_i^2 \, \vert \vec L\vert},
\end{equation}
where  $\vec S_{1,2}$  are the spins (intrinsic angular momenta) of the two individual black holes, $\vec L$ is the orbital angular momentum, and  $m_{1,2}$ are the masses of the two black holes.
We also define the total mass $M=m_1+m_2$, and the symmetric mass ratio $\eta = m_1 m_2 /M^2$.
An approximate map from the non-precessing to the precessing parameter space \cite{Schmidt:2012rh,Hannam:2013oca,Khan:2018fmp} can then be used  to extend the model to include the leading precession effects. 

The paper is organized as follows.
In Sec.~\ref{sec:prelim} we collect some preliminaries: our conventions, notes on waveform phenomenology which motivate our modelling approach, and a brief description of our plan of fitting numerical data.
In Sec.~\ref{sec:input} we briefly describe our input data set of hybrid waveforms, and the underlying numerical relativity and perturbative  Teukolsky waveforms.
The construction of our model is discussed in Secs.~\ref{sec:inspiral}-\ref{sec:ringdown} for the inspiral, intermediate region, and ringdown respectively. 
The accuracy of the model is evaluated in Sec.~\ref{sec:quality}, and we conclude with a summary and discussion in
Sec.~\ref{sec:conclusions}. 
Appendix \ref{appendix:spheroidal_spherical} discusses the conversion from spheroidal to spherical-harmonic modes. 
In appendix \ref{appendix:HMconventions} we describe our method to test tetrad conventions in  multi-mode waveforms. Some technical details of our \texttt{LALSuite}
\cite{lalsuite} implementation are presented in appendix \ref{appendix:LAL}.
Details regarding the rescaling of the inspiral phase are presented in
appendix \ref{app:inspiral_hm}, and appendix \ref{appendix:FPN} summarizes post-Newtonian results on the Fourier domain amplitude.

%%%%%%%%%%%%%%%%%%%%%%%%%%%%%%%%%%%%%%%%%%%%%%%%%%%%%%%
\section{Preliminaries} \label{sec:prelim}

\subsection{Conventions} \label{sec:conventions}

Our waveform conventions are consistent with those chosen for the \phX model in \cite{phenX} and our catalogue of multi-mode hybrid
waveforms \cite{hybrids}, which we introduce in Sec.~\ref{sec:input}.
We use a standard spherical coordinate system $(r,\theta,\phi)$ and spherical harmonics  $Y^{-2}_{\ell m}$ of spin-weight $-2$ (see e.g.~\cite{Wiaux:2005fm}).
The black holes orbit in the plane $\theta=\pi/2$. Due to the absence of spin-precession the spacetime geometry exhibits equatorial symmetry, i.e. the northern hemisphere $\theta < \pi/2$ is isometric to the southern hemisphere $\theta > \pi/2$, and in consequence the same holds for the gravitational-wave signal.

The gravitational-wave strain $h$ depends on an inertial time coordinate $t$, the angles $\theta, \phi$ in the sky of the source, and the source parameters $(\eta,\chi_1,\chi_2)$.
We can write the strain in terms of the polarizations as
 $h=h_+(t, \theta , \varphi )  - i \, h_{\times} \, (t, \theta , \varphi )$, 
 or decompose it into spherical harmonic modes $h_{\ell m}$ as
\begin{equation}\label{eq:h_harmonics}
    h(t,\theta,\varphi) = \sum_{\ell=2,m=-\ell}^{4,\ell}   h_{\ell m} \, (t) \; _{-2}Y_{\ell m} (\theta,\varphi).
\end{equation}
The split into polarizations (i.e.~into the real and imaginary parts of the time domain complex gravitational wave strain) is ambiguous due to the freedom to perform tetrad rotations, which corresponds to the freedom to choose an arbitrary overall phase factor. As discussed in detail in \cite{Bustillo:2015ova,hybrids} and in
appendix \ref{appendix:HMconventions}, only two inequivalent choices are consistent with equatorial symmetry, and for simplicity we adopt the convention that for large separations (i.e. at low frequency) the time domain phases satisfy
\begin{equation}\label{eq:tetrad_convention}
    \Phi_{\ell m} \approx  \frac{m}{2}\Phi_{2 2} .
\end{equation}
This differs from the convention of Blanchet et al. \cite{Blanchet:2008je} by overall factors of $(-1) (-i)^m$ in front of the $h_{\ell m}$ modes. In appendix \ref{appendix:HMconventions} we discuss how to test a given waveform model for the tetrad convention that is used.

The equatorial symmetry of non-precessing binaries implies
\begin{equation}\label{eq:neg_m}
    h_{\ell m}(t) = (-1)^\ell h^{*}_{\ell -m}(t), \quad 
\end{equation}
it is thus sufficient to model just one spherical harmonic for each value of $\vert m \vert$.

We adopt the conventions of the LIGO Algorithms Library \cite{lalsuite} for the Fourier transform,
\begin{equation}\label{eq:defFT}
\tilde h(f)=\int_{-\infty}^{\infty} h(t) \, e^{-i \, 2 \pi f t}\, dt .
\end{equation}
% was:
%\begin{align}
%\tilde{h} (f ; \vth , \theta , \phi ) &= \tilde{h}_+ \, (f ; \vth , \theta , \phi )  - i \, \tilde{h}_{\times} \, (f ; \vth , \theta , \phi ) \\
%&= \displaystyle\sum_{m = -2,2} \tilde{h}_{2m} \, (f ; \vth) %\; _{-2}Y_{2m} (\theta,\phi) .
%\end{align}
%
With these conventions the time domain relations between modes (\ref{eq:neg_m}) that express equatorial symmetry can be converted to the Fourier domain, where they read   
\begin{equation}\label{eq:htilde_mode_relations}
   \tilde h_{\ell m}(f) =  (-1)^\ell \tilde h^{*}_{\ell -m}(-f).
\end{equation}
%,
The definitions above then also imply that $\tilde h_{\ell m}(f)$ (with $m>0$) is concentrated
in the negative frequency domain and $\tilde h_{\ell -m}(f)$ 
in the positive frequency domain. For the inspiral, this can be checked against the stationary phase approximation (SPA), see e.g. \cite{Damour:2000gg}.

As we construct our model in the frequency domain, it is convenient to model $\tilde h_{\ell -m}$, which is non-zero for positive frequencies. The mode $\tilde h_{\ell m}$, defined for negative frequencies, can then be computed from (\ref{eq:htilde_mode_relations}).
We model the Fourier amplitudes $A_{\ell m}(f > 0)$, which are non-negative functions for positive frequencies, and zero otherwise, and the Fourier domain phases $\Phi_{\ell m} (f > 0)$, defined by 
\begin{equation}
\label{eq:amp_and_phase}
\tilde{h}_{\ell -m} (f) = A_{\ell m} (f) \, e^{-i \Phi_{\ell m} (f)}. 
\end{equation}

The contribution to the gravitational wave polarizations of {\em both} positive and negative modes and for positive frequencies is then given by
\begin{eqnarray}\label{def:polarizations_FD}
    \tilde h_{+}(f)         &=& \frac{1}{2} \left( Y_{\ell-m} + (-1)^\ell Y_{\ell m}^*  \right) \tilde{h}_{\ell-m}(f), \\ 
    \tilde h_{\times}(f) &=& \frac{i}{2} \left( Y_{\ell-m} - (-1)^\ell Y_{\ell m}^*  \right) \tilde{h}_{\ell-m}(f).
\end{eqnarray}

If we are only interested in the contribution of just one mode for positive frequencies then the polarizations read as:
\begin{eqnarray}\label{def:polarizations_FD2}
    \tilde{h}_{+}^{\ell,m}(f) &=& \frac{1}{2} (-1)^\ell Y_{\ell m}^* \tilde{h}_{\ell-m}(f),\\
    \tilde{h}_{\times}^{\ell,m}(f) &=& -\frac{i}{2} (-1)^\ell Y_{\ell m}^* \tilde{h}_{\ell-m}(f),\\
    \tilde{h}_{+}^{l,-m}(f) &=& \frac{1}{2} Y_{\ell-m} \tilde{h}_{\ell-m}(f),\\
    \tilde{h}_{\times}^{l,-m}(f) &=& \frac{i}{2} Y_{\ell-m} \tilde{h}_{\ell-m}(f).
\end{eqnarray}
For the ${\ell=2, \pm  2}$ modes these equations correspond to our \phX model \cite{phenX}.

In appendix \ref{appendix:LAL} we discuss conventions which are specific to our \texttt{LALSuite} implementation, in particular how to specify a global rotation, and the time of coalescence.

\subsection{Perturbative waveform phenomenology: inspiral and ringdown} \label{sec:prelim_phenomenology}

The phenomenology of the oscillating subdominant modes $\vert m \vert >0$ is largely similar to the dominant modes
$\ell=\vert m \vert  = 2$, which has been discussed in detail in \cite{Husa:2015iqa,phenX} - with some important exceptions that lead to both simplifications and complications when modelling these modes, as opposed to modelling $\ell=\vert m \vert  = 2$.

The main simplification is that at low frequencies post-Newtonian theory, combined with the stationary phase approximation, predicts
an approximate relation between the phases $\Phi_{\ell m}$ of different harmonics, which with our choice of tetrad takes the simple form of eq.~(\ref{eq:tetrad_convention}).
This approximation is not exact, and becomes less accurate for higher frequencies.
We have studied this in detail in \cite{Bustillo:2015ova,hybrids}, and in \cite{hybrids} we find that for comparable mass binaries we can neglect the error of the approximation (\ref{eq:tetrad_convention}) before a binary system reaches its minimal energy circular orbit (MECO) as defined in
\cite{Cabero:2016ayq}. As in our \phX model, we will use the MECO to guide the choice of transition frequency between the inspiral and intermediate frequency regions.
In the mass ratio range where we have numerical relativity data ($q\leq 18$) it is thus not necessary to model the 
inspiral phase, but we can use the scaling relation (\ref{eq:tetrad_convention}), as has been done in \cite{London:2017bcn}.

For the time domain amplitude, approximate scaling relations have been discussed in \cite{Schnittman:2007ij,Baker:2008mj}, and in the frequency domain they have been used in the \phHM model \cite{London:2017bcn}. Unlike for the phase, however, even in the inspiral the errors are too large for our purposes, and we will need to model the amplitude for each spherical harmonic in a similar way as for \phX, including in the inspiral.

\begin{figure}[ht]
    \centering
    \includegraphics[height=3.8cm]{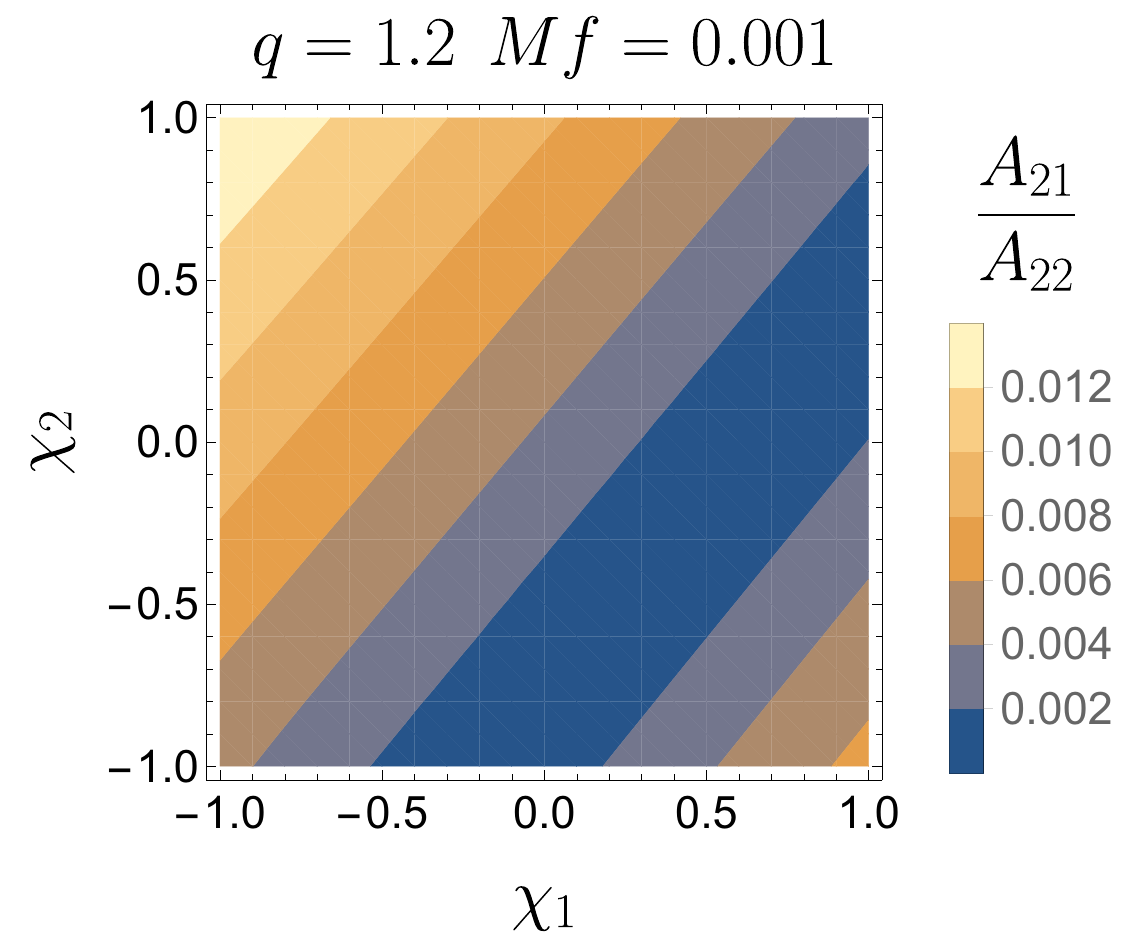}\includegraphics[height=3.8cm]{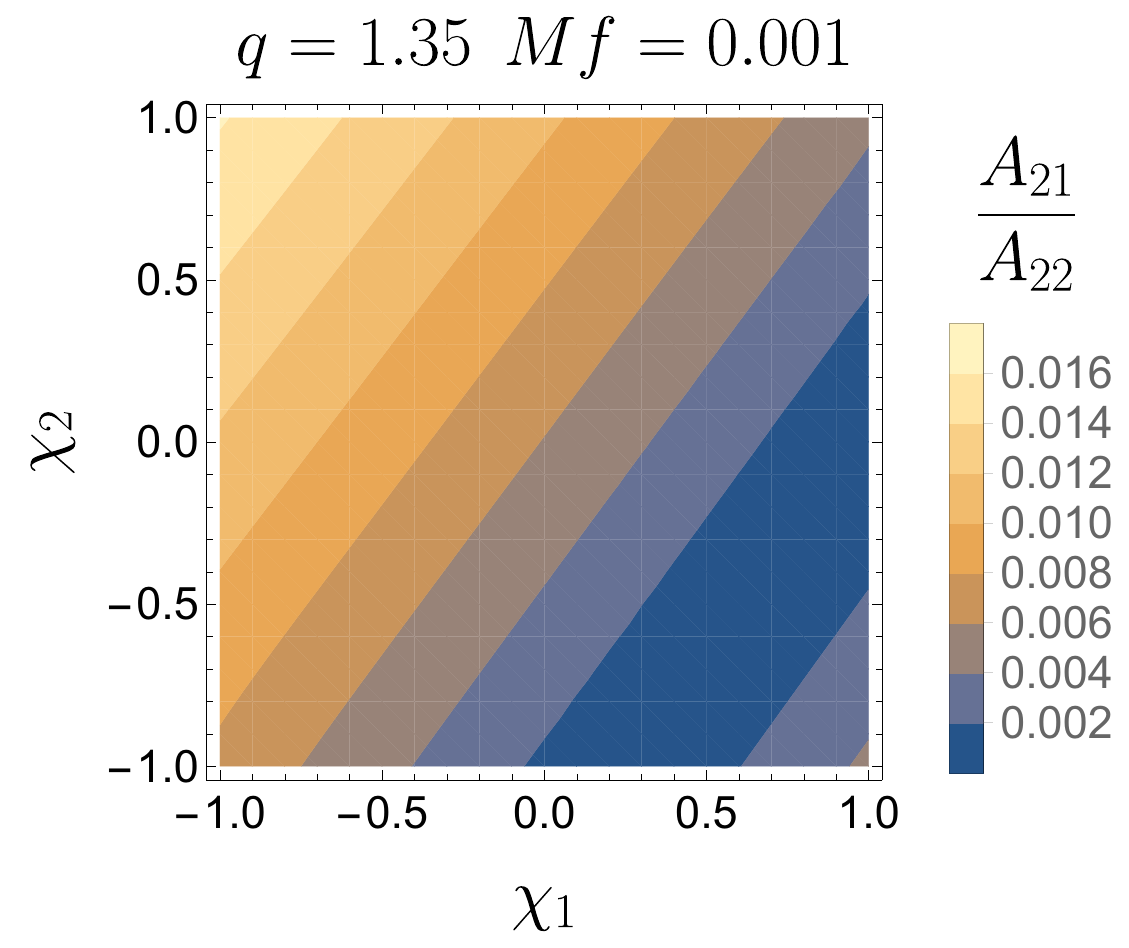}
    \includegraphics[height=3.8cm]{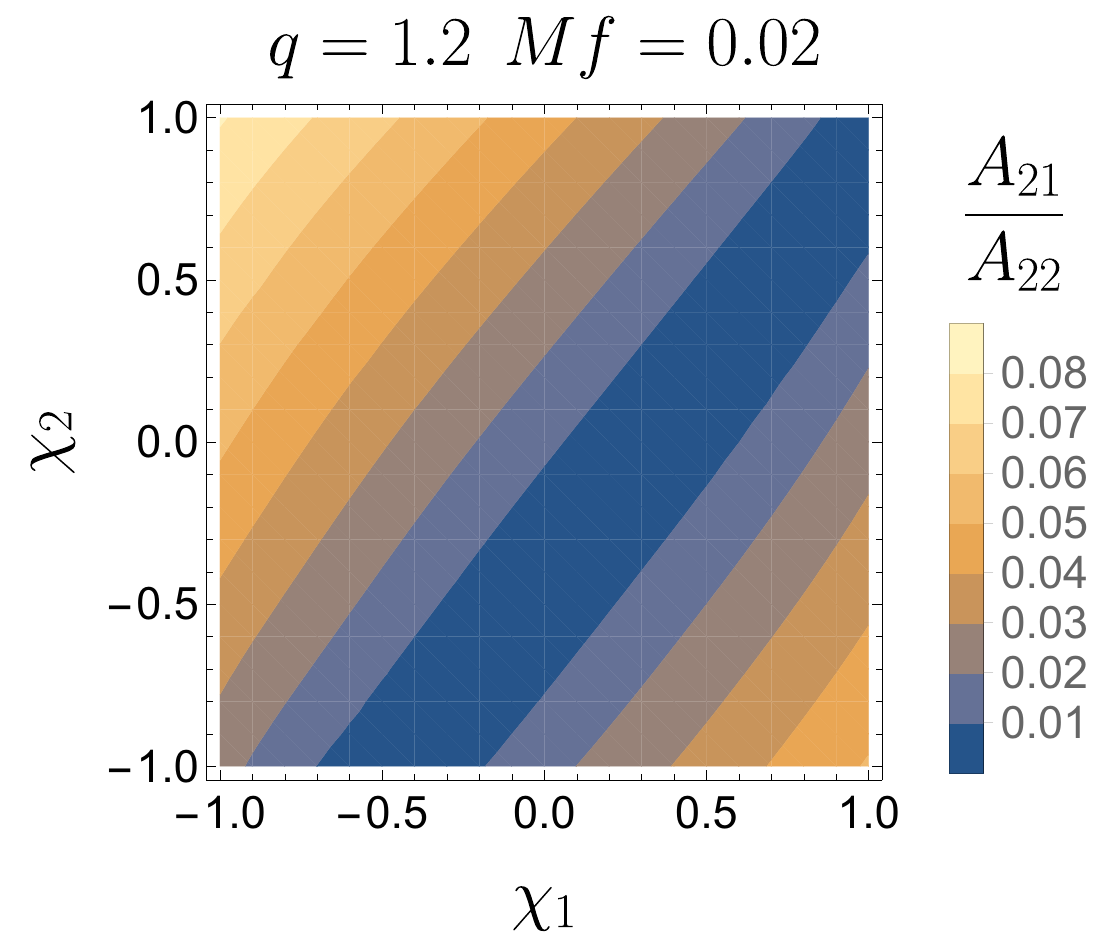}\includegraphics[height=3.8cm]{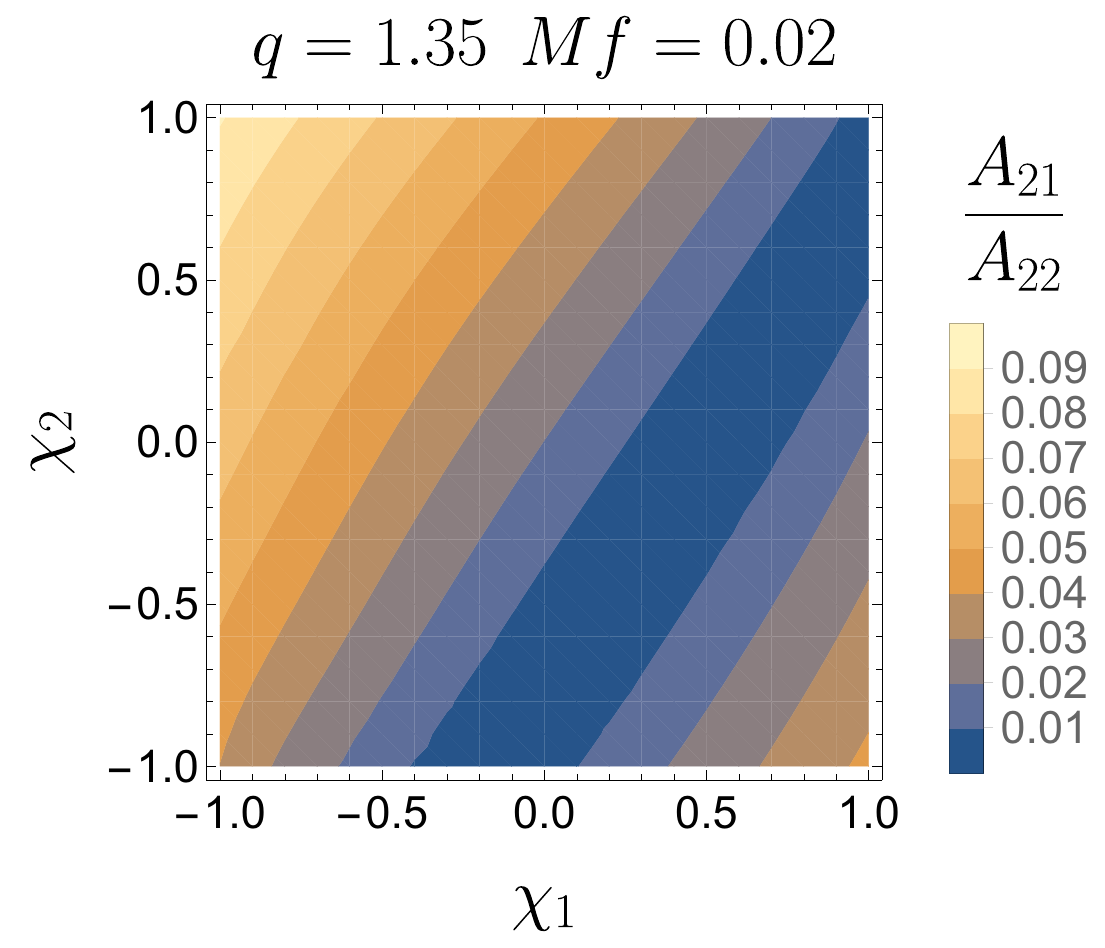}
    \caption{Relative amplitude of the $(2,1)$ mode respect to the $(2,2)$ mode as a function of the spins. The top row shows the amplitude ratio for a frequency of $Mf=0.001$ while the bottom one shows it for $Mf=0.02$. Left-side panels refer to a $q=1.2$ binary and right-side ones to a $q=1.35$ one. There exists a region (blue diagonal) where the amplitude of the 21 mode tends to zero. This diagonal moves toward the bottom-right corner of the plot as one increases the mass ratio and typically disappears for $q \gtrsim 2$.
    %Two top panels: Amplitude of the $(2,1)$ mode relative to the $(2,2$ at a fixed frequency of $Mf=0.001$ as a function of the spins for mass ratios 1.2 (top) and 1.35 (middleL). We can see the regions where the $(2,1)$ amplitude tends to zero. The minimum amplitude region consists in a diagonal that is moving towards one of the corners as we increase the mass ratio. For $q\sim1.7$ it has already disappeared. Bottom panel: A similar behaviour can be observed for higher frequencies ($Mf=0.02$).
    }
    \label{fig:small21amp}
\end{figure}

Rotations in the orbital plane by an angle $\varphi$ transform the spherical harmonic modes as %
\begin{equation}\label{eq:mode_rotation}
h_{\ell m} \rightarrow h_{\ell m} e^{i m  \varphi}.
\end{equation}
Interchange of the two black holes thus corresponds to a rotation by $\varphi=\pi$, and modes with odd $m$ vanish for equal black hole systems.
A problem can arise in regions of the parameter space where the amplitude is close to zero, even in the inspiral, as discussed in \cite{Cotesta:2018fcv,hybrids}: 
For black holes with very similar masses, the amplitude can be very small, with the sign depending on the mass ratio, spins, and the frequency -- which can lead to sign changes with frequency. In such cases the amplitude does become oscillatory, and the approximate relation (\ref{eq:tetrad_convention}) can not be expected to be satisfied. This happens in particular for the $(2,\pm 1)$ modes. We do not currently model the amplitude oscillations, and thus for a certain region of parameter space our model does not properly capture the correct waveform phenomenology. This region does depend on the frequency, but roughly corresponds to very similar masses and anti-aligned spins, for a particular example for the $(2,\pm1)$ modes see Fig.~\ref{fig:small21amp}. However, as phenomenon happens precisely when the amplitude of the  $(2,\pm1)$ modes is very small, this is not expected to be a significant effect for the current generation of detectors. 

During the inspiral and merger, gravitational wave emission is dominated by the direct emission due to the binary dynamics.
As the final black hole relaxes toward a stationary Kerr black hole, the gravitational wave emission is eventually dominated
by a superposition of quasinormal modes before the late time polynomial tail falloff sets in (for an overview see e.g.~
\cite{Andersson:1996cm,Kokkotas1999}).
As is common in waveform modelling targeting applications in GW data analysis, we neglect the late-time power-law tail falloff, and focus our description of the ringdown on the quasinormal mode (QNM) emission, where the strain can be written as a sum of exponentially damped oscillations,
\begin{equation}\label{eq:QNM}
h(t,\theta,\varphi)  \approx \sum_{\ell m n} a_{\ell m n} e^{i (\omega_{\ell m n} t) + \phi_{\ell m n}} \tensor[_{-2}]{Y}{_{\ell m}} (\theta,\phi),
\end{equation}
where the complex frequencies $\omega_{\ell m n}$ are functions of the black hole spin and mass and their real and imaginary parts define the ringdown and damping frequencies as 
\begin{equation}
f_{ring}^{\ell m} = \mathrm{Re}(\omega_{\ell m}), \hspace{1cm} f_{damp}^{\ell m} = |\mathrm{Im}(\omega_{\ell m})|.
\end{equation}
The functions 
 $\tensor[_{-2}]{Y}{_{\ell m}} (\theta,\phi)$ are the {\em spheroidal} harmonics of spin weight $-2$ \cite{Teukolsky1972,Teukolsky1973}.
The amplitude parameters $a_{\ell m n}$ and phase offsets $\phi_{\ell m n}$ will in general have to be fitted to numerical relativity data.

It has long been known that representing the spheroidal harmonic ringdown modes can result in \textit{mode mixing} for modes with approximately the same real part of the ringdown frequency $\omega_{\ell m n}$ \cite{BertiMixingCoeffs,KB2012,London:2020uva}. This happens in particular for modes with the same value of $m$, where then values with larger $\ell$ are much weaker, and do not show the usual exponential amplitude drop, but a more complicated phenomenology. In our case this happens for the $(3,2)$ mode.
In this work we will model mixing only between two modes, specifically the $(3,2)$ with the $(2,2)$, as these are the two most strongly coupled modes (the coupling of the $(3,2)$ mode to other $m=2$ modes is in fact suppressed). While for the modes that do not show mixing it is sufficient to model their spherical harmonics, for the $(3,2)$ we will model the spheroidal harmonic, and then transform to the spherical harmonic basis, as discussed in Sec.~\ref{sec:ringdown}. 
For a recent non-spinning model of mode-mixing see
\cite{Mehta:2019wxm}.

A key challenge of accurately modeling multi-mode waveforms is to preserve the relative  time and phase difference between the individual modes, say as measured at the peaks of the modes. In the frequency domain time shifts are encoded in a phase term that is linear in frequency:
the Fourier transformation of a time shifted function 
$h_\tau = h(t - \tau)$ will be given by 
$\tilde h_\tau = \tilde h e^{-i 2 \pi f \tau}$.
In GW data analysis the quality of a model is typically evaluated in terms of how well two waveforms match, up to time shifts and global rotations, e.g. in terms of match integrals.
Adding a linear term in the phase leaves such match integrals invariant.
In order to improve the conditioning of the model calibration it has thus been common for phenomenological frequency domain models to subtract the linear part in frequency before calibrating the model to improve the conditioning of numerical fits, and then add back a linear in frequency term at the end, which approximately aligns the waveforms in time, e.g.~by approximately aligning the amplitude peak at a certain time. This strategy has also been followed in our construction of the \phX model, i.e.~for the $\ell=\vert m\vert= 2$ modes,
whereas for the other modes we directly model a given alignment in time.

More specifically, our strategy of aligning the different spherical harmonic modes in time and phase has been the following: our hybrid waveforms are aligned in time and phase such that the peak of the $\ell=\vert m\vert= 2$ modes of $\psi_4$ in the time domain is located at $t=0$,
and the corresponding phase $\Phi_{22}(\psi_4, t=0) = 0 $, which corresponds to a time $\Delta t = 500 M$ before the end of the waveform. 
For modes with odd $m$ this leaves an ambiguity of a phase shift by multiples of $\pi$. %We do not use the odd-$m$ modes of the numerical waveforms to resolve this ambiguity in order to not depend on the poor quality of many odd mode data sets, and rather resolve the ambiguity in our model by making a smooth choice of the inspiral phase across our parameter space.
We do not use the odd-$m$ modes of the numerical waveforms to resolve this ambiguity in order not to depend on the poor quality of many odd-$m$ data sets, and simply require a smooth transition of the inspiral phase into the merger-ringdown. This ambiguity in the phase will be harmless, as the relative phases among the modes can be unambiguously fixed using post-Newtonian prescriptions, as we will explain in Subsec. \ref{Subsec:insp_phase} below.
Our subdominant modes are calibrated to agree with this alignment of our hybrid waveforms.
The $\ell=\vert m\vert= 2$ mode is aligned a posteriori to the same alignment,
similar to what has been done in previous phenomenological frequency domain models. A difference here is that this a posteriori time alignment is achieved via an additional parameter space fit.

\subsection{Strategy for fitting our model to numerical data} \label{sec:fitting_plan}

As discussed above, following \phX our model is constructed in terms of closed form expressions for the frequency domain amplitude and phase of spherical or spheroidal harmonic modes, which are each split into three frequency regimes.
We will refer to a model for the amplitude or phase for one of the frequency regimes as a {\em partial model}.
We thus need to construct a total of six partial models for each mode. For the inspiral phase, we can use the scaling relation (\ref{eq:tetrad_convention}) for comparable masses, and only need to model the extreme mass ratio case.
For each of the six partial models the ansatz will be formulated in terms of some coefficients, which for example in the inspiral will be the pseudo-PN coefficients that correspond to yet unknown higher post-Newtonian orders.

We thus employ two levels of fits to our input numerical data:
First, for each waveform in our hybrid data set we perform fits of the six partial models for each mode to the data. This yields a set of coefficients for each mode, quantity (amplitude or phase), and frequency interval. We call this first level the direct fit of the model to our data.
Second, we fit each coefficient across the 3D parameter space of mass ratio and component spins. We call this second level the parameter space fit.
In the direct fit we usually collect redundant information: such as the model coefficients, values and derivatives at certain frequencies, and other quantities.
This redundancy provides for some freedom when
reconstructing the model waveform after the parameter space fits. We make extensive use of this freedom when tuning our model, while the final model uses a particular reconstruction, which is what will be described below.

Fitting the coefficients of a particular partial model across the parameter space may not turn out to be a well-conditioned procedure, e.g.~for the inspiral the pseudo PN coefficients have alternating signs, the PN series converges slowly, and different sets of PN coefficients
can yield very similar functions. We will thus sometimes transform the set of coefficients we need to model to an alternative representation, in particular collocation points, following
\cite{Husa:2015iqa,Khan:2015jqa,phenX}. In this approach, one constructs fits for the values of the amplitude and phase at specific frequency nodes. The coefficients of the phenomenological ansätze are then obtained by solving linear systems that take such values as input. This method has been adopted to avoid fitting directly the phenomenological coefficients, which would result in a worse conditioned problem. We thus fit the values or derivatives at certain frequencies, and use the freedom in reconstructing the final phase or amplitude in tuning the model as mentioned above.

In order to perform the 3D parameter space fits in symmetric mass ratio $\eta$ and the two black-hole spins $\chi_{1,2}$ we use the hierarchical fitting procedure described in \cite{Jimenez-Forteza:2016oae}, which we have also used for the underlying \phX model.
The goal of this procedure is to avoid both underfitting
and overfitting our data set. In order to simplify the problem we split the 3D problem into a hierarchy of 
 lower dimensional fits to some particular subsets of all data points. For each lower dimensional problem it is significantly easier to choose an ansatz that avoids  underfitting and overfitting, and finally we combine the lower-dimensional fits into the full 3D fit and check the global quality of the fit. In order to check fit quality we compute residuals and compute the RMS error, 
 and we employ different information criteria to penalize models with more parameters
 as discussed in \cite{Jimenez-Forteza:2016oae} as an approximation to a full Bayesian analysis.
 
 As a first step of our hierarchical procedure we perform a 1-dimensional fit for non-spinning subspace, choosing the symmetric mass ratio $\eta$ as the independent variable.
 We can then identify two further natural one-dimensional problems: First, for the extreme mass ratio case we can neglect the spin of the smaller black hole, and consider the mass ratio as a scaling parameter, and we thus consider a one-dimensional problem in terms of the spin of the larger hole. Second, at fixed mass ratio we can fix a relation between the spins. For quantities such as the final spin and mass, or the coefficients of the \phX model for the $(2,2)$ mode, it has been natural to consider equal black holes, e.g. equal mass and equal spin for this one-dimensional problem.
 
 It is then useful to express the results for the one-dimensional spin fits in terms of a suitably chosen effective spin, such as 
 \begin{equation}\label{def:chieff}
    \chi_{\mathrm{eff}}=\frac{m_1 \chi_1 + m_2 \chi_2}{m_1 + m_2}, 
 \end{equation}
 which is typically measured in parameter estimation (see e.g.~\cite{LIGOScientific:2018mvr}), and which has also been the choice in the early phenomenological waveform models \phB \cite{Ajith:2009bn} and \phC \cite{Santamaria:2010yb}.
 A judicious choice of effective spin parameter can minimize the errors when approximating functions of the 3D parameter space by functions of $\eta$ and effective spin, and can be sufficient for many applications, since spin-differences are a sub-dominant effect.
 For \phD \cite{Husa:2015iqa,Khan:2015jqa} two effective spin parameters have been used:
 For the inspiral calibration to hybrid waveforms (\ref{def:chieff}) has been augmented by extra terms motivated by post-Newtonian theory \cite{Ajith:2011ec}
 \begin{equation}\label{def:chiPN}
     \chi_{PN} =  \chi_{\mathrm{eff}} - \frac{38 \eta}{113}\left(\chi_1 + \chi_2\right) 
 \end{equation}
The final spin and mass, and thus the ringdown frequency have been fit to numerical data in terms of the rescaled total angular momentum of the two black holes 
 \begin{equation}
    \hat S = \frac{m_1^2 \chi_1 + m_2^2 \chi_2}{m_1^2 + m_2^2}.
 \end{equation}

We model the dominant spin-contributions to the amplitude as functions of $\chi_{PN}$ during the inspiral and as functions of $\hat S$ during the merger-ringdown.
For the phase inspiral we use the scaling relation (\ref{eq:tetrad_convention})
for comparable masses, such that a phase inspiral calibration is only necessary
for large mass ratios, which we treat in the same way as other phase coefficients, where we again use $\hat S$.
We will denote a generic effective spin parameter by $\chi$ in general equations involving the effective spin, implying that this refers either to $\chi_{PN}$ or $\hat S$ as appropriate.

We thus perform three
one-dimensional fits, one for the non-spinning sub-space (depending on $\eta$), one for equal black holes (depending on $\chi$), and one for extreme mass ratios (again depending on $\chi$).
Then a 2D ansatz depending on $\eta$ and $\chi$ is built such that it reduces to the 1D fits for those particular cases. The 2D fit is then performed for all data points and from it we get the best-fit function of $\eta$ and $\chi$ minimizing the sum of squared residuals.
 
In order to extend the hierarchical method to the full 3D parameter space, a second spin parameter needs to be chosen, which incorporates spin difference effects. For small spin difference effects, we can simply choose
 \begin{equation}
    \Delta \chi = \chi_1 - \chi_2
 \end{equation}
 without worrying about a particular mass-weighting of the spins, since differences in mass-weighting could be absorbed into higher order terms.
 The spin difference effects are then modelled with a function $f_{\Delta\chi}(\eta)$ as a term
  \begin{equation}
    f_{\Delta\chi}(\eta) \Delta \chi,
 \end{equation}
and the modelling of small spin difference effects is reduced to the 1-dimensional problem of fitting a function of $\eta$. Note that this is reminiscent of the structure of post-Newtonian expansions (see for instance \cite{PNamps}), where spin-difference effects are usually described in terms of the anti-symmetric spin combination $\chi_a=(\chi_1-\chi_2)/2$.
For larger spin difference effects, we will however need higher order terms, and in \cite{Jimenez-Forteza:2016oae}
a term quadratic in $\Delta \chi$ and a term proportional to 
$\chi \, \Delta\chi$ were included, and again these terms can be modelled as one-dimensional problems in terms of functions of $\eta$.

Extensions to this procedure are needed to model the behaviour of sub-dominant modes, in particular for the $(3,2)$ mode and for odd $m$ modes. For the $(3,2)$, we need to model mode mixing in the ringdown, as discussed briefly above, and in detail in Sec.~\ref{sec:ringdown}. While this requires a transformation from spheroidal to spherical harmonics, it does not directly affect our strategy for carrying out the direct fits and the parameter space fits. For odd $m$ modes, changes are required due to the change of sign in the amplitude when rotating by an angle of $\pi$, see eq.~(\ref{eq:mode_rotation}), corresponding to interchanging the two black holes.
For even $m$ modes, rotations by $\pi$ correspond to the identity, and as for \phX it is natural to work with non-negative gravitational-waveform amplitudes. 
For odd $m$ modes however, restricting the amplitude to positive values will make it a non-smooth function in the two-dimensional spin parameter space, where the amplitude $A_{\ell m}(f)$, as defined through Eq.\,(\ref{eq:amp_and_phase}), corresponds to the absolute value of a function that can change sign. 

\begin{figure}[ht]
  \begin{center}
    \includegraphics[width=0.8\columnwidth]{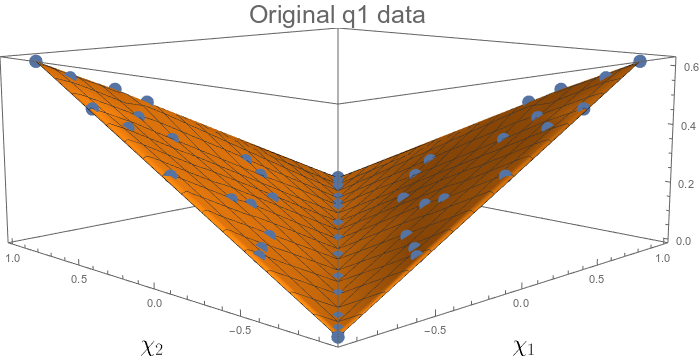} 
    \includegraphics[width=0.8\columnwidth]{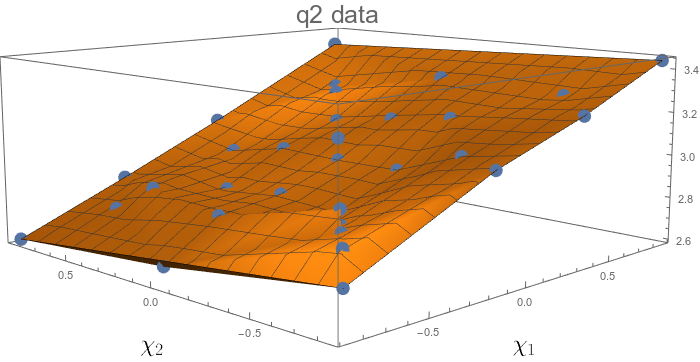}
    \includegraphics[width=0.8\columnwidth]{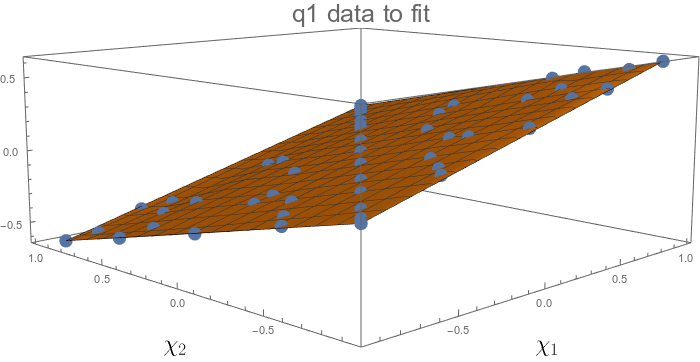}
    \end{center}
    \caption{Example of how the amplitude data set is modified for the parameter space fit. The data shown correspond to the first intermediate collocation point of the $(3,3)$ mode. One of the leaves of the original $q=1$ data (top plot) is flipped in sign so we get a flat surface that is easier to fit. The leaf to be flipped is chosen such that the final behaviour is consistent with the data for other mass ratios, e.g. $q=2$  (intermediate plot). After the fit we take the absolute value of the final fit to return back a positive amplitude.}
    \label{fig:q1flip}
\end{figure}

This can be best understood by plotting the values of the amplitude at a collocation point as a function of the two BH's spins, for a given mass ratio (see Fig.~\ref{fig:q1flip}). It can be seen that the two-dimensional data in the spin parameter space exhibit a crease along a line which corresponds to a vanishing amplitude.
For equal masses, this line appears for equal spin systems,
as shown in the top panel. In order to work with smooth surfaces, we allow the amplitude to take negative values when fitting our numerical data. Such sign flips occur at the level of the full gravitational-wave strain and we choose to fold them into the amplitude for mere convenience, to simplify the fitting procedure. Since we require our reconstructed amplitudes to be non-negative functions, we then take the absolute value of the fits, whenever we allow such sign flips (see Eq.~\ref{eq:amp_and_phase} and discussion therein). 
Notice that, for sufficiently large mass ratios, the spin dependence of the sign of the phase can be neglected, we thus choose which part of the crease we flip in sign to be consistent with the behaviour for higher mass ratios data.

When modelling the amplitude of odd $m$ modes we can not use equal masses as one of our one-dimensional fitting problems, since the amplitude vanishes there, and instead we use a different mass ratio, typically $q=3$. Applying appropriate boundary conditions for equal black hole systems is then essentially straightforward -- we simply demand that the amplitude vanishes.
Setting appropriate boundary values for odd mode phases
for equal black hole is however a complicated problem, since the phase will in general not vanish as one takes the limit toward the boundary. The numerical data typically become very noisy and inaccurate for modes with very small amplitude, and thus one can not in general expect to model odd $m$ modes for close-to-equal black holes with small relative errors.
%
%To make it smooth and easy to fit we just extend one of the halves of the leaf to be negative. There is an ambiguity when choosing which leaf we do negative and we take the one that makes the behaviour similar to the one of the higher mass ratios data. 
When building a parameter space ansatz for odd mode amplitudes we are adding a minus sign when exchanging the spins, the non-spinning and effective-spin parts of the ansatz must be manifestly set to zero because they pick up a minus sign when exchanging the BHs, while at the same time they are invariant by symmetry. We implement this
 by adding a multiplicating factor $\sqrt{1-4\eta}\:$ that cancels these parts for equal mass  systems. The even modes behave in the opposite way: since when exchanging the two spins they remain the same, it is the spin-difference part which has to vanish because it would introduce a minus sign.

%%%%%%%%%%%%%%%%%%%%%%%%%%%%%%%%%%%%%%%%%%%%%%%%
\section{Calibration data set} \label{sec:input}
%%%%%%%%%%%%%%%%%%%%%%%%%%%%%%%%%%%%%%%%%%%%%%%%

Our input data set coincides with the data we have used for the \phX model \cite{phenX}, and
comprises a total of 504 waveforms: 466 for 
comparable masses (with $1\leq q \leq 18$) and 38 for extreme mass-ratios (with $q=\{200,1000\})$.    
These waveforms are ``hybrids'', constructed by appropriately gluing a computationally inexpensive inspiral waveform to a computationally expensive waveform, which covers the late inspiral, merger and ringdown (IMR).
For comparable masses, the $\ell=\vert m\vert$ inspiral is taken
from the SEOBNRv4 (EOB) model \cite{Taracchini:2014zpa}, and the subdominant modes are constructed from the phase of the $\ell=\vert m\vert$ mode and post-Newtonian amplitudes as described in \cite{hybrids} along with other details of our hybridization procedure. The IMR part of the waveform is taken from numerical relativity simulations summarized below.

For extreme mass ratios the IMR part is taken from
numerical solutions of the perturbative Teukolsky
equation, and the inspiral part is taken from a consistent EOB description, as discussed below in
Sec.~\ref{subsec:emr_waveforms}.

Due to the poor quality of many of the numerical relativity waveforms, and the fact that our extreme mass ratio waveforms are only approximate perturbative solutions, we do not use all of the waveforms of our input data set for the calibration of all the quantities we need to
model across the parameter space. Already for
\phX (see \cite{phenX}) we had to carefully select outliers, which lacked sufficient quality for model calibration. Higher-modes waveforms are typically even noisier and more prone to exhibit pathological features than the dominant quadrupolar ones. This can result in a large number of outliers in the parameter-space fits, which can introduce unphysical oscillations in the fit surfaces.  To attenuate this problem, we developed a system of annotations that stores relevant information about the quality of all the waveforms in the calibration set. 
%The database is conceived in a modular way, and it is therefore possible to retrieve a list of potential outliers for each fitting region. 
A careful analysis of data quality is needed, separately for each quantity that we fit, such as the value of the amplitude (phase) at a given collocation point for each particular mode. We will not document these procedures in detail, instead, we will discuss outliers in Sec.~\ref{sec:quality}, where we evaluate the quality of our model by comparing to the original hybrid data. We will see that this comparison has less stringent quality criteria: pathologies which prohibit the use of a some waveform accurate fit for a particular coefficient may in the end not significantly contribute to the waveform mismatch. We will thus only discuss those waveforms which we excluded from
the model evaluation, because of doubts in their quality.

\subsection{Numerical relativity waveforms} \label{subsec:nr_waveforms}

\begin{figure}[ht]
  \begin{center}
    \includegraphics[width=\columnwidth]{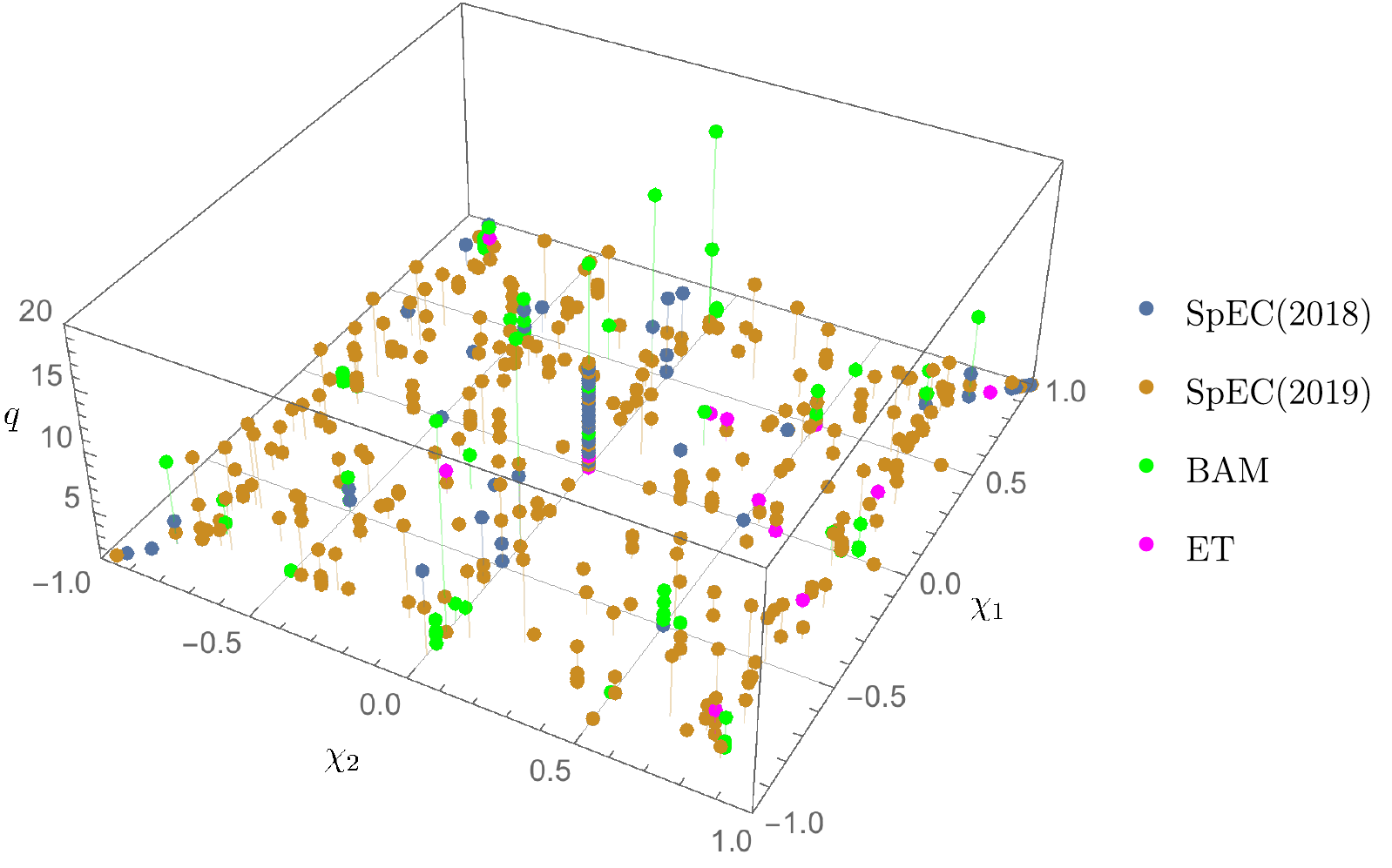} 
    \end{center}
        \caption{Mass ratios and Kerr parameters for the comparable-mass cases in our waveform catalogue. We also indicate the NR codes used to carry out the simulations.}
        \label{fig:NR_parspace}
\end{figure}

The NR simulations used in this work were produced using three different codes to solve the Einstein equations: for the amplitude calibration we used 186 waveforms \cite{SXS:catalog} 
from the public SXS collaboration catalog, as of 2018 \cite{SXS:code} obtained with the SpEC code \cite{PFEIFFER2003253},
95 waveforms
\cite{Husa:2015iqa,Jimenez-Forteza:2016oae}
obtained with the BAM code \cite{Bruegmann2008,Husa:2007rh}, and 16 waveforms from simulations we have performed with the Einstein-Toolkit \cite{L_ffler_2012} code. After the release of the latest SXS collaboration catalog \cite{Boyle:2019kee}, we extended the data set to include 355 SpEC simulations, and updated the parameter-space fits for the phase accordingly. We chose not to update the amplitude fits, as their recalibration was expected to have a smaller effect on the overall quality of the model waveforms. 
The parameters of our waveform catalogue are
visualized in Fig.~\ref{fig:NR_parspace}.
Note that to the same set of $(\eta,\chi_1,\chi_2)$ can correspond multiple NR waveforms: this allows to compare at the same point in parameter-space different resolutions and/or numerical codes and it is therefore important for data-quality considerations. 
A detailed list of the waveforms we have used can be found in our paper on the hybrid data set \cite{hybrids}.

%%%%%%%%%%%%%%%%%%%%%%%%%%%%%%%%%%%%%%%%%%%%%%%%%%
\subsection{Extrapolation to the test-mass limit}
\label{subsec:emr_waveforms}
%%%%%%%%%%%%%%%%%%%%%%%%%%%%%%%%%%%%%%%%%%%%%%%%%%

Due to the computational cost of high mass-ratio simulations, our catalogue of NR waveforms extends only up to $q=18$, which would leave the test-particle limit of our model poorly constrained.  Here, following \cite{Keitel:2016krm}, we chose to pin down the large-q boundary of the parameter-space using Extreme Mass-Ratio Inspiral (EMRI) waveforms. We produced two sets of such waveforms, one with $q=200$ and the other with $q=10^3$. The spin on the primary spans the interval $[-0.9,0.9]$ in uniform steps of $0.1$, while the secondary is assumed to be non-spinning. The waveforms in the calibration set were generated by hybridizing a longer inspiral EOB waveform with a shorter numerical waveform computed using the code of Ref. \cite{Harms_2014}. The code solves the $2+1$ Teukolsky equations for perturbations that can be freely specified by the user.
In our case the gravitational perturbation was sourced by a particle governed by an effective-one-body dynamics, with radiation-reaction effects included up to 6PN order for all the multipoles modelled in this work \cite{EOB_RadiationReaction}. The EOB and Teukolsky waveforms, being both extracted at future null infinity, can be consistently hybridized, following the same hybridization routine used for comparable-mass cases. 

% \sascha{Final wording on which Teukolsky waveforms we have used depends on what plots we show about 
% the consistency (or lack thereof) between comparable masses and EMR.}

%%%%%%%%%%%%%%%%%%%%%%%%%%%%%%%%%%%%%%%%%%%%%%%%%%%%%%%%

%%%%%%%%%%%%%%%%%%%%%%%%%%%%%%%%%%%%%%%%%%%%%%%%%%%%%%%%
\section{Inspiral model} \label{sec:inspiral}

In the inspiral region we work under the assumption that the SPA approximation (see e.g.~\cite{Damour:2000gg} and our discussion in the context of \phX \cite{phenX}) is valid. The frequency domain strain of each mode will therefore take the form
\begin{align}
\label{eq:PNstrain}
\tilde h_{\ell m}(f)=A_{\ell m}\sqrt{\frac{2\pi}{m \ddot{\phi}}}e^{i \left(2\pi f t_c-\phi_{\ell m}-\pi/4+\psi_0\right)}:=A_{\ell m}^{\mathrm{SPA}}e^{i \Psi_{\ell m}},
\end{align}
where $t_c$ is a time shift parameter, $\ddot\phi$ is the second derivative respect to time of the orbital phase (expressed as a function of the frequency), $\psi_0$ is an overall phase factor that depends on the choice of tetrad conventions and 
\begin{align}
A_{\ell m}^{\mathrm{SPA}}:&=A_{\ell m}\sqrt{\frac{2\pi}{m \ddot{\phi}}}, \\
\Psi_{\ell m}:&=2\pi f t_c-\phi_{\ell m}-\pi/4+\psi_0.
\label{eq:psiLM}
\end{align} 
Notice that, in our tetrad-convention, $\psi_0=\pi$ (see Eq.\,(\ref{eq:tetrad_convention}) and discussion therein). 

Furthermore, we will assume that we can work within the post-Newtonian framework, and that we can model currently unknown higher-order terms in the PN-expansions with NR-calibrated coefficients.
Let us stress that in \phXHM the phase and amplitude are treated in different ways: while the latter is fully calibrated to NR, the former is built with a reduced amount of calibration, as we will explain below.

Following the approach taken in \phX, we set the end of the inspiral region around the frequency of the MECO (minimum energy circular orbit), as defined in Ref. \cite{Cabero:2016ayq}. For the amplitude, the default end-frequency of the inspiral is taken to be:
\begin{align}
\label{eq:fcutIns}
f_{\mathrm{Ins}}^{\ell m} = 
\frac{m}{2}\left(f^{22}_{\mathrm{ MECO}} \, \epsilon^{\ell m}_{\mathrm{Ins}}(\eta,\chi_{eff})+|f^{22}_{\mathrm{ISCO}}-f^{22}_{\mathrm{MECO}}| \, \delta^{\ell m}_{\mathrm{Ins}}(\eta,\chi_{eff})\right),
\end{align} 
where $f^{22}_{\mathrm{ISCO}}$ and $f^{22}_{\mathrm{MECO}}$ are the gravitational-wave frequency of the $22$-mode evaluated at the ISCO (innermost stable circular orbit) and MECO respectively. 
The mode-specific expressions for the functions $\delta^{\ell m}_{\mathrm{Ins}}$ and $\epsilon^{\ell m}_{\mathrm{Ins}}$ are given in Tab. \ref{tab:deltaInsp}. In the large-mass-ratio regime, the transition frequency of Eq.\,(\ref{eq:fcutIns}) is replaced by that of a local maximum in the amplitude of $\psi_4$, as we explain in Subsec.\,\ref{subsec:emr_inter_amp} below. 

While the fully NR-calibrated amplitude requires careful tuning of the above transition frequencies, we find that for the phase we can simply set $f_{\mathrm{Ins}}^{\ell m}=\frac{m}{2}f^{22}_{\mathrm{MECO}}$.

The start frequencies of our hybrid waveforms \cite{hybrids} are set up such that
$Mf \geq 0.001453 \:m/2 $ for comparable masses, and 
$Mf \geq 0.001872 \:m/2 $ for extreme-mass-ratio waveforms (depending on the spherical harmonic index $m$). We start the amplitude calibration at a higher minimum frequency of $f_{\mathrm{min}}=0.002 \:m/2 $
to avoid contamination from Fourier transform artefacts. Note that a higher cutoff frequency was chosen for \phX due to the higher accuracy requirements in that case. For the phase, only a small and simple (linear) correction term (\ref{eq:deltaphiIns}) is calibrated, and the same minimum frequency cutoff is applied in this case.

\begin{table}[h!]
\caption{Explicit expressions for the coefficients $\delta^{\ell m}_{\mathrm{Ins}}$ and $\epsilon^{\ell m}_{\mathrm{Ins}}$ entering the inspiral cutting frequencies of the amplitude reconstruction, according to the notation of Eq.~(\ref{eq:fcutIns}). $f^{\ell m}_{ring}$ is the fundamental quasi-normal mode frequency of the $(\ell,m)$ mode.}
\label{tab:deltaInsp}
\begin{center}
{\renewcommand{\arraystretch}{1.5}
\begin{tabular}{| c || c | c | c | c |}
\hline
 &\multicolumn{2}{c|}{Amplitude} \\
\hline
$(\ell m)$      & $\delta_{\mathrm{Ins}}^{\ell m} $    & $\epsilon_{\mathrm{Ins}}^{\ell m}$  \\   
\hline                        
$21$     & $ \left(3/4-0.235 \:\chi_{\mathrm{eff}} - 5/6 \:\chi_{\mathrm{eff}}^2 \right) \:$  & 1. \\
\hline
$33$ & $ \left(3/4-0.235 \:\chi_{\mathrm{eff}} - 5/6 \:\chi_{\mathrm{eff}} \right) \:$ & 1.  \\
\hline
$32$ & $ \left(3/4-0.235 \:\vert\chi_{\mathrm{eff}}\vert  \right) \:  \frac{f^{32}_{ring}}{f^{32}_{ring}}$ & $\frac{f^{32}_{ring}}{f^{22}_{ring}}$ \\
\hline
$44$ & $ \left(3/4-0.235 \:\chi_{\mathrm{eff}} \right) \: $ & 1. \\
\hline
\end{tabular}
}
\end{center}
\end{table}

\subsection{Amplitude} \label{sec:ins_amp}

In the inspiral region, the amplitude ansatz of \phXHM augments a post-Newtonian expansion 
with terms up to 3PN oder with three NR-calibrated coefficients, 
which correspond to higher-order PN terms. 
A 3PN-order expansion for the Fourier domain amplitudes is computed in \cite{PNamps}. 
% We found however significant discrepancies with our numerical data, which we resolved by re-deriving the Fourier domain
% amplitudes from the time domain expressions given in \cite{Arun:2008kb}, see appendix \ref{appendix:FPN},
% which lists the expressions we use in this work.
However, some of these expressions exhibit significant discrepancies with our numerical data and with analogous expressions we derived independently. The authors of \cite{PNamps} acknowledged a mistake in their derivation and agreed with our results. We gather the correct 3PN-order Fourier domain amplitudes in appendix \ref{appendix:FPN} and advise the reader to use the expressions reported there.
%The original time domain amplitudes contain spin terms up to 2PN order, and non-spinning terms up to 3.5PN order.

%Using the notation of Eq.~(\ref{eq:PNstrain}), at low-frequency one has
At low frequency, the
leading order post-Newtonian behavior of the Fourier-amplitude of the $(2,2)$ mode is
\begin{align}\label{eq:AlmScaling}
%A_{\ell m}^{\mathrm{SPA}}\propto 
A_{22}^{0}:= \pi \sqrt{\frac{2\eta}{3}}\left(\pi f\right)^{-7/6},
\end{align} 
while the higher modes present milder divergences. 
Such a divergent behaviour is expected to negatively impact the conditioning of our amplitude fits. Therefore, we do not model the SPA amplitudes directly, but similarly to \phX we rather work with the quantities 
\begin{align}
\mathcal{H}_{\ell m}:=\frac{|A_{\ell m}^{\mathrm{SPA}}|}{A_{22}^0},
\end{align} 
which are non-negative by construction and non-divergent in the limit $f\rightarrow 0$.
%Note that the leading power law in $A_{\ell m}^{\mathrm{SPA}}$ for a given $(\ell, m)$ depends on the spin, it is for this reason that we normalize with the amplitude of the 22-mode.
\subsubsection{Default reconstruction}
Currently the highest known PN-term in the expansion of the $\mathcal{H}_{\ell m}$ is proportional to $f^{2}$. In order to model currently unknown higher-order effects, we introduce up to three pseudo-PN terms $\{\alpha, \beta, \gamma\}$ that depend only on the intrinsic parameters of the source, i.e. mass ratio and spins. The ansatz employed for the inspiral amplitude is given by 
\begin{equation}
\label{eq:insp_amp}
\mathcal{H}_{\ell m}(f)=\frac{|A_{\ell m}^{PN}(f)|}{A_{22}^0(f)} + \alpha \left(\frac{f}{f^{\mathrm{Ins}}_{\ell m}}\right)^{\frac{7}{3}} + \beta \left(\frac{f}{f^{\mathrm{Ins}}_{\ell m}}\right)^{\frac{8}{3}} + \gamma \left(\frac{f}{f^{\mathrm{Ins}}_{\ell m}}\right)^{\frac{9}{3}}.
\end{equation}  
%During the direct fit, the values of the coefficients $\{\alpha, \beta, \gamma\}$ are obtained across parameter space however, 
Following \cite{Khan:2015jqa}, we do not perform parameter-space fits of $\{\alpha, \beta, \gamma\}$.
%we do not calibrate $\{\alpha, \beta, \gamma\}$ during the direct fit: 
Instead, we compute parameter-space fits of the hybrids' amplitudes evaluated at three equispaced frequencies  $\left[0.5 \: f_{\ell m}^{\mathrm{Ins}}, 0.75\: f_{\ell m}^{\mathrm{Ins}},f_{\ell m}^{\mathrm{Ins}}\right]$, which we refer to as ``collocation points''.

We require that the reconstructed inspiral amplitude
%  the ansatz of Eq.~(\ref{eq:insp_amp}) 
 go through the three collocation points given by the parameter-space fits. This yields a system of three equations that can be solved to obtain the values for $\{\alpha, \beta, \gamma\}$.
%By default, the inspiral amplitude is reconstructed by requiring that the ansatz of Eq.~(\ref{eq:insp_amp}) passes through the three collocation points given by the parameter space fits. 
We observe however, that in some regions of
parameter space this leads to oscillatory behaviour of the reconstructed amplitude for the $(2,1)$ and $(3,2)$ modes, and a lower order polynomial, calibrated to a smaller number of collocation points, gives more robust results. 
This problem arises in regions of the parameter space where the model is poorly constrained due to the lack of NR simulations, such as in cases with very high positive spins (where in addition the correct functional form is rather simple and a higher order polynomial is not required), and where the amplitude of the waveform is very small (see e.g. Fig.~\ref{fig:small21amp}). For this reason we apply a series of vetoes that remove collocation points and allow for a smooth reconstruction, as discussed next.

\subsubsection{Vetoes and non-default reconstruction}\label{subsubsec:insp_vetoes}
The removal of a collocation point implies the modification of the ansatz used in the reconstruction. For each collocation point removed, we set to zero one of the coefficients of the pseudo-PN terms, starting from the highest order one. The removal of the three collocation points would lead us to an ansatz without any pseudo-PN term.

For the $(2,1)$ mode, when $q<8$ we remove the collocation points which gives a Fourier domain amplitude $|\tilde{h}_{21}|$ below $0.2$ (in geometric units), which is a typical value for the ringdown for comparable masses. Furthermore, we check whether the amplitude values at the three collocation points form a monotonic sequence, otherwise we remove the middle point to avoid oscillatory behaviour. 

For the $(3,2)$ mode we drop the middle collocation point if it is not consistent with monotonic behaviour. In addition, for this mode we have isolated two particular regions of the parameter space where 
we drop collocation points from our reconstruction due to the poor quality of the reconstruction. The first one is given by $q>2.5$, $\chi_1<-0.9$, $\chi_2<-0.9$, where we do not use any of the collocation points and just reconstruct with the PN ansatz. The second is given by $q>2.5$, $\chi_1<-0.6$, $\chi_2>0$ where we just remove the highest frequency collocation point.
For the 33 mode we remove the last collocation point in the region $q \in (1,1.2)$, $\chi_1<-0.1$, $\chi_2>0$.
%If the case does not belong to any of these regions then we check for the wavy behaviour of the middle point as it is done for the $(2,1)$ mode. 

In the future we will revisit this problem when recalibrating the amplitude against the recently released new SXS catalogue of NR simulations \cite{Boyle:2019kee}, which we expect to mitigate some of the issues we observe.

\subsection{Phase} 
\label{Subsec:insp_phase}
\begin{figure*}[ht]
  \begin{center}
    \includegraphics[width=2\columnwidth]{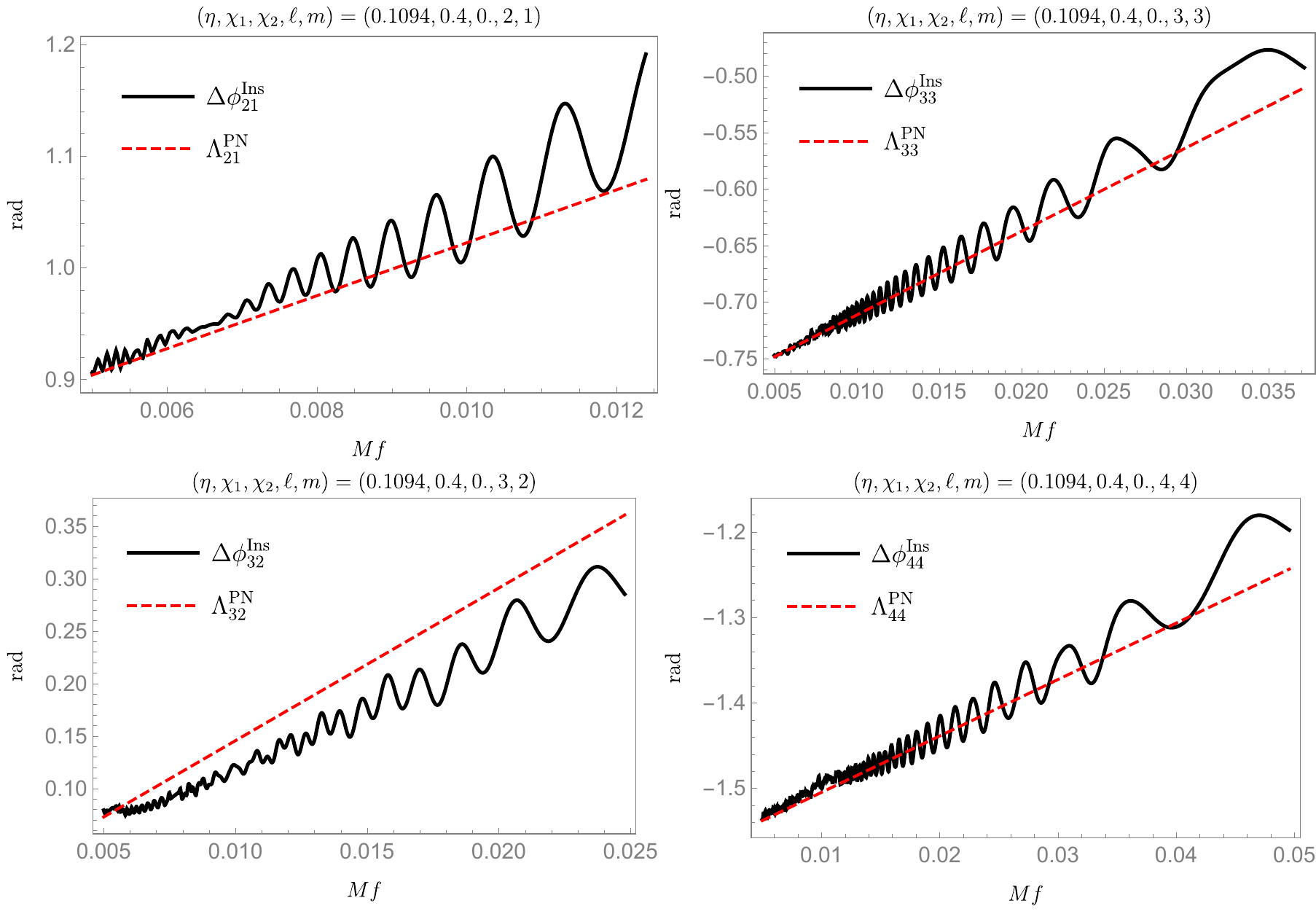}
    \caption{We compare the quantities  $\Lambda_{\ell m}^{PN}$ defined in Eq.\,(\ref{lambdaPN_eq}) (red dashed lines), with the $\Delta\phi_{\ell m}^{\mathrm{Ins}}$ computed from a Fourier transformed hybrid waveform with parameters $(\eta,\chi_1,\chi_2)=\left\{0.1094, 0.4, 0\right\}$ (black solid lines). Each plot is truncated around the end of the inspiral region corresponding to the selected mode.}
    \label{fig:lambdas}
  \end{center}
\end{figure*}

%According to PN, the individual multipoles of the inspiral waveform can be written in frequeny domain as
% \begin{align}
% \tilde{h}^{\ell m}= \frac{M^2}{D_{L}} \pi \sqrt{\frac{2\eta}{3}}\left(\frac{2 \pi M f}{m}\right)^{-7/6} \mathcal{H}^{\ell m} e^{-\iota m \Psi},
% \end{align}
% where $D_L$ is the luminosity distance of the binary, $M$ its total mass, $\Psi$ its orbital phase computed in the stationary phase approximation and $\mathcal{H}^{\ell m}$ are some (generally complex) amplitudes. 
For the inspiral phase, we start from the consideration that, with good accuracy, the NR data satisfy the relation \cite{Bustillo:2015ova}, 
\begin{align}
\label{inspiral_scaling}
\phi_{\ell m}(f)\approx\frac{m}{2}\phi_{22}
\left(\frac{2}{m} f\right).
\end{align}
As discussed above, our amplitude fits return a real quantity, but one must be mindful that the PN expansions of the $A_{\ell m}^{\mathrm{SPA}}$ are, in general, complex (see, for instance, \cite{Blanchet2014, PNamps}). 
For each mode, we re-expand to linear order in the frequency the quantities
\begin{align}
\label{lambdaPN_eq}
\Lambda_{\ell m}^{PN}=\arctan\left(\Im(A_{\ell m}^{PN})/\Re(A_{\ell m}^{PN})\right).
\end{align}  
%and add them to Eq.(\ref{inspiral_scaling}), obtaining the following ansatz:
%\begin{align}
%\phi_{\ell m}(f)=\frac{m}{2}\phi_{22}^{X}(2/m f)+\Lambda_{\ell m}^{\mathrm{PN}}+\mathrm{const}.
%\end{align}
%where $\phi_{22}^{X}$ denotes the quadrupole's inspiral phase returned by the \phX model. 
We evaluated the quantity $\Delta\phi_{\ell m}^{\mathrm{Ins}}:=\phi_{\ell m}(f)-\frac{m}{2}\phi_{22}(2/m f)$ for all the hybrids in our catalogue and found that 
\begin{align}
\label{eq:deltaphiIns}
\Delta\phi_{\ell m}^{\mathrm{Ins}}\approx \Lambda_{\ell m}^{PN}
\end{align}
In Fig.~\ref{fig:lambdas} we show the behaviour of this approximation for an example case of comparable mass ratio, as compared to the result obtained from the hybrids.

\begin{figure*}[ht]
  \begin{center}
    \includegraphics[width=2\columnwidth]{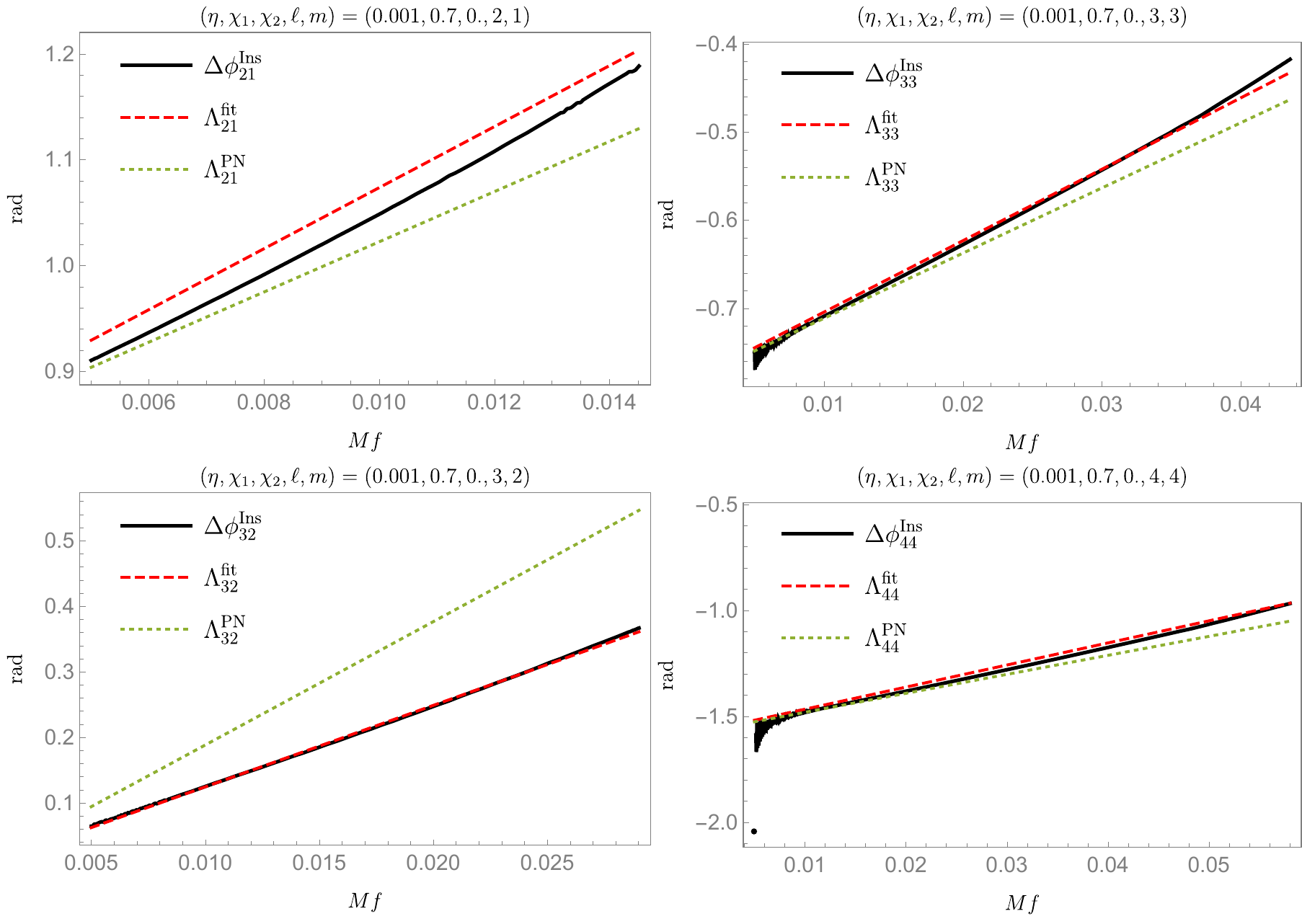}
    \caption{We show the quantities $\Delta\phi_{\ell m}^{\mathrm{Ins}}$ computed from a hybrid FD waveform with parameters $(\eta,\chi_1,\chi_2)=\left\{0.001, 0.7, 0.\right\}$ (black solid lines). Red dashed lines mark our parameter-space fits (corresponding to the $\Lambda_{\ell m}^{\mathrm{fit}}$ of Eq.\,\ref{eq:Lambda_lm}), while dotted green lines the corresponding analytical PN approximations (Eq.\,(\ref{lambdaPN_eq})). Each plot is truncated around the end of the inspiral region corresponding to the selected mode.}
    \label{fig:emrlambdas}
  \end{center}
\end{figure*}

The accuracy of the above approximation tends to degrade for high-mass ratios, high-spins configurations and the PN relative phases are recovered only at sufficiently low frequencies, as illustrated in Fig.\,\ref{fig:emrlambdas}. Therefore, we compute parameter-space fits to capture the leading order behaviour of each $\Delta\phi_{\ell m}^{\mathrm{Ins}}$, and use them to build our final inspiral ansatz in the extreme-mass ratio regime. 

Based on the above discussion, we express the inspiral phase of each multipole as 
\begin{align}
\label{eq:phase_ins_ansatz}
\phi_{\ell m}(f)=\frac{m}{2}\phi_{22}^{X}(2/m f)+\Lambda_{\ell m}(f)+d\phi^{\mathrm{Ins}}_{\ell m} f+\phi^{\mathrm{Ins}}_{\ell m},
\end{align}
where $\phi_{22}^{X}$ is the quadrupole phase reconstructed with \phX, whose coefficients need to be rescaled as detailed in appendix \ref{app:inspiral_hm}, and 
\begin{align}
\label{eq:Lambda_lm}
\Lambda_{\ell m} =
  \begin{cases}
  \Lambda_{\ell m}^{PN} & \text{if $q<100$} \\
  \\                                   
  \Lambda_{\ell m}^{fit} & \text{if $q\geq 100$} 
  \end{cases}
\end{align}
The constant $d\phi^{\mathrm{Ins}}_{\ell m}$ in Eq.~(\ref{eq:phase_ins_ansatz}) is determined by continuity with the intermediate-region ansatz. 

Once a smooth phase derivative, defined with respect to the Fourier variable $f$, has been reconstructed, the remaining constant, $\phi^{\mathrm{Ins}}_{\ell m}$, is fixed by requiring that, at low frequencies, one has
\begin{align}
\lim_{f\rightarrow 0}\left(\Psi_{\ell m}-\frac{m}{2}\Psi_{22}\right) =\frac{3}{4}\pi\left(1-\frac{m}{2}\right),
\label{eq:phase_alignment}
\end{align}
which follows from Eq.\,(\ref{eq:psiLM}) and from our choice of tetrad convention (see also \,(\ref{eq:tetrad_convention}).

\section{Intermediate region} \label{sec:intermediate}

The intermediate region connects the inspiral regime to the ringdown. It is the only region where \phXHM is fully calibrated, both in amplitude and phase. 
While for the amplitude this is the last region to be attached to the rest of the reconstruction, for the phase this is the central piece of the model, to which inspiral and ringdown phase derivatives will be smoothly attached. Physically, this implies that, in \phXHM, the relative time-shifts among different modes are entirely calibrated around merger. 
%Failing to accurately reproduce these time shifts will not change the match between spherical harmonic modes of the hybrid versus the modes of the model, but it will affect matches involving the full waveform polarizations, as discussed in Sec.~\ref{sec:quality}, where we will also directly check the accuracy of preserving the time shifts.
We have found that a good practical way of testing whether time shifts are consistent between modes is to compute the recoil of the merger remnant, which is also of 
astrophysical interest, and which we discuss in Sec.~\ref{subsec:recoil}.

The intermediate region covers the range of frequencies 
\begin{align}
\label{eq:frange_int}
f\in\left[f_{\mathrm{Ins}}^{\ell m},f_{RD}^{\ell m}\right],
\end{align}
where the inspiral cutting frequencies $f_{\mathrm{Ins}}^{\ell m}$ are those of Eq.~(\ref{eq:fcutIns}) and  the ringdown cutting frequencies are defined as:
\begin{align}
\label{eq:fcutRD}
f_{RD}^{\ell m}=\delta^{\ell m}_{RD}f_{ring}^{\ell m}+\epsilon_{RD}^{\ell m}.
\end{align}
In the above equation, $f_{ring}^{\ell m}$ is the fundamental quasi-normal mode (QNM) frequency of the $(\ell,m)$ mode and the default coefficients $\delta^{\ell m}_{RD}, \epsilon_{RD}^{\ell m}$ are given in Tab.\,\ref{tab:fcutRD}. We have computed fits for the real and imaginary parts of the QNMs as a function of the final dimensionless spin, based on publicly available data \cite{bertiQNMs}, see also \cite{Berti:2009kk}. Notice that, in the $(3,2)$ reconstruction, high-mass, high-spin cases require an adjustment of the default cutting frequencies, as we explain in Subsecs.\,\ref{subsec:amp_intermediate} and \ref{subsec:phase_intermediate} below. 

The rationale behind our choice of cutting frequencies is simple: in general, the QNM frequencies $f_{\mathrm{ring}}^{\ell m}$ mark the onset of the ringdown region and it is therefore natural to terminate our intermediate region slightly before those. This does not apply to the $(3,2)$-mode, where, due to mode-mixing, new features appear in the waveforms already around $f\approx f_{\mathrm{ring}}^{22}<f_{\mathrm{ring}}^{32}$.

In the following subsections, we will describe in more detail our choice of ansätze and collocation points, based on the specific features of our numerical waveforms in the intermediate region. 
 
\begin{table}[h!]
\caption{Choices for the coefficients $\delta^{\ell m}_{RD}$ and $\epsilon^{\ell m}_{RD}$ entering the default ringdown cutting frequencies in Eq.~(\ref{eq:fcutRD}). $f^{\ell m}_{damp}$ is the quasi-normal mode damping frequency of the $(\ell,m)$ mode.}
\label{tab:fcutRD}
\begin{center}
{\renewcommand{\arraystretch}{1.5}
\begin{tabular}{| c || c | c | c | c |}
\hline
 &\multicolumn{2}{c|}{Amplitude} & \multicolumn{2}{c|}{Phase}\\
\hline
$(\ell m)$      & $\delta_{RD}^{\ell m} $    & $\epsilon_{RD}^{\ell m}$   & $\delta_{RD}^{\ell m} $    & $\epsilon_{RD}^{\ell m}$    \\  
\hline                        
$21$     & 0.75  & 0 & 1 & -$f_{damp}^{21}$ \\
\hline
$33$ & 0.95 & 0 & 1 & -$f_{damp}^{33}$ \\
\hline
$32$ &  0 & $f_{ring}^{22}$ & 0 & $f_{ring}^{22}-0.5 f_{damp}^{22}$\\
\hline
$44$ & 0.9 & 0 & 1 & $-f_{damp}^{44}$\\
\hline
\end{tabular}
}
\end{center}
\end{table}

\subsection{Amplitude}
\label{subsec:amp_intermediate}

\subsubsection{Default reconstruction}
In the intermediate frequency regime we model the amplitude
with the inverse of a fifth-order polynomial as
\begin{equation}\label{eq:AmpInter}
\frac{A_{\ell m}^{\mathrm{Inter}}}{A_0} = \frac{1}{\delta_0 + \delta_1 f + \delta_2 f^2 + \delta_3 f^3 + \delta_4 f^4 + \delta_5 f^5},
\end{equation} 
where $A_0 =  \pi \sqrt{\frac{2\eta}{3}}$. This function has six free parameters, which are determined by imposing the value of the amplitude at two collocation points, together with four boundary conditions (two on the amplitude itself, and two on its first derivative), so that the final IMR amplitude is a $C^1$ function. 
We use two collocation points at the equally spaced frequencies $f_{\mathrm{Int}_1}^{\ell m}=f_{\mathrm{Ins}}^{\ell m} + (f_{\mathrm{RD}}^{\ell m}-f_{\mathrm{Ins}}^{\ell m})/3$ and $f_{\mathrm{Int}_2}^{\ell m}=f_{\mathrm{Ins}}^{\ell m} + 2/3(f_{\mathrm{RD}}^{\ell m}-f_{\mathrm{Ins}}^{\ell m})$.
%\begin{align}{
%f_{\mathrm{Int}_1}^{\ell m}&=f_{\mathrm{Ins}}^{\ell m} + (f_{\mathrm{RD}}^{\ell m}-f_{\mathrm{Ins}}^{\ell m})/3,\nonumber\\ f_{\mathrm{Int}_2}^{\ell m}&=f_{\mathrm{Ins}}^{\ell m} + 2/3(f_{\mathrm{RD}}^{\ell m}-f_{\mathrm{Ins}}^{\ell m}).}
%\end{align}

\begin{table}[h!]
\centering
\caption{Conditions imposed to determine the parameters of the fourth-order polynomial used in the pre-intermediate region of the EMR amplitude reconstruction.}
\label{tab:emr_inter_eqs}
{\renewcommand{\arraystretch}{1.5}
\begin{tabular}{|c|c|c|}
\hline
Collocation Points                              & Value                      & Derivative                            \\ \hline
$f_1=f_{\mathrm{Ins}}^{\ell m} $                                 & $v_1 = A^{\rm{Inter}}_{\ell m} (f_1)/A_0$ & $d_1 = (A^{\rm{Inter}}_{\ell m}/A_0)^{\prime} (f_1)$   \\
$f_2=f_{\mathrm{Ins}}^{\ell m}+(f_{\mathrm{Int_1}}^{\ell m}-f_{\mathrm{Ins}}^{\ell m})/3$ & $v_2 = A^{\rm{Inter}}_{\ell m} (f_2)/A_0$ & $d_2 = (A^{\rm{Inter}}_{\ell m}/A_0)^{\prime} (f_2)$                               \\
$f_3=f_{\mathrm{Int_1}}^{\ell m}$                            & $v_3 = A^{\rm{Inter}}_{\ell m} (f_3)/A_0$ &   \\ \hline
\end{tabular}}
\end{table}

\begin{table}[ht]
\centering
\caption{Conditions imposed to determine the parameters of the fifth-order polynomial used in the default intermediate amplitude reconstruction, see Eq.\,(\ref{eq:AmpInter}).}
\label{tab:default_inter_eqs}
{\renewcommand{\arraystretch}{1.5}
\begin{tabular}{|c|c|c|}
\hline
Collocation Points                              & Value                      & Derivative                            \\ \hline
$f_1=f_{\mathrm{Ins}}^{\ell m} $                                 & $v_1 = A^{\rm{Inter}}_{\ell m} (f_1)/A_0$ & $d_1 = (A^{\rm{Inter}}_{\ell m}/A_0)^{\prime} (f_1)$   \\
$f_2=f_{\mathrm{Ins}}^{\ell m} + (f_{\mathrm{RD}}^{\ell m}-f_{\mathrm{Ins}}^{\ell m})/3$ & $v_2 = A^{\rm{Inter}}_{\ell m} (f_2)/A_0$ &                                \\
$f_3=f_{\mathrm{Ins}}^{\ell m} + 2(f_{\mathrm{RD}}^{\ell m}-f_{\mathrm{Ins}}^{\ell m})/3$                            & $v_3 = A^{\rm{Inter}}_{\ell m} (f_3)/A_0$ &   \\ 
$f_4=f_{\mathrm{RD}}^{\ell m}$ &$v_4 = A^{\rm{Inter}}_{\ell m} (f_4)/A_0$ &$d_2 = (A^{\rm{Inter}}_{\ell m}/A_0)^{\prime} (f_4)$
\\\hline
\end{tabular}}
\end{table}

\subsubsection{Extreme-mass-ratio reconstruction} \label{subsec:emr_inter_amp}
For the EMR regime we refine the model to adapt to the steep amplitude drop at the end of the inspiral part, which is associated with the sharp transition from inspiral to plunge for extreme mass ratios. As one would expect, this drop is deeper for very negative spins. We find that the ansatz of Eq.\,(\ref{eq:AmpInter}) is not suited to this regime and we introduce a pre-intermediate region that ranges from the inspiral cutting frequency up to the frequency of the first collocation point. We add an extra collocation point at the frequency $f_{\mathrm{Int_0}}^{\ell m} = f_{\mathrm{Ins}}^{\ell m}+(f_{\mathrm{Int_1}}^{\ell m}-f_{\mathrm{Ins}}^{\ell m})/3$, where we calibrate the value of the amplitude and its derivative. We then use the inverse of a fourth order polynomial to model the amplitude in this new region. The five free coefficients of the polynomial are specified by imposing the conditions listed in Tab.\,\ref{tab:emr_inter_eqs}.

%The 5 free coefficients of the polynomial are specified by the value and derivative of the inspiral at the transition frequency $f_{\mathrm{Ins}}$, by the value and derivative of the new collocation point, and by the value of the first collocation point at $f_{Int1}$. 
We then apply the default reconstruction procedure in the region $f\in\left[f_{\mathrm{Int_0}}^{\ell m},f_{\mathrm{RD}}^{\ell m}\right]$, imposing the conditions listed in Tab.\,\ref{tab:default_inter_eqs}, with the replacement $f_{\mathrm{Ins}}^{\ell m}\rightarrow f_{\mathrm{Int_0}}^{\ell m}$.

\begin{figure}[ht]
\includegraphics[scale=0.5]{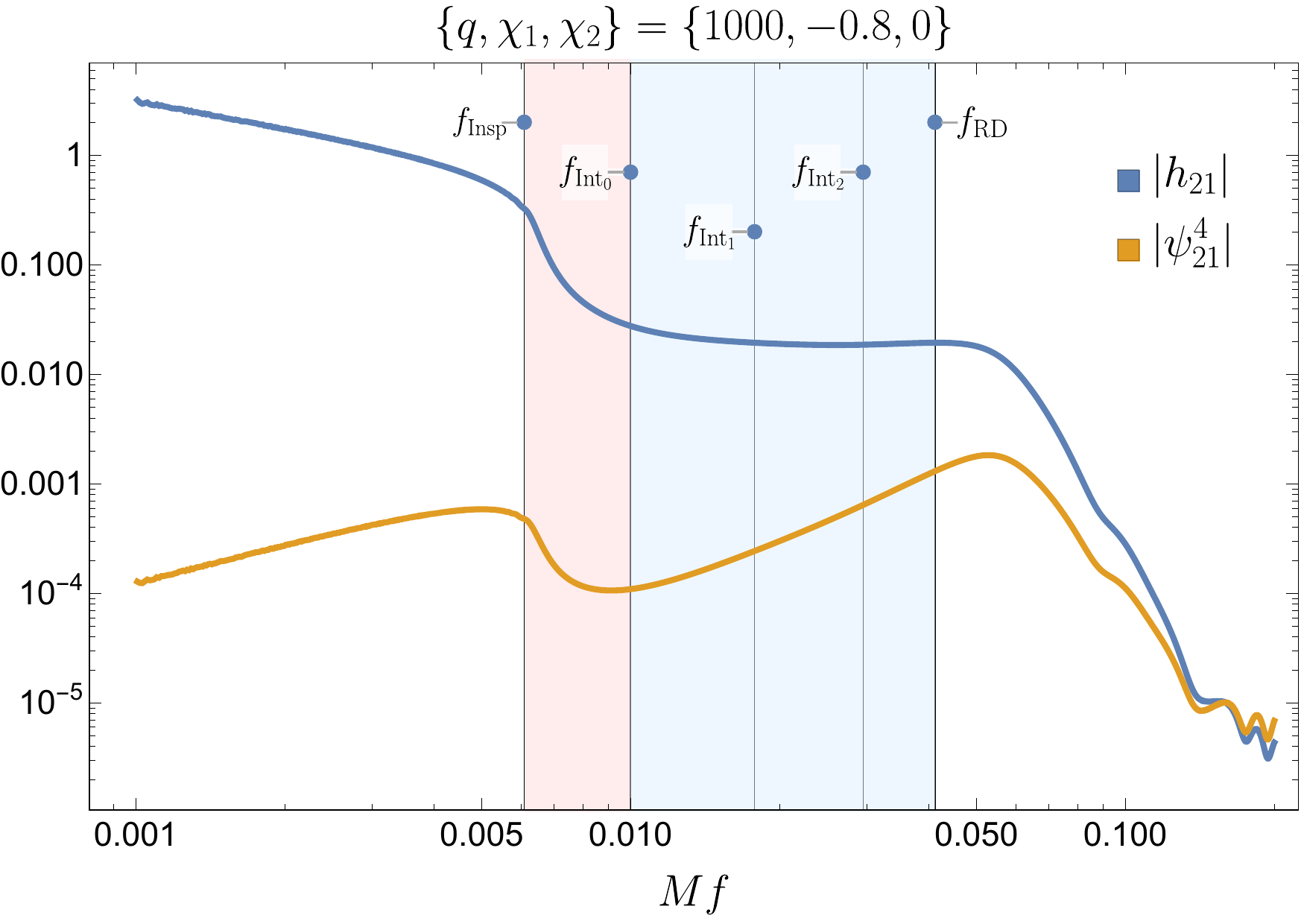}
\caption{$(2,1)$ Fourier domain amplitude of an extreme-mass ratio waveform. The strain amplitude shows a deep drop after the inspiral. This feature is preceded by a local maximum in the amplitude of $\psi_4$. The red-shaded area corresponds to the pre-intermediate region mentioned in the text, where the amplitude is reconstructed using a fourth-order polynomial. The blue-shaded area corresponds to the intermediate region, where the default reconstruction procedure applies.}
\label{fig:StrainvsPsi4}
\end{figure}

The two calibration regions are shown in Fig.\,\ref{fig:StrainvsPsi4}, where have shaded the pre-intermediate region in red and the intermediate one in blue. We find it convenient to terminate the inspiral region just before the amplitude drop. Therefore, we replace the cutting frequency of Eq.\,(\ref{eq:fcutIns}) with that of a local maximum in the amplitude of $\psi_4$, which  always precedes the drop (see Fig. \ref{fig:StrainvsPsi4}). We carried out a fit over the EMR parameter space of the frequency at which this maximum occurs and used it to replace the default inspiral cutting frequency when $q>70$.

%The second deviation from the standard procedure appears when we evaluate the model in some regions outside of the calibration regime where the intermediate collocation points may not be well behaved and they can induce a zero-crossing of the 5th order polynomial.
\subsubsection{Vetoes and non-default reconstruction}
While for comparable masses the ansatz of Eq.\,(\ref{eq:AmpInter}) is well suited to model the intermediate amplitude, in other regions of the parameter space modelling errors can result in a zero-crossing of the fifth order polynomial. We resolve this problem using a strategy akin to that of Subsec.\,\ref{subsubsec:insp_vetoes} above, i.e. by removing collocation points and switching to a lower order polynomial, the minimum order being one. %In the worst case we would reconstruct with the inverse of a first order polynomial that is guaranteed to not cross zero. 
The regions of the parameter space affected are typically those where the amplitude is very small, or the high-spin regime. 
We have also isolated some regions where, due to the poor quality of the reconstruction, we drop both intermediate collocation points. This system of vetoes is summarized in Tab.~\ref{tab:inter_veto}.

We describe now in more detail the adjustments made to the default reconstruction procedure mode-by-mode. A summary of the rules applied can be found in Tab.~\ref{tab:inter_veto_summary}.

For the $(2,1)$ mode, we remove the intermediate collocation points at which the strain Fourier domain amplitude $|\tilde{h}_{21}|$ is below 0.2. This happens when the $(2,1)$ amplitude is very small and in consequence the current accuracy of the parameter space fits is not sufficient. The advantage is that for those cases the $(2,1)$ mode does not contribute significantly to the total waveform and we can afford to simplify the reconstruction. It can be seen from Fig.~\ref{fig:small21amp} that the ratio between the $(2,1)$ and $(2,2)$ amplitude is in some cases well below $1\%$. If the amplitude at the ringdown cutting frequency is below a threshold of $0.01$, we remove the two intermediate collocation points.
%The threshold is chosen such that the total amplitude and that frequency is below 0.01. 
If the intermediate collocation points have passed these preliminary tests, we check whether they form a monotonic sequence, and if not we remove $f_{\mathrm{Int}_{2}}^{\ell m}$. Finally, we apply the parameter-space vetoes indicated in Tab.~\ref{tab:inter_veto}.

For the $(3,2)$ mode, we require that the amplitude at the ringdown cutting frequency is above the same threshold applied to the $(2,1)$ mode. If this condition is not satisfied, we remove the two intermediate collocation points. If it is, we check whether the collocation points form a monotonic sequence. If not, we drop $f_{\mathrm{Int}_{2}}^{\ell m}$. Finally, we apply our set of parameter-space vetoes.%If this condition is met, we check if the two collocation points together with the point at the ringdown cutting frequency have an oscillatory behaviour (as done for the inspiral), if so the middle point is removed. Then we apply the vetoes in table \ref{tab:inter_veto}.

The $(3,3)$ and $(4,4)$ modes are typically less problematic. However, we find that in the high-spin, high-mass-ratio region ($q>7$, $\chi_1>0.95$)
the inspiral is very long and there is a sharp transition to the ringdown, without a specific merger signature. For that reason we remove the two intermediate collocation points and connect inspiral and ringdown with a third-order polynomial. Once again, we apply the vetoes of Tab.\,\ref{tab:inter_veto}.

After applying all the mode-specific vetoes, we check whether the denominator of our polynomial ansatz ever crosses zero in the frequency range of the intermediate region. If so, we lower the order of the polynomial by iteratively relaxing the boundary conditions until we obtain a well-defined ansatz. 
%%%%%

\begin{table}[h!]
\caption{Parameter-space regions where the two intermediate collocation points at $f_{\mathrm{Int_1}}$ and $f_{\mathrm{Int_2}}$ are removed. "Still alive" means if the collocation point has not been removed yet by the previous vetoes.}
\label{tab:inter_veto}
\begin{center}
\begin{tabular}{| c | c | c |}
\hline
$(\ell m)$ &  Region   &   Veto applied if   \\  
\hline                        
$21$ & $\eta<0.23 ~\&~ \chi_1>0.7 ~\&~ \chi_2<-0.5$       & Always \\
     & $q>40 ~\&~ \chi_1>0.9$                         & $f_\mathrm{Int_{1,2}}^{\ell m}$ still alive  \\
\hline
$33$ & $q>40  ~\&~ \chi_1>0.9$                         & $f_\mathrm{Int_{1,2}}^{\ell m}$ still alive \\
\hline
$32$ & $q>2.5  ~\&~ \chi_1<-0.6  ~\&~\chi_2>0$               & Always \\
     & $\chi_1<-0.9 ~\&~\chi_2<-0.9$                  & Always \\ 
     & $q>40 ~\&~ \chi_1>0.9$                         & $f_\mathrm{Int_{1,2}}^{\ell m}$ still alive \\  
\hline
$44$ & $q>40 ~\&~ \chi_1>0.9$                         & $f_\mathrm{Int_{1,2}}^{\ell m}$ still alive\\ 
\hline
\end{tabular}
\end{center}
\end{table}

\begin{table*}[htbp]
%\begin{center}
\caption{Summary of the sanity checks used in the intermediate amplitude reconstruction. The vetoes are sorted in order of application. The coefficient $a_{\lambda}$ will be presented in Eq.~\ref{eq:ringdown_ansatz}.}
\label{tab:inter_veto_summary}
\begin{tabular}{| c | c | c | c |}
\hline
Veto description &  Applied to modes & Region where applied  &  Collocation point removed  \\  
\hline
\hline                        
Amplitude at  $f_\mathrm{Int_{1,2}}^{\ell m}$ $<0.2$ &  21 & $q<8$  &  $f_\mathrm{Int_{1,2}}^{\ell m}$   \\
\hline
Amplitude at $f^{\ell m}_{RD}$ $<0.01$ & 21 & Always  &  $f_{\mathrm{Int_1}}^{\ell m}$ \& $f_{\mathrm{Int_2}}^{\ell m}$   \\
 &  32 & & $f_{\mathrm{Int_1}}^{\ell m}$, $f_{\mathrm{Int_2}}^{\ell m}$ \& derivatives at boundaries \\
\hline
Monotonicity (if $f_{\mathrm{Int_1}}^{\ell m}$ and $f_{\mathrm{Int_2}}^{\ell m}$ &  21, 32 & Always  & $f_{\mathrm{Int_2}}^{\ell m}$  \\have passed the previous checks) &   &   &   \\
\hline
$a_{\lambda}$ badly behaved & 33, 44 & $q>7, \chi_1>0.95$  &  $f_{\mathrm{Int_1}}^{\ell m}$ \& $f_{\mathrm{Int_2}}^{\ell m}$ \\
\hline
Parameter-space vetoes & 21, 33, 32, 44 & see Tab.~\ref{tab:inter_veto} &  $f_{\mathrm{Int_1}}^{\ell m}$ \& $f_{\mathrm{Int_2}}^{\ell m}$   \\
\hline
\hline
\multicolumn{4}{|c|}{Check that the denominator of the resulting ansatz does not cross zero, if so remove derivatives at boundaries}
\\
\hline
\end{tabular}
%\end{center}
\end{table*}
%\end{widetext}

\subsection{Phase} 
\label{subsec:phase_intermediate}

In the intermediate region our most general ansatz for the phase-derivative of each mode reads:
\begin{align}
\label{eq:inter_phase_ansatz}
\frac{d\phi_{\ell m}^{\mathrm{Int}}}{df}&=a_\lambda^{\ell m}\frac{f_{\mathrm{damp}}^{\ell m}}{(f_{\mathrm{damp}}^{\ell m})^2+(f-f_{\mathrm{ring}}^{\ell m})^2}+\sum_{k=0}^{4}\frac{a_k^{\ell m}}{f^{k}}.
\end{align}
For the modes that do not show significant mode-mixing, namely $(2,1),(3,3),(4,4)$, we set $a_{3}^{\ell m}=0$ and retain all the remaining coefficients. This was done for consistency with the \phX model (see Subsec. VII B. of \cite{phenX}). For these modes, we do not impose any boundary condition, which leaves us with a total of five free coefficients, which we determine by solving the linear system
\begin{equation}
\frac{d\phi_{\ell m}^{\mathrm{Int}}}{df}(f_{\ell m}^{i})=\mathcal{F}^{i}_{\ell m}, \quad i \in\left[1,5\right].
\end{equation}
In the above equation, $f_{\ell m}^{i}$ are the frequencies of the intermediate-region collocation points, and $\mathcal{F}^{i}_{\ell m}$ are the values of the phase-derivative evaluated at each $f_{\ell m}^{i}$, as reconstructed through our parameter-space fits. 

\begin{figure}[h!]
  \begin{center}
    \includegraphics[width=\columnwidth]{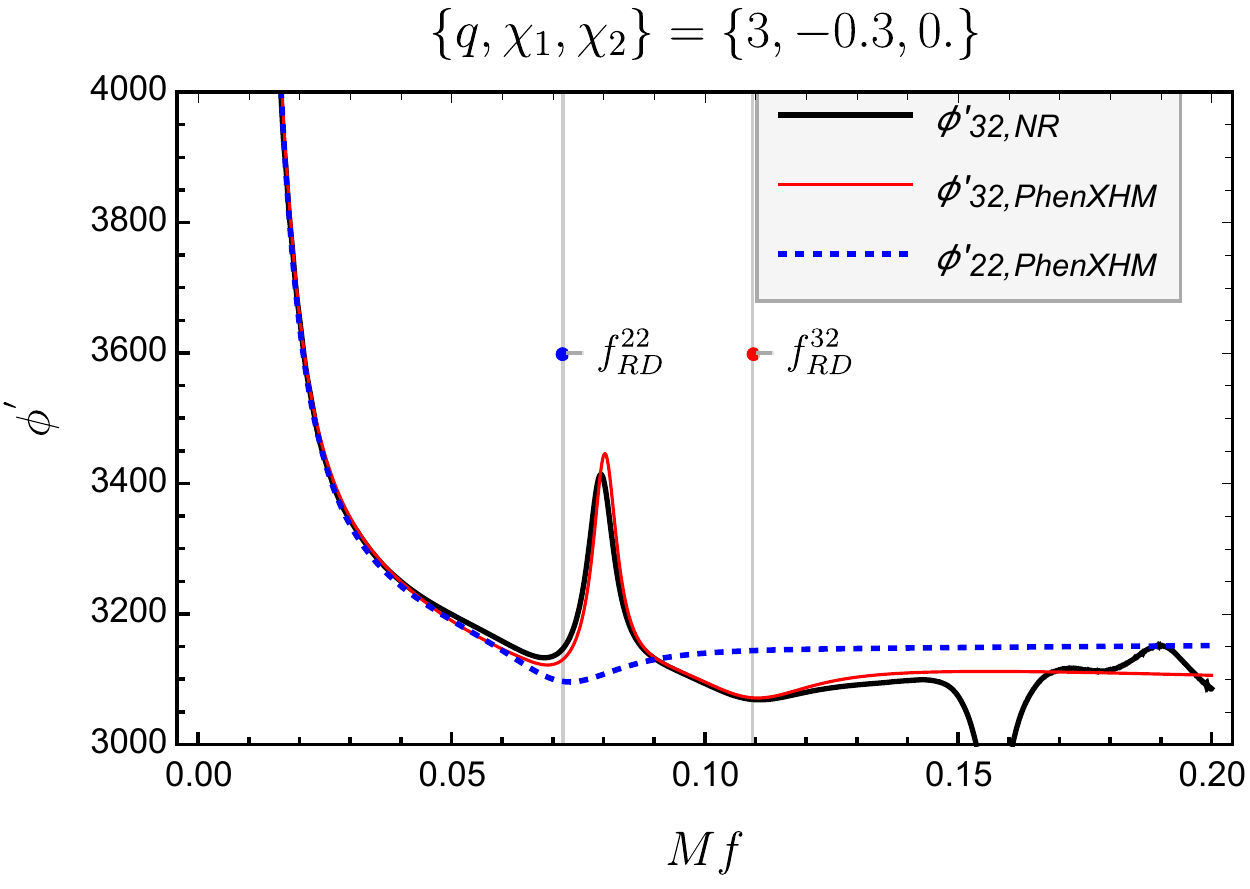}
    \caption{The plot shows the phase derivatives of the $(3,2)$ mode of a FD hybrid waveform with parameters $(q,\chi_1,\chi_2)=\left\{3, -0.3, 0.\right\}$, compared with the reconstructed $(2,2)$ and $(3,2)$ modes. The $(3,2)$ phase derivative does not show the usual fall-off in the ringdown region, due to mode-mixing with the $(2,2)$ mode.}
    \label{fig:32dphidf}
  \end{center}
\end{figure}

In the reconstruction of the $(3,2)$-mode, we allow $a_{3}^{32}$ to be non-zero: this extra degree of freedom allows to have better control on the effects caused by mode-mixing (see Fig. \ref{fig:32dphidf}). In this case only, we impose two boundary conditions coming from the ringdown-region, where the $(3,2)$ phase is also fully calibrated \footnote{Notice that, when mode-mixing is absent, the ringdown is built through an appropriate rescaling of the quadrupole's phase and does not contain any information about the physical relative time-shifts among the modes.}. We determine the six free coefficients of Eq.\,(\ref{eq:inter_phase_ansatz}) by solving the system
\begin{align}
\frac{d\phi_{3\,2}^{\mathrm{Int}}}{df}(f_{32}^{i})&=\mathcal{F}^{i}_{32}, \quad i \in\left[1,4\right], \nonumber \\
\frac{d\phi_{3\,2}^{\mathrm{Int}}}{df}(f_{RD}^{32})&=\frac{d\phi_{3\,2}^{\mathrm{RD}}}{df}(f_{RD}^{32}), \nonumber \\
\frac{d^2\phi_{3\,2}^{\mathrm{Int}}}{df}(f_{RD}^{32})&=\frac{d^2\phi_{3\,2}^{\mathrm{RD}}}{df}(f_{RD}^{32}) .
\end{align}

The explicit expressions for the frequencies of our intermediate-region collocation points are given in Tab.\,\ref{tab:inter_phase_cpoints}. These values result from taking a mixture of equidistant and Gauss-Chebyshev nodes in the interval $\left[\beta(\eta)f_{\mathrm{Ins}}^{\ell m},f_{End}^{\ell m}\right]$, where
\begin{align*}
f_{End}^{\ell m} =
  \begin{cases}
  f_{\mathrm{ring}}^{\ell m} & \text{if $(\ell,m)\neq(3,2)$} \\
  \\                                   
   f_{\mathrm{ring}}^{22}-0.5 f_{\mathrm{damp}}^{22} & \text{if $(\ell,m)=(3,2)$} 
  \end{cases}
\end{align*}
 and $\beta(\eta)$ is a monotonically decreasing function of $\eta$ that shifts forward the frequency of the first collocation points for small $\eta$, thus reducing the steepness of the parameter-space fit surfaces in this limit, chosen here as
$\beta(\eta):=(1.+0.001(0.25/\eta-1))$.

\begin{table}[h!]
\caption{Frequencies of the collocation points used in the reconstruction of the intermediate phase derivative.}
\label{tab:inter_phase_cpoints}
\begin{center}
{\renewcommand{\arraystretch}{1.5}
\begin{tabular}{| c | c | }
\hline
& Collocation point frequencies \\
\hline
   $f^{1}_{\ell m}$ & $\beta(\eta)f_{\mathrm{Ins}}^{\ell m}$ \\
\hline
   $f^{2}_{\ell m}$ & $\frac{1}{4} \left(\left(\sqrt{3}+2\right) \beta(\eta)f_{\mathrm{Ins}}^{\ell m}-\left(\sqrt{3}-2\right) f^{\ell m}_{\mathrm{End}}\right)$ \\  
   \hline
   $f^{3}_{\ell m}$ & $\frac{1}{4} ( f_{\mathrm{End}}^{\ell m}+3 \beta(\eta)f_{\mathrm{Ins}}^{\ell m})$ \\  
   \hline
   $f^{4}_{\ell m}$ &$\frac{1}{2} ( f^{\ell m}_{\mathrm{End}}+\beta(\eta)f_{\mathrm{Ins}}^{\ell m}) $ \\  
   \hline
   $f^{5}_{\ell m}$ &$\frac{1}{4} (3 f_{\mathrm{End}}^{\ell m}+\beta(\eta)f_{\mathrm{Ins}}^{\ell m})$ \\  
   \hline
   $f^{6}_{\ell m}$ &$\frac{1}{8} (7 f_{\mathrm{End}}^{\ell m}+\beta(\eta)f_{\mathrm{Ins}}^{\ell m})$ \\  
   \hline
\end{tabular}
}
\end{center}
\end{table}

%As one can see, Gauss-Chebyshev nodes have the disadvantage of providing somehow redundant information close to the ringdown, while the equidistant ones are too coarse in the vicinity of the inspiral region, where the phase derivative changes more dramatically.  

\begin{figure}[h!]
\includegraphics[width=\columnwidth]{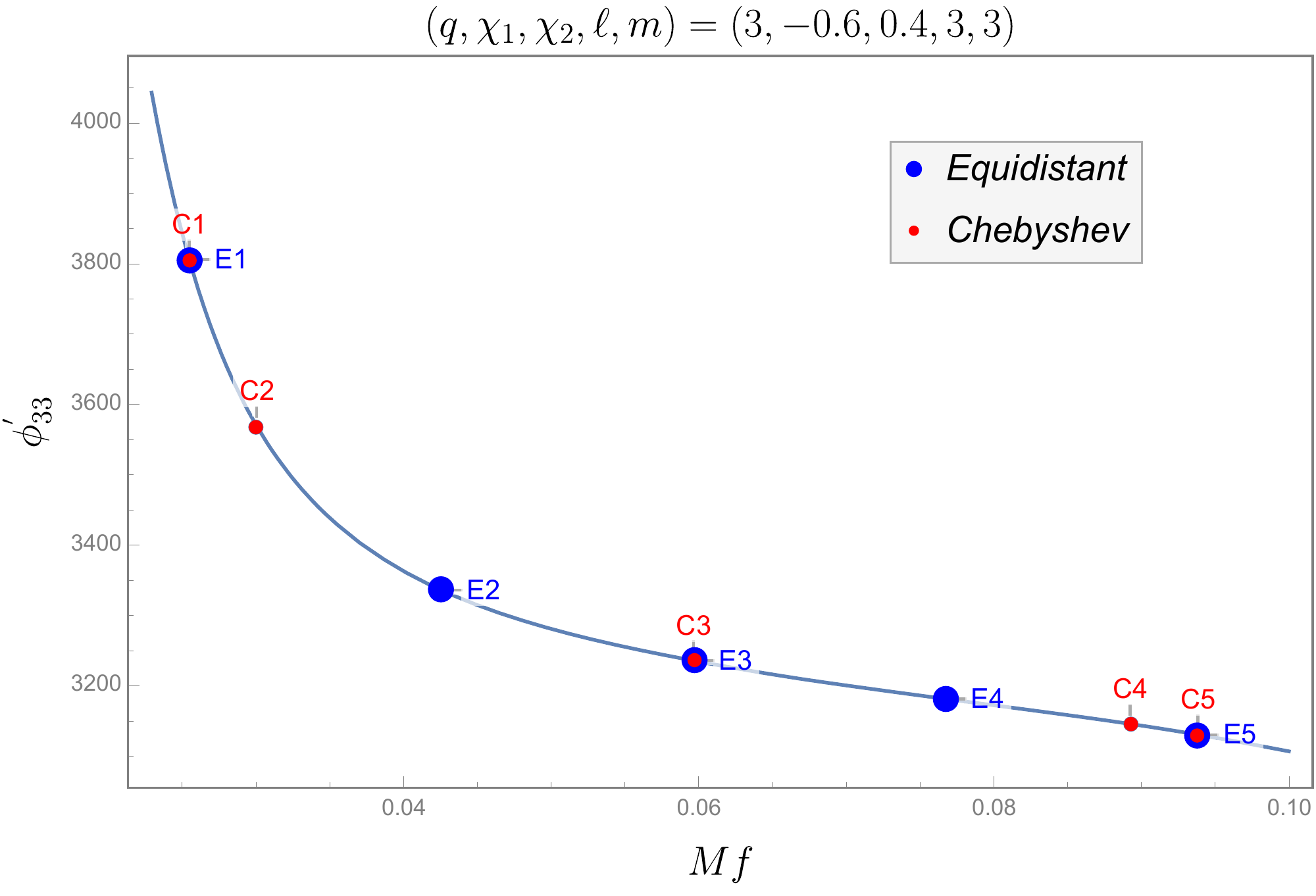}
\caption{We compare two different sets of collocation points in the intermediate region of the $(3,3)$ phase-derivative. Gauss-Chebyshev nodes are marked in red, while equidistant nodes are marked in blue. We compute parameter-space fits of the phase-derivative evaluated at the points $\left[\mathrm{C1,C2,E2,E3,E4,E5}\right]$. The default set of collocation points is $[\mathrm{C1,C2,E2,E3,E5}]$; in regions with fewer calibration waveforms, we switch to the subset $[\mathrm{C1,C2,E3,E4,E5}]$, which contains more points close to the ringdown. Here, the phase-derivative becomes flatter and our parameter-space fits are more robust.}
\label{fig:cpoints_compare}
\end{figure}

We find it convenient to model one more collocation point than what is strictly needed, in order to add some flexibility to the calibration. Our standard choice of collocation points can result in a badly-behaved reconstruction in regions of the parameter space where we have fewer calibration waveforms, such as the high spin and/or low-$\eta$ regime. In such cases, we drop one of the collocation points close to inspiral, where the phase derivative has a steeper slope and is harder to model accurately, and replace it with a point in the flatter near-ringdown region, as we illustrate in Fig.~\ref{fig:cpoints_compare}.

%%%%%%%%%%%%%%%%%%%%%%%%%%%%%%%%%%%%%%%%%%%%%%%%%%%%%%%%
\section{Ringdown model} \label{sec:ringdown}
The ringdown region covers the frequency range \begin{equation}
    M f\in\left[f_{\mathrm{RD}}^{\ell m},0.3\right],
\end{equation} 
where $f_{\mathrm{RD}}^{\ell m}$ was defined in Eq.\,(\ref{eq:fcutRD}).

In \phXHM, the modes $(2,1),(3,3),(4,4)$ have a fully calibrated amplitude, while their phase is built by appropriately rescaling the quadrupole's phase, along the lines of \phHM. 

\begin{center}
\begin{figure}[h!]
\includegraphics[width=\columnwidth]{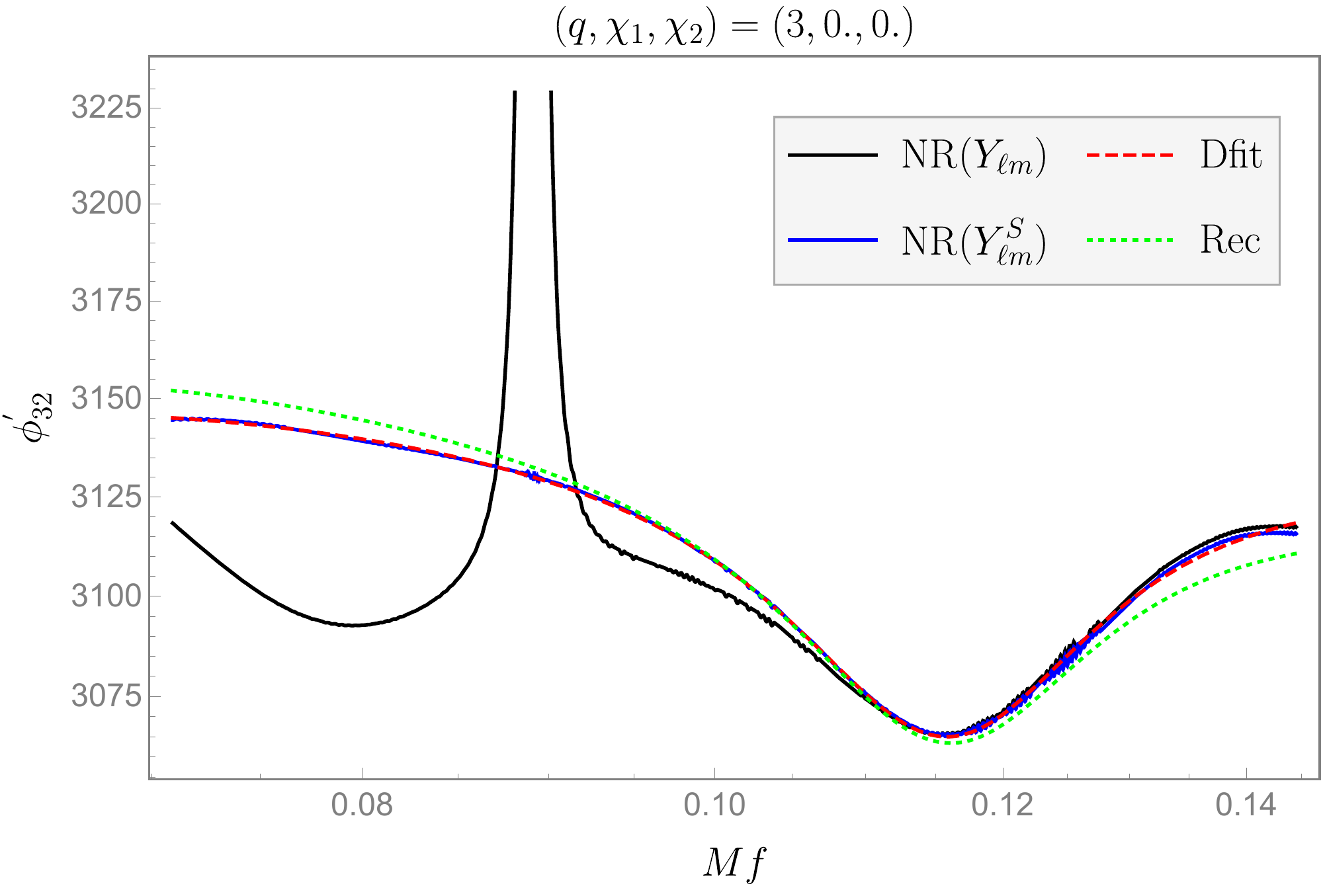}
\caption{The phase derivative of the $(3,2)$ mode can exhibit sharp features when plotted in the original spherical-harmonic basis (black solid line). However, the same signal written in terms of spheroidal harmonics is much simpler (blue solid line) and amenable to be fitted with the same ansatz used in \phX and \phD. Red and green lines mark the direct fit to the data and the final reconstruction.}
\label{fig:mixing_phase}
\end{figure}
\end{center}

\begin{figure}[h!]
\includegraphics[width=\columnwidth]{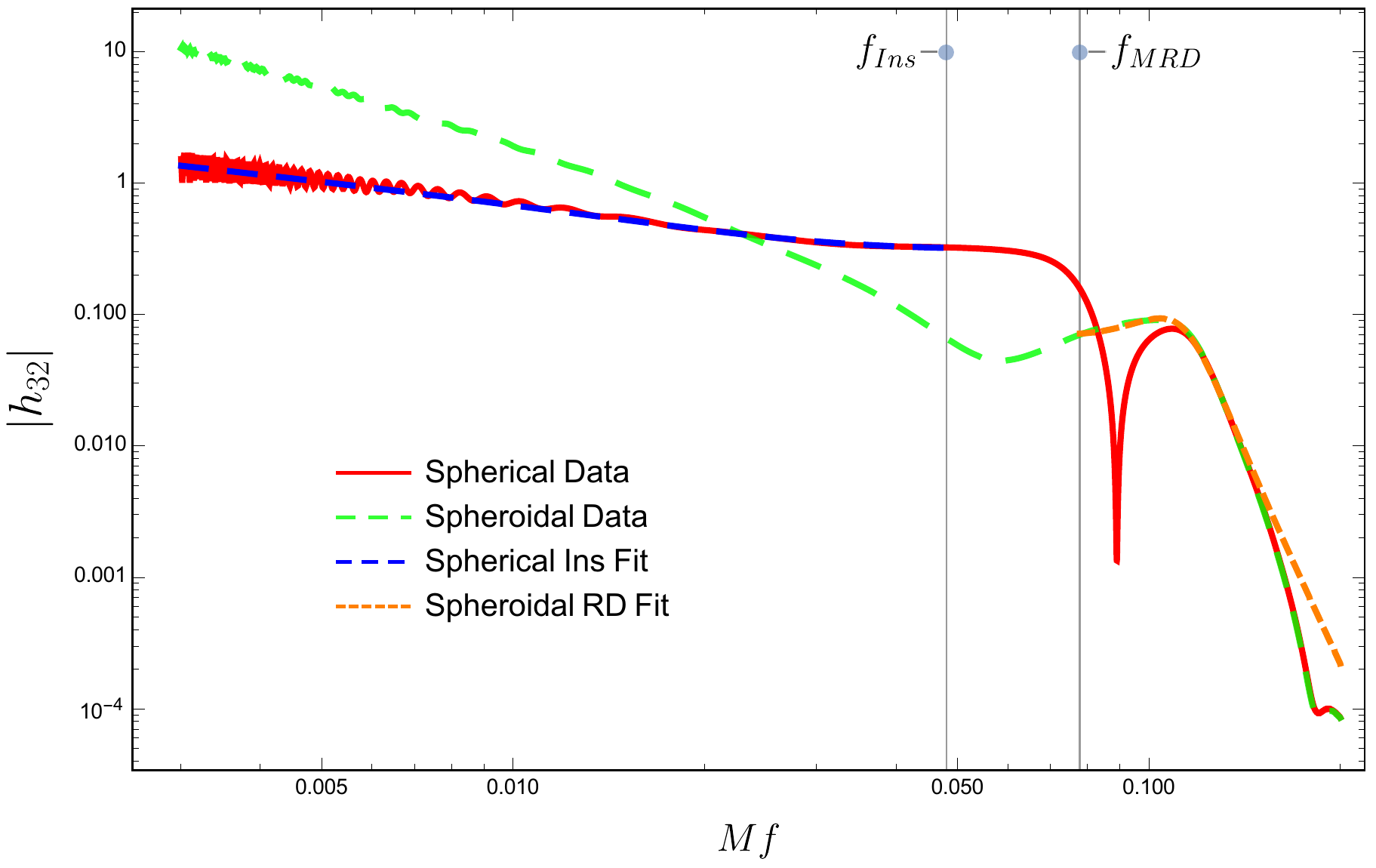}
\includegraphics[width=\columnwidth]{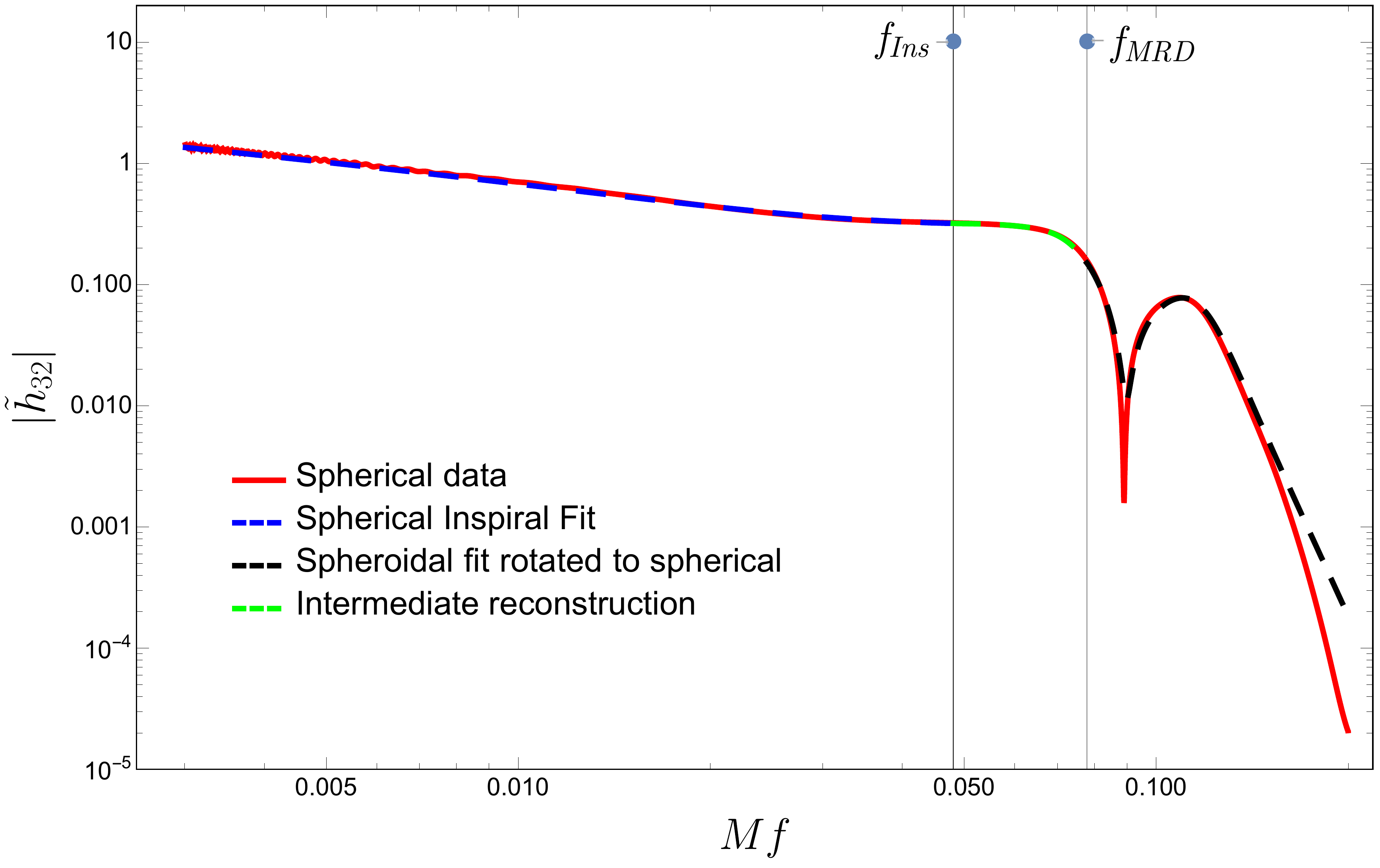}
\caption{Top: The amplitude of the (3,2) mode expressed in a \textit{spherical} (red) and \textit{spheroidal} (green) basis for a $\{q,\chi_1,\chi_2\}=\{3,0,0\}$ case. The latter can be easily fitted using the ansatz \ref{eq:ringdown_ansatz} (orange curve). The inspiral portion of the amplitude is fitted in a spherical-harmonic basis (blue dashed line), therefore the ringdown waveform must be transformed back to the original basis before being smoothly attached to the rest of the model. Bottom: NR data and final reconstruction in spherical-harmonics for the same case plotted in the top panel. The black line is the transformation to a spherical-harmonic basis of the orange curve in the top plot (which is computed instead in a spheroidal-harmonic basis) . The green line is a smooth connection between the inspiral and ringdown fits (blue and black lines respectively) that goes through the two collocation points in the intermediate region. }
\label{fig:mixing_amp}
\end{figure}

When mode-mixing visibly affects the ringdown waveform (i.e. in the $(3,2)$-mode reconstruction), the model is instead fully calibrated to NR. In this case, the key observation is that the signal is much simpler when expressed in terms of spin-weighted spheroidal harmonics, as we illustrate in Figs.\,\ref{fig:mixing_phase} and \ref{fig:mixing_amp}. This can be traced back to the fact that the Teukolsky equation is fully separable only in a basis of spheroidal harmonics, and not in a spherical-harmonic one \cite{1972:Teukolsky}. 

Under the simplifying assumption that the $(3,2)$-mode interacts only with the $(2,2)$-mode, the spherical-harmonic strains can be projected onto a spheroidal-harmonic basis by means of a simple linear transformation, which we describe in App.\,\ref{appendix:spheroidal_spherical}. We reconstruct amplitude and phase of the signal in a spheroidal-harmonic basis and then transform the full waveform back into the original basis. After doing so, the ringdown reconstruction can be smoothly connected to the inspiral-merger waveform. 

In the following subsections, we provide further details about the ansätze used in this region. 

\subsection{Amplitude} \label{sec:ring_amp}
% this part has been moved elsewhere
%The ringdown part comprise the region from a frequency $f_{MRD}$ slighty lower than the ringdown frequency of the corresponding mode up to the end of the waveform. The precise values of these frequencies are choosen in a heuristical way and are:
%\begin{equation}
%f_{MRD}^{21} = 0.75 \:f^{21}_{ring}, \;
%f_{MRD}^{33} = 0.95 \:f^{33}_{ring},\;
%f_{MRD}^{44} = 0.9 \:f^{44}_{ring}.
%\end{equation}
The ansatz we adopt is similar to the one used in \phX:
\begin{equation} \label{eq:ringdown_ansatz}
\frac{A_{RD}^{\ell m}}{A_0^{22}}= \frac{1}{f^{\frac{1}{12}}} \frac{|a_\lambda| \: f^{\ell m}_{damp}\:\sigma}{\left(f-f^{\ell m}_{ring}\right)^2+\left(f^{\ell m}_{damp}\:\sigma\right)^2}e^{-\frac{ \left(f-f^{\ell m}_{ring}\right)\:\lambda}{f^{\ell m}_{damp}\sigma}},
\end{equation}
except for the factor $f^{-1/12}$ used here for historical reasons and for the replacement of the $(2,2)$ ringdown and damping frequencies with those of the corresponding $(\ell,m)$ mode. This ansatz is used for all the modes calibrated in the model. Notice, however, that for the $(3,2)$ mode the ansatz is fitted to the data expressed in a spheroidal-harmonic basis, see Fig.~\ref{fig:mixing_amp}.

We first fit the free coefficients $\{a_{\lambda},\lambda,\sigma\}$ to NR data in a ``primary'' direct fit and then perform a parameter space fit of each coefficient for every mode. We find that the coefficient $\sigma$ shows a very small dynamic range for the modes $(3,3)$, $(3,2)$ and $(4,4)$, and
we thus take it as a constant. We then perform a ``secondary'' direct fit where we redo the direct fits using the constant values for $\sigma$ shown in Tab.~\ref{tab:sigmavalues}. Finally we repeat the parameter space fits for $a_{\lambda}$ and $\lambda$.

\begin{figure}[H]
\centering
\includegraphics[scale=0.3]{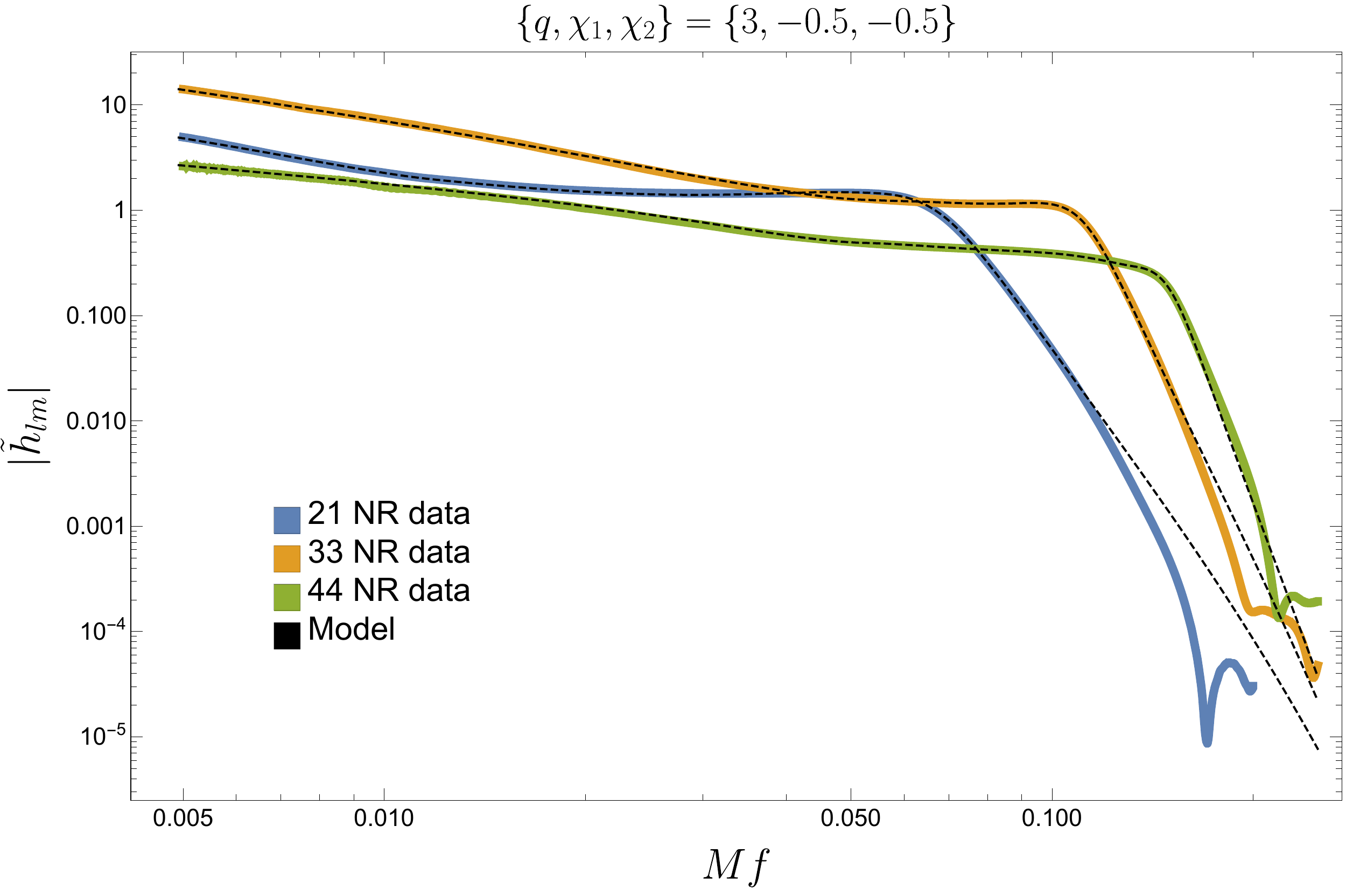}
\caption{Comparison between the NR data and the final model for the amplitude of three modes for a system with $\{q,\chi_1,\chi_2\}=\{3,0,0\}$.}
\label{fig:nomixing}
\end{figure}

The final reconstruction of the amplitude through inspiral, merger and ringdown for the modes without mixing can be seen in Fig.~\ref{fig:nomixing}.

\begin{table}[H]
\caption{Mode-specific values for the parameter $\sigma$ appearing in Eq.\,(\ref {eq:ringdown_ansatz}). In the final model $\sigma$ is taken to be fixed across parameter space execpt for the 21 mode. Here we show these fixed values, which correspond to an average across the parameter-space of the values obtained through direct fits where $\sigma$ is not specified a priori.}
\label{tab:sigmavalues}
\begin{center}
{\renewcommand{\arraystretch}{1.5}
\begin{tabular}{| c | c |}
\hline
  \; Mode \; & \; $\sigma$ value \; \\ [5pt]
\hline
   33 & 1.3 \\                        
\hline
   32 & 1.33 \\  
   \hline
   44 & 1.33 \\  
   \hline
  
\end{tabular}
}
\end{center}
\end{table}

%For some particular modes (32, 44, 43) the $\sigma$ parameter looked quite noisy but in a very small dynamic range. That means that this coefficient is basically constant over the parameter space, so we keeped it fixed for these modes and redid the fit with only two free parameters. 

\subsection{Phase}
\label{sec:ring_phase}

\subsubsection{Modes without mode-mixing} 
To model the $(2,1),(3,3),(4,4)$ modes, for which mode-mixing is negligible, we rescale a simplified reconstruction of the quadrupole's ringdown phase, much in the spirit of \phHM. 

Our ansatz for the phase-derivatives in this case reads:
\begin{align}
\frac{d\phi_{\ell m}^{\mathrm{RD}}}{df}=\alpha_2^{\ell m} \frac{(f^{\ell m}_{ring})^2} {f^2}+\alpha_\lambda^{\ell m} \frac{f_{\mathrm{damp}}^{\ell m}}{(f_{\mathrm{damp}}^{\ell m})^2+(f-f^{\ell m}_{ring})^2}+d\phi_{\mathrm{RD}}^{\ell m},
\end{align}
which, integrated, gives:
\begin{align}
\phi_{\ell m}^{\mathrm{RD}}=-\alpha_2^{\ell m} \frac{(f^{\ell m}_{ring})^2} {f}+\alpha_\lambda^{\ell m} \tan ^{-1}\left(\frac{f-f^{\ell m}_{ring}}{f_{\mathrm{damp}}^{\ell m}}\right)+d\phi_{\mathrm{RD}}^{\ell m} f+\phi_{\mathrm{RD}}^{\ell m},
\end{align}

We first compute parameter-space fits of the quantities $\alpha_{\lambda}^{22}$ and $\alpha_{2}^{22}$ and rescale them to obtain their higher-modes counterparts.
We set 
\begin{align}
\alpha_{\lambda}^{\ell m}&=\alpha_{\lambda}^{22}, \nonumber\\ \alpha_2^{\ell m}&=w_{\ell m}\frac{f_{\mathrm{damp}}^{22}}{f_{\mathrm{damp}}^{\ell m}} \alpha_2^{22},
\end{align}
where $w_{\ell m}$ are some constants, which only depend on $\ell,m$ and not on the intrinsic parameters of the binary. We verified that the above equalities hold, albeit approximately, for the parameters of the direct fits to each mode's phase derivative. 

The shifts $d\phi_{\mathrm{RD}}^{\ell m}$ and $\phi_{\mathrm{RD}}^{\ell m}$ are fixed by requiring a smooth connection to the intermediate-region reconstruction. 

\subsubsection{Modes with mode-mixing}
As we outlined at the beginning of this section, the morphology of the $(3,2)$-mode ringdown signal is significantly affected by mode-mixing. In this case, we first build a reconstruction of the phase derivative in a spheroidal-harmonics base, using the ansatz below
\begin{align}
\label{eq:dphiS_ansatz}
\frac{d\phi_{32,S}}{df}= \frac{\alpha_2^{32}}{f^2} +\frac{\alpha_4^{32}}{f^4} + \alpha_\lambda^{32} \frac{f_{\mathrm{damp}}^{32}}{(f_{\mathrm{damp}}^{32})^2+(f-f^{32}_{ring})^2}+d\phi_{\mathrm{RD}}^{32}.
\end{align}
Integrating the above equation, one obtains
\begin{align}
\phi_{32,S}=-\frac{\alpha_2^{32}} {f}-\frac{\alpha_4^{32}} {3 f^3}+\alpha_\lambda^{32} \tan ^{-1}\left(\frac{f-f_{ring}^{32}}{f_{damp}^{32}}\right)+d\phi_{\mathrm{RD}}^{32} f +\phi_{\mathrm{RD}}^{32},
\end{align}
where the subscript $S$ is a reminder that we are now working in a spheroidal-harmonic basis. The four free coefficients of Eq.\,(\ref{eq:dphiS_ansatz}) are determined by solving the linear system
\begin{equation}
\frac{d\phi_{32,S}^{\mathrm{RD}}}{df}(f_{32}^{i})=\mathcal{G}^{i}_{32},\ \ \ \ \ \ i \in\left[1,4\right],
\end{equation}
where $\mathcal{G}^{i}_{32}$ are some parameter-space fits of the value of the phase derivative, evaluated at four collocation points $f_{RD}^{i}, i\in[1,4]$, given in Tab.\,\ref{tab:RD_cpoints_phase}.

\begin{table}[h!]
\caption{Frequencies of the collocation points used in the reconstruction of the $(3,2)$-mode phase derivative.}
\label{tab:RD_cpoints_phase}
\begin{center}
{\renewcommand{\arraystretch}{1.5}
\begin{tabular}{| c | c |}
\hline
\multicolumn{2}{|c|}{Collocation points for $\phi_{32,S}^{'}$}  \\[5pt]
\hline
   $f_{32}^{1}$ & $f_{\mathrm{ring}}^{22}$ \\                        
\hline
   $f_{32}^{2}$ & $f_{\mathrm{ring}}^{32}-3/2 f_{\mathrm{damp}}^{32} $ \\  
   \hline
   $f_{32}^{3}$ & $f_{\mathrm{ring}}^{32}-1/2 f_{\mathrm{damp}}^{32} $ \\  
   \hline
   $f_{32}^{4}$ &$f_{\mathrm{ring}}^{32}+1/2 f_{\mathrm{damp}}^{32} $\\  
   \hline

\end{tabular}
}
\end{center}
\end{table}

One must ensure that $\phi_{32,S}$ has the correct relative time and phase shift with respect to the $(2,2)$ mode that is being used, or else the transformation back to the original spherical-harmonic basis will produce an incorrect result. Therefore, we compute two extra fits
\begin{align}
\label{eq:RD_shifts}
\Delta T_{32,S}&=\phi_{32,S}^{'}(f_{0})-\phi_{22}^{'}(f_0)\\
\Delta\phi_{32,S}&=\phi_{32,S}(f_{1})-\phi_{22}(f_1),
\end{align} 
at some suitable reference frequencies $f_0,f_1$ in the ringdown region, and use them to correctly align our spheroidal-harmonic reconstruction to the quadrupole's phase given by \phX.  

\begin{figure}
  \begin{center}
    \includegraphics[width=\columnwidth]{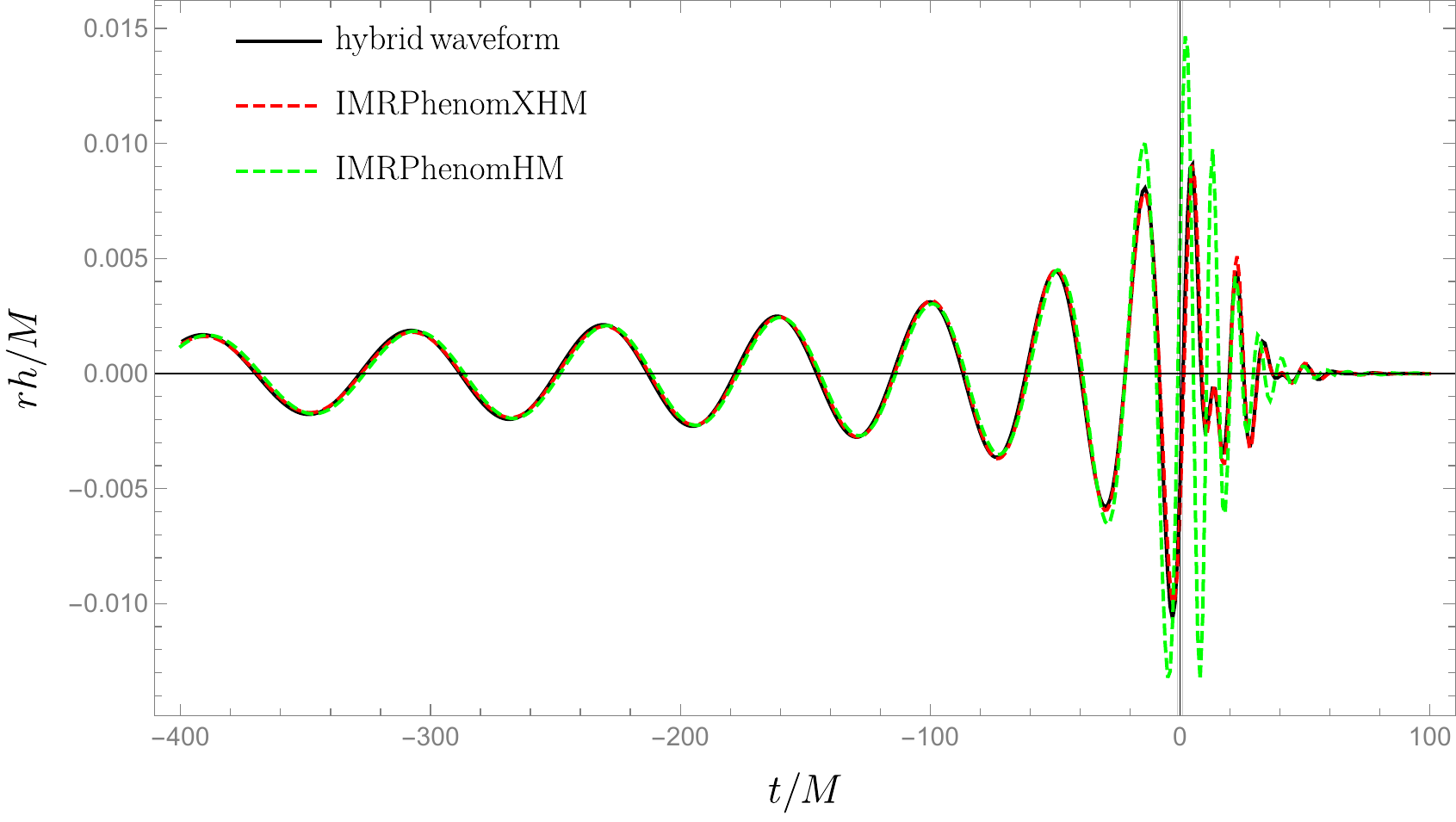}
    \caption{The plot shows the last few cycles of a hybrid $(3,2)$ mode waveform built hybridizing the SXS simulation \textsc{SXS:BBH:0271} with SEOBNRv4 (black solid line), together with the corresponding \phXHM and \phHM
    reconstructions (red and green dashed lines respectively). The $(2,2)$ modes of all waveforms have been previously time-shifted so that their amplitudes peak at $t=0$.}
    \label{fig:32TD}
  \end{center}
\end{figure}

In Fig.\,\ref{fig:32TD} we plot the $(3,2)$ mode of a hybrid waveform built from the SXS simulation \textsc{SXS:BBH:0271} and show the corresponding time-domain reconstructions resulting from \phXHM and \phHM. It can seen that our model can better capture the effects of mode-mixing on the ringdown waveform.

%%%%%%%%%%%%%%%%%%%%%%%%%%%%%%%%%%%%%%%%%%%%%%%%%%%%%%%%
\section{Quality control} \label{sec:quality}

\subsection{Single Mode Matches}

To quantify the agreement between two single-mode waveforms (reals in time domain) we use the standard definition of the inner product (see e.g.~\cite{Cutler:1994ys}),
\begin{equation}
\Braket{h_1, h_2} = 4 Re \int_{f_{min}}^{f_{max}} \frac{\tilde{h_1}(f) \:\tilde{h^*_2}(f)}{S_{n}(f)},
\end{equation}
where $S_{n}(f)$ is the one-sided power spectral-density of the detector. 
The \textit{match} is defined as this inner product divided by the norm of the two waveforms and maximised over relative time and phase shifts between both of them,
\begin{equation}
\mathcal{M}(h_1,h_2) = \mathop{max}_{t_0, \phi_0} \frac{\Braket{h_1, h_2}}{\sqrt{\Braket{h_1, h_1}}\sqrt{\Braket{h_2, h_2}}}.
\end{equation}
Accordingly, we define the mismatch between two waveforms as
\begin{equation}
\mathcal{MM}(h_1,h_2)=1-\mathcal{M}(h_1,h_2).
\end{equation}
For our match calculations we use the Advanced-LIGO design sensitivity Zero-Detuned-High-Power Power Spectral Density (PSD) \cite{TheLIGOScientific:2014jea,LIGOPSD} with a lower cutoff frequency for the integrations of 20 Hz. 

\begin{figure}[htbp]
    \centering
    \includegraphics[width=0.5\columnwidth]{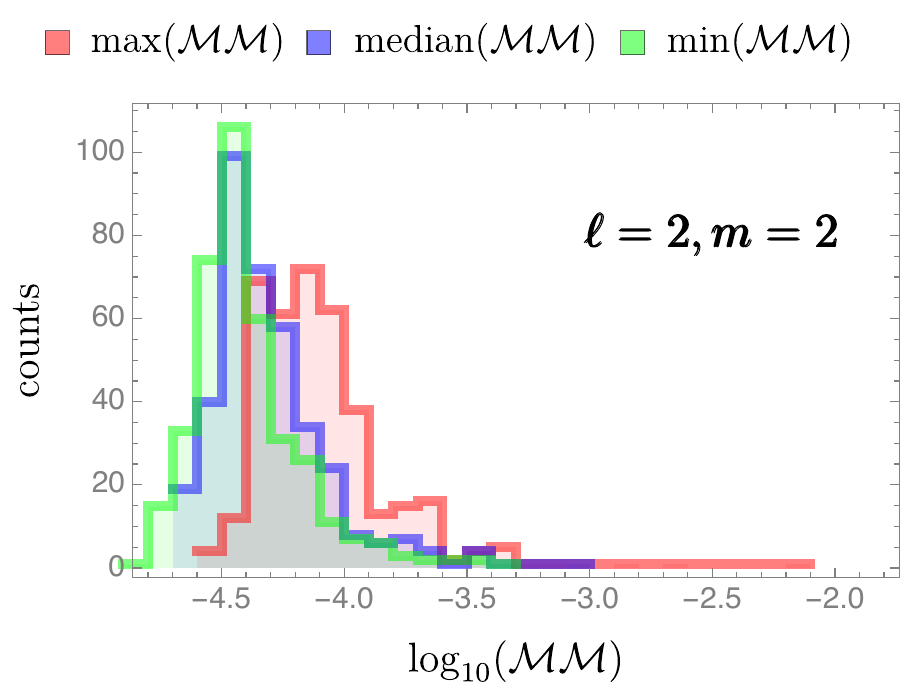}\\
    \includegraphics[width=0.5\columnwidth]{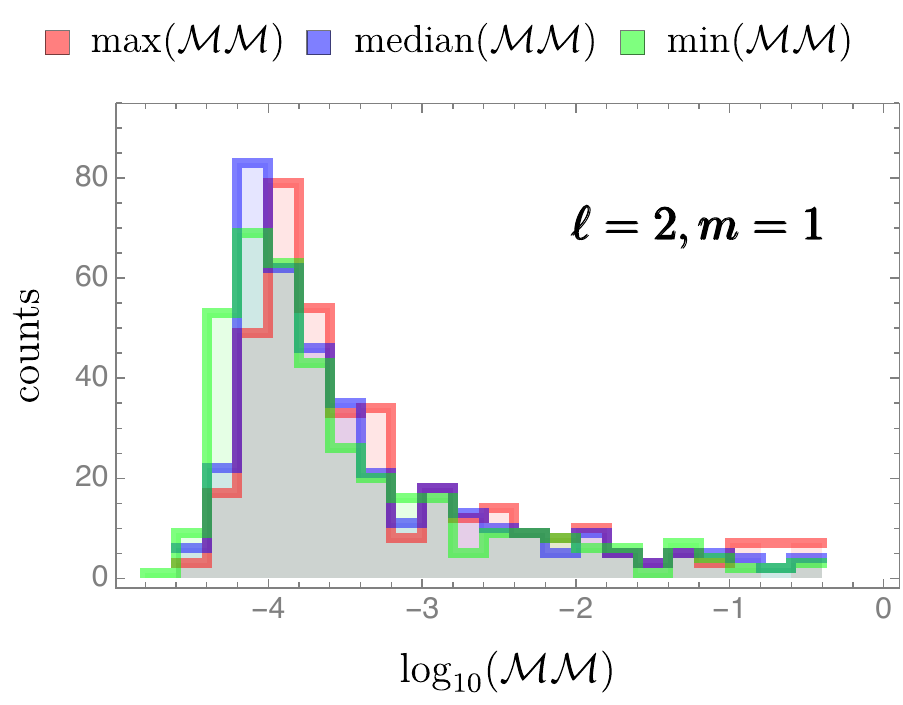}\includegraphics[width=0.5\columnwidth]{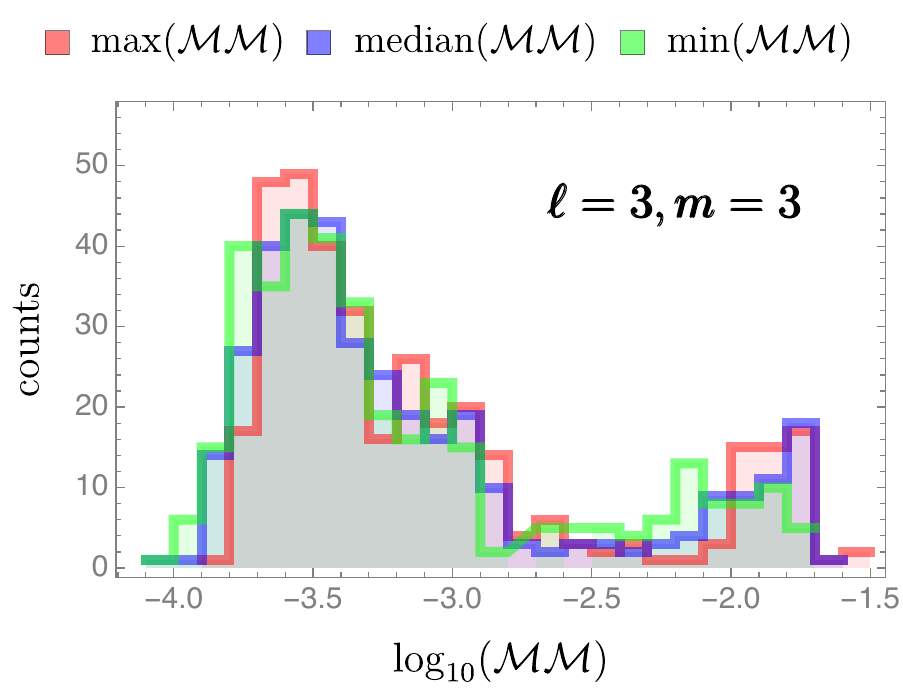}\\
    \includegraphics[width=0.5\columnwidth]{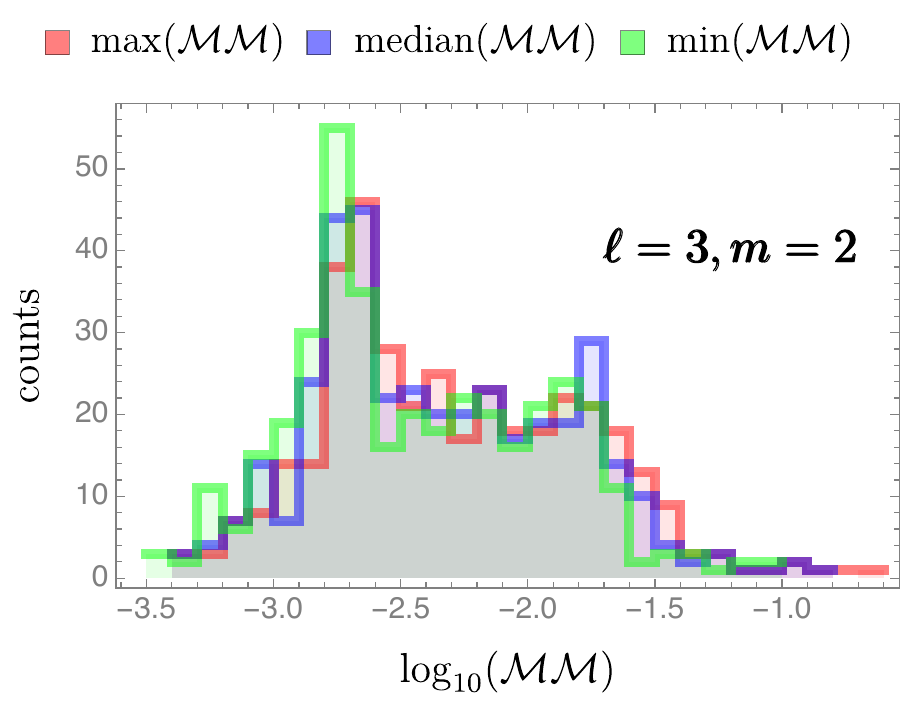}\includegraphics[width=0.5\columnwidth]{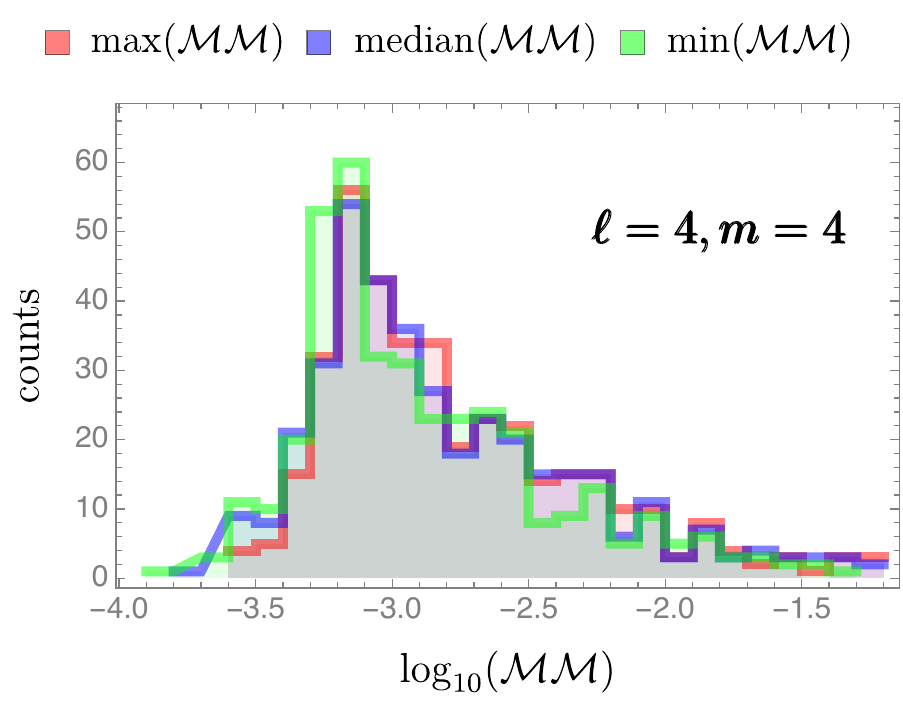}
\caption{Mode-by-mode mismatches between \phXHM and a validation set of hybrids built using the latest release of the SXS collaboration catalog. Each plot shows the maximum (red), median (blue) and minimum (green) mismatch over a range of total masses between 20 and 300 solar masses.}
\label{fig:mode_matches_hyb}
\end{figure}

In Fig. \ref{fig:mode_matches_hyb} we first show single-mode mismatches against a validation set consisting of 387 of our hybrid waveforms built from the latest SXS collaboration catalog, where we have discarded 152 hybrids, which show up as outliers in our calibration procedure for at least one of the modes, which we suspect to be due to quality problems with the waveforms. The list of SXS waveforms we have used is provided as supplementary material. The matches were computed for masses between $20 M\odot$ and $300 M\odot$, with a spacing of $10 M\odot$ between subsequent bins. 

\begin{figure}[htbp]
    \centering
    \includegraphics[width=0.5\columnwidth]{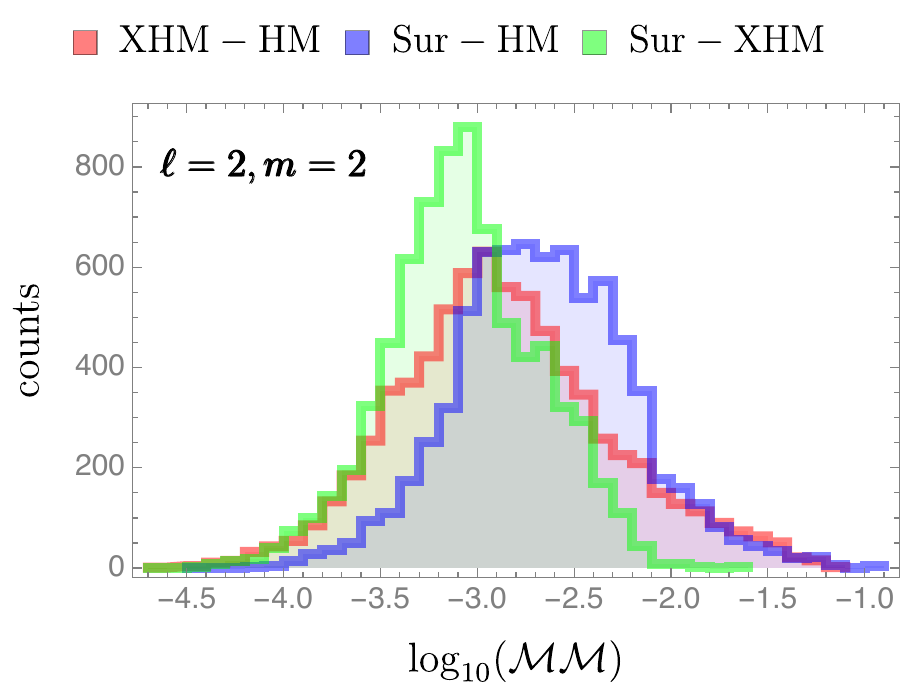}\\
    \includegraphics[width=0.5\columnwidth]{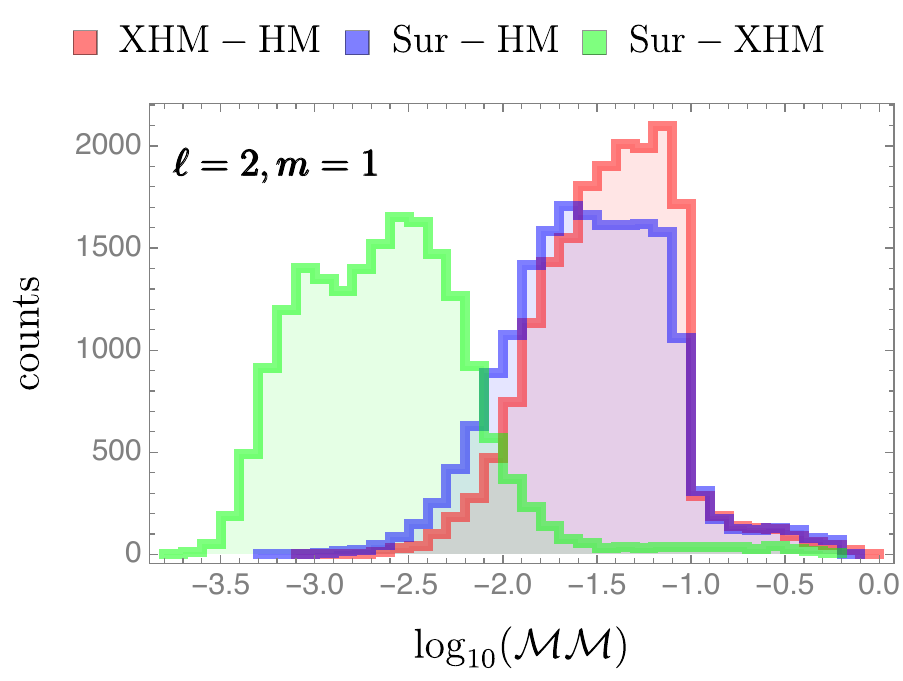}\includegraphics[width=0.5\columnwidth]{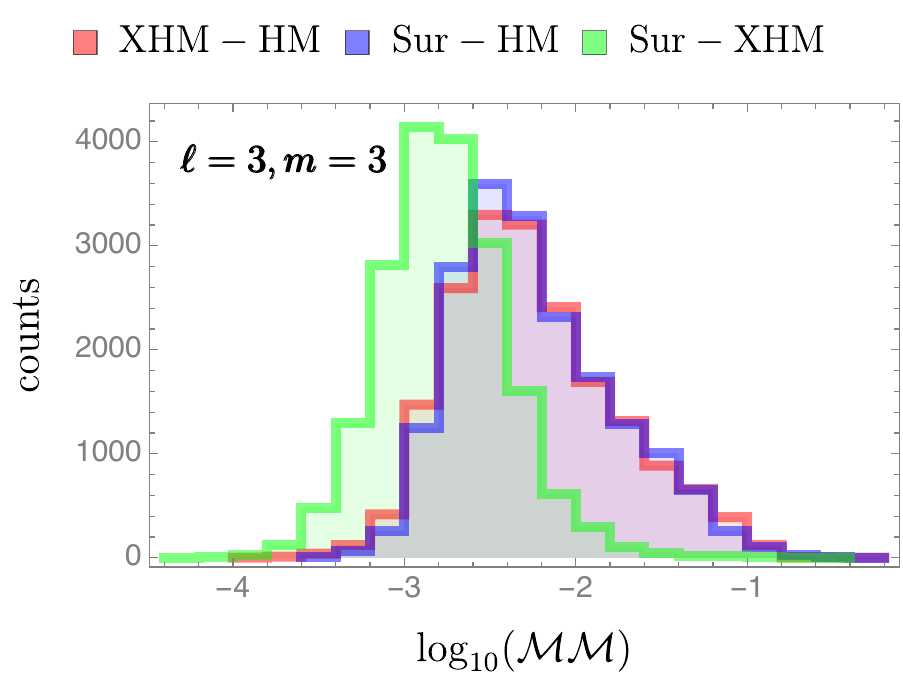}\\
    \includegraphics[width=0.5\columnwidth]{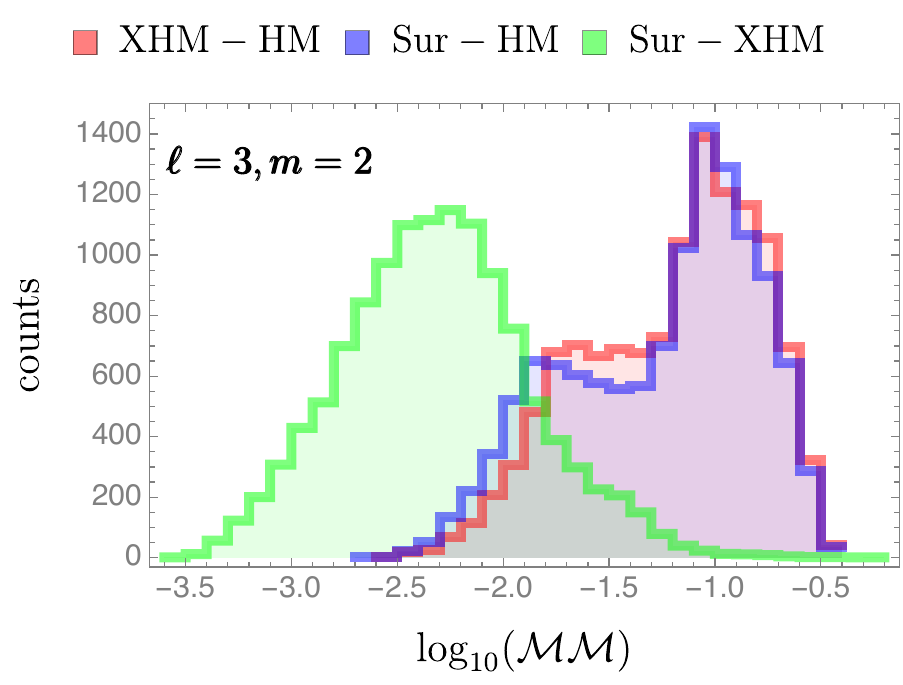}\includegraphics[width=0.5\columnwidth]{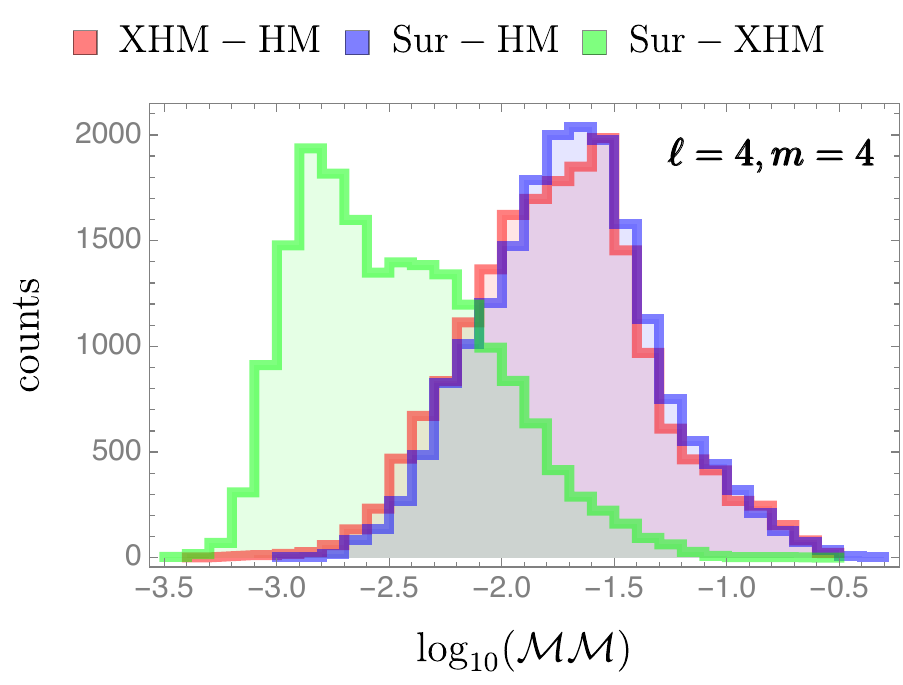}
\caption{Mismatches between different models for all the modes modelled by \phXHM with the aLIGOZeroDetHighPower PSD (see the main text for further details). The parameter range is restricted to avoid extrapolation as detailed in the main text. }
    \label{fig:mode_matches_noExtrap}
\end{figure}

\begin{figure}[htbp]
    \centering
    \includegraphics[width=0.5\columnwidth]{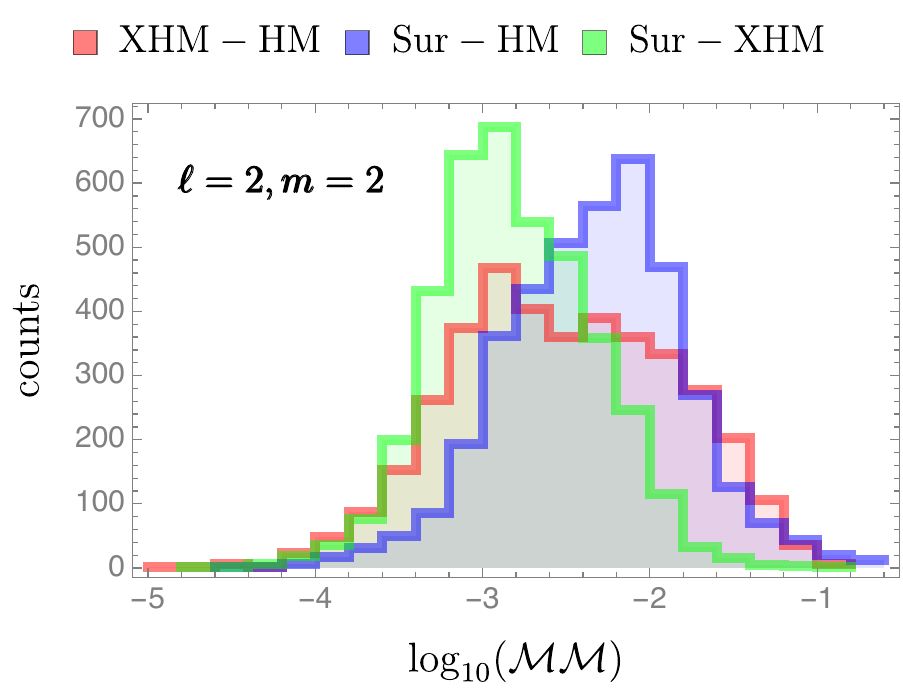}\\
    \includegraphics[width=0.5\columnwidth]{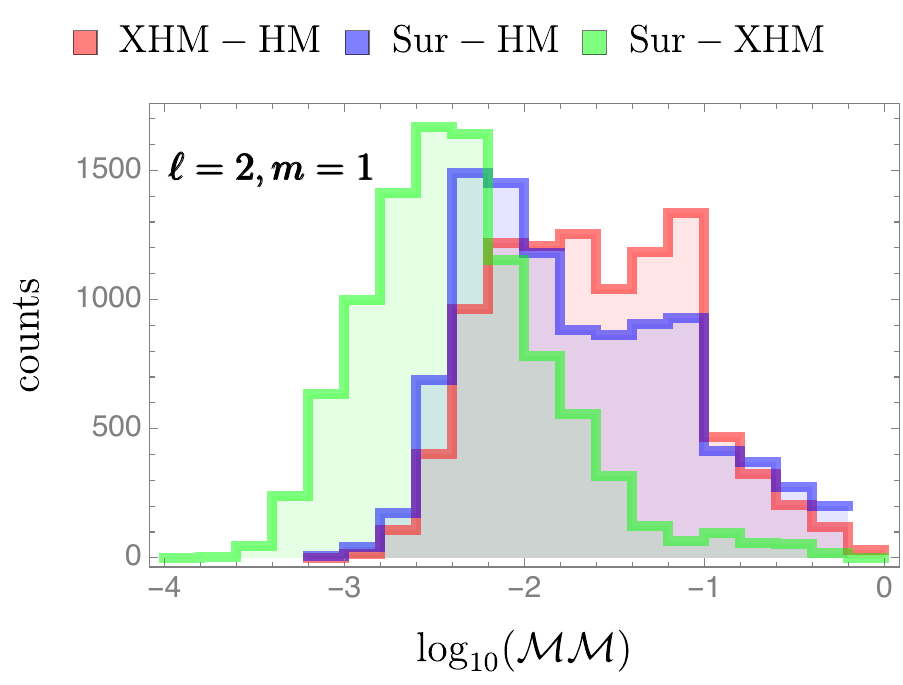}\includegraphics[width=0.5\columnwidth]{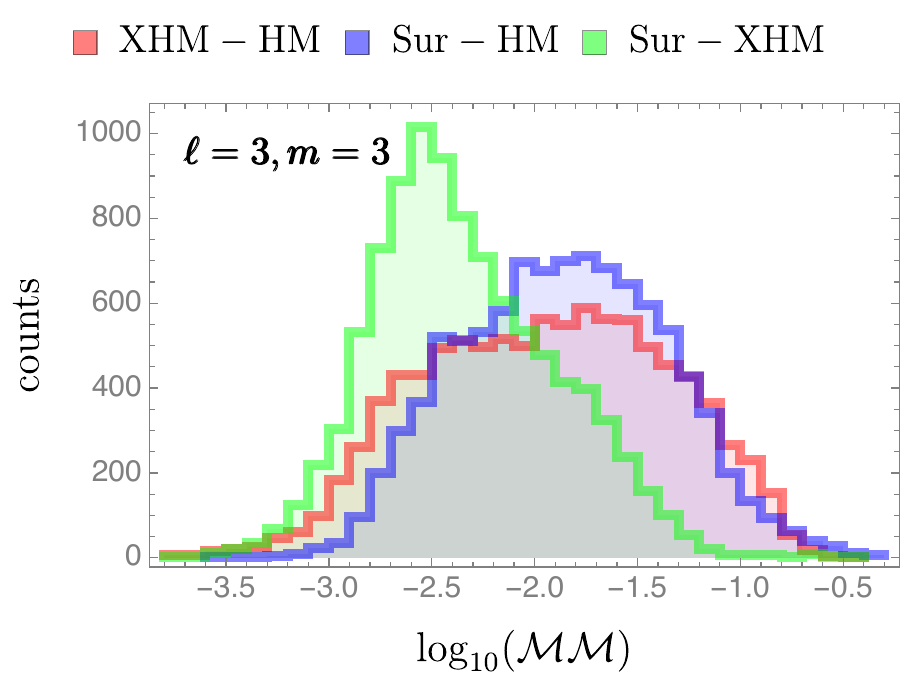}\\
    \includegraphics[width=0.5\columnwidth]{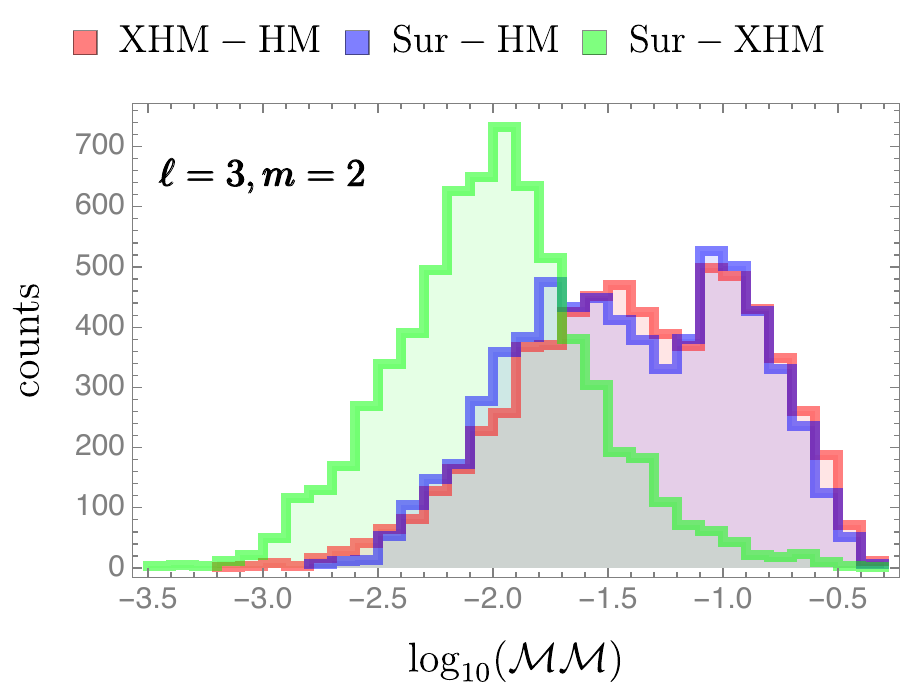}\includegraphics[width=0.5\columnwidth]{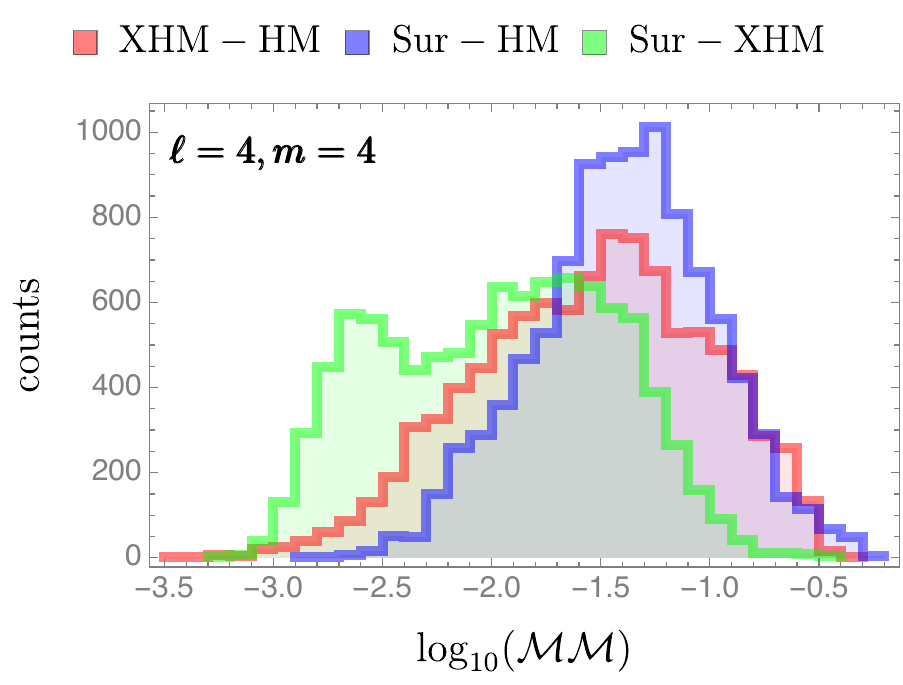}
\caption{Mismatches between different models for all the modes modelled by \phXHM with the aLIGOZeroDetHighPower PSD (see the main text for further details). The parameter range is chosen to test the extrapolation region of parameter space detailed in the main text.}
    \label{fig:mode_matches_rest}
\end{figure}

We also show mismatches among \phXHM, the previous \phHM model and the independent \surro surrogate model. 
Total masses are log-uniformly distributed in the range $\left[3 M_\odot,150 M_\odot\right]$ (with individual masses not smaller than $1 M_\odot$).
In Fig.~\ref{fig:mode_matches_noExtrap} we show the mismatches for the calibration region of \surro, for mass ratios below $9.09$ and dimensionless spin magnitudes up to $0.8$, and up to $0.5$ in the neutron star sector of total masses up to $3 M_\odot$. We carry out three sets of comparisons, in red we have the mismatches between \phXHM and \phHM, in blue \phHM versus \surro and finally in green \phXHM versus \surro. The results show that \phXHM is in a much better agreement with the surrogate model that the previous version \phHM, the improvement is particularly remarkable for the $(3,2)$ mode due to the modelling of the mode-mixing. 
In Fig.~\ref{fig:mode_matches_rest} we show matches for cases outside of the spin region defined before to assess the effects of extrapolation beyond the calibration region.

\subsection{Multi-Mode Matches}
\begin{figure}
    \centering
    \includegraphics[width=0.5\columnwidth]{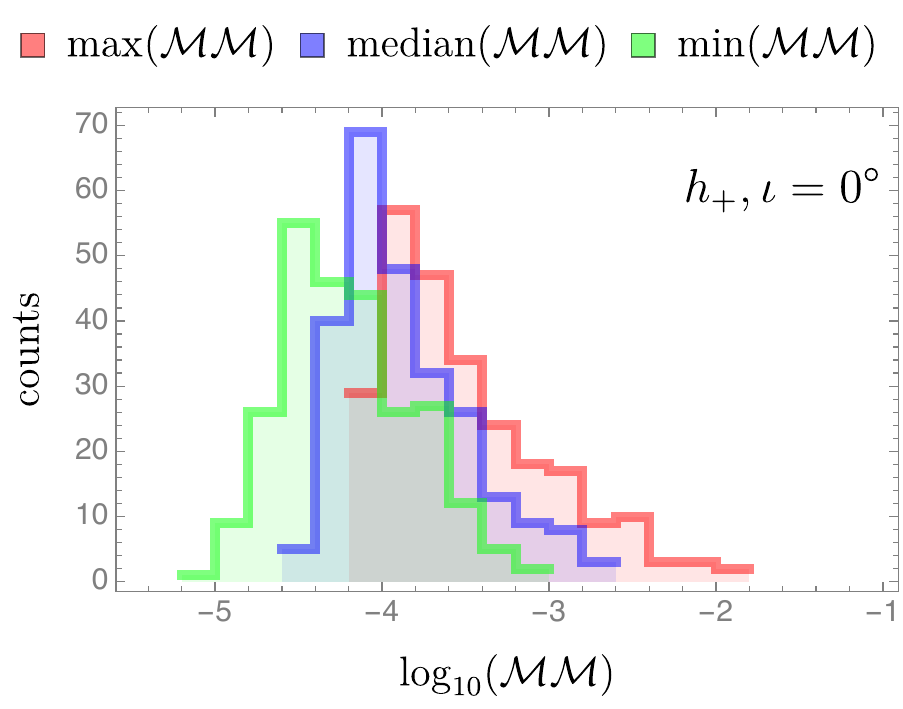}\includegraphics[width=0.5\columnwidth]{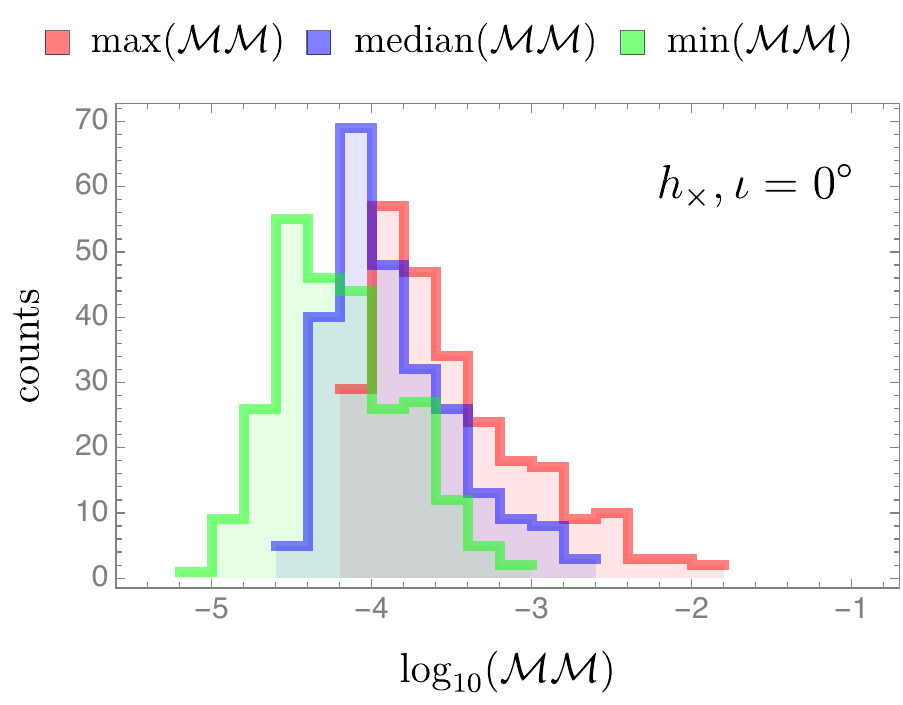}
   \includegraphics[width=0.5\columnwidth]{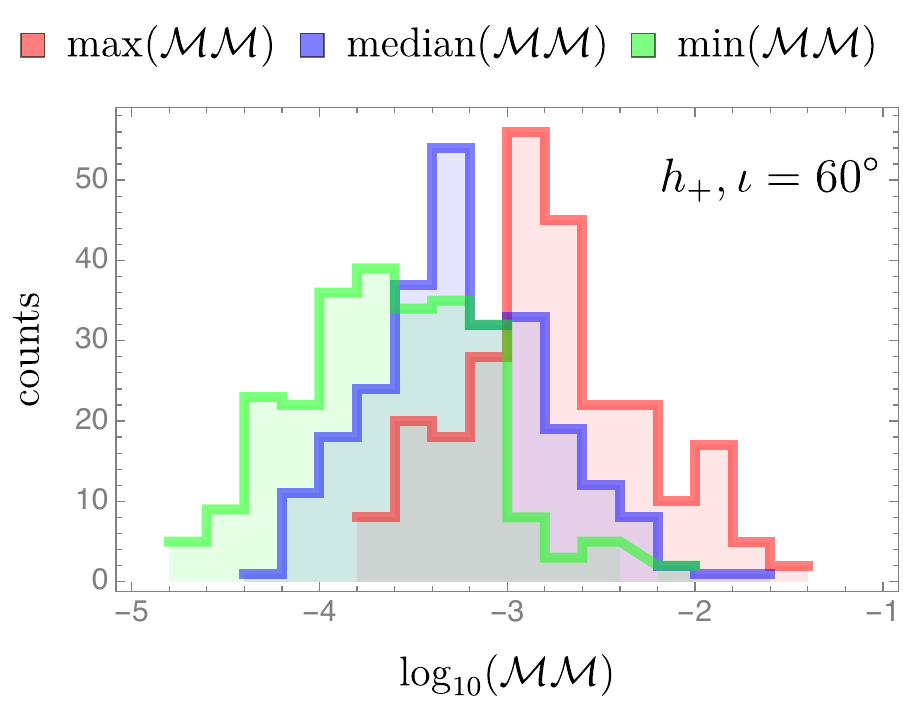}\includegraphics[width=0.5\columnwidth]{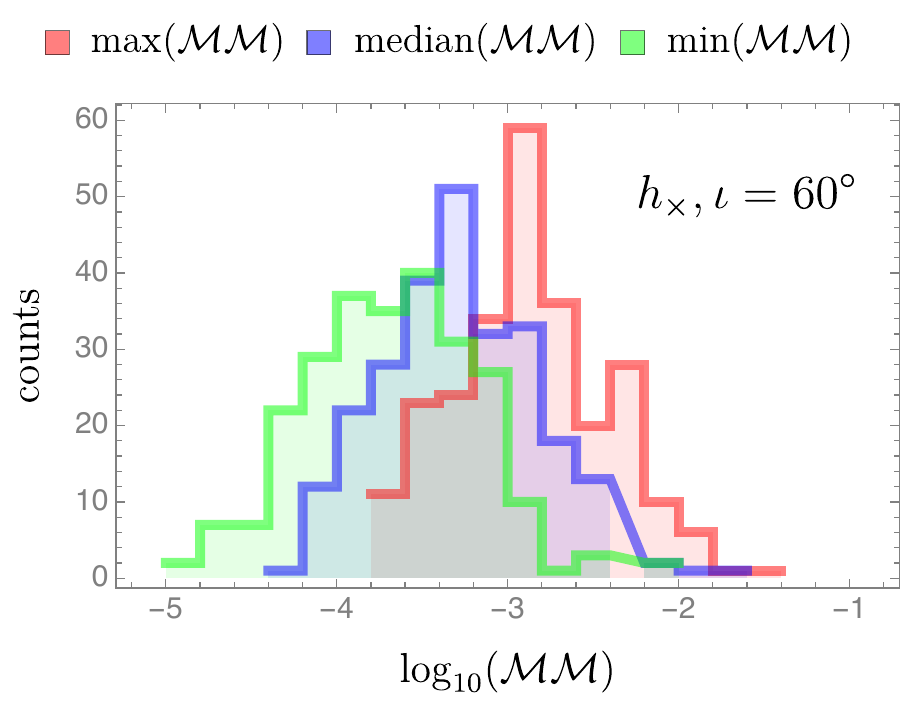}
   \includegraphics[width=0.5\columnwidth]{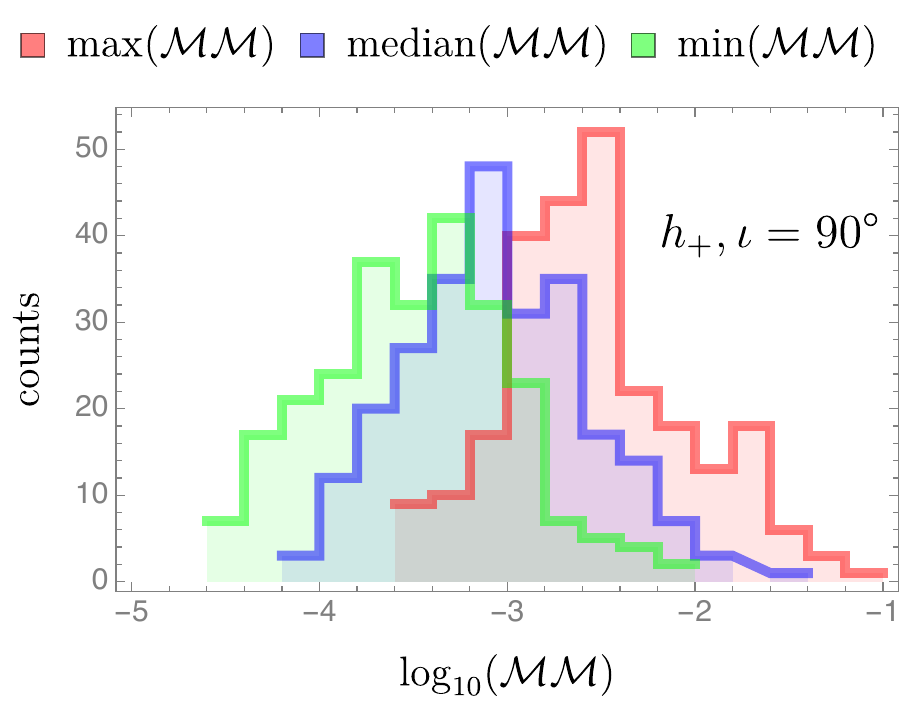}%\includegraphics[width=0.5\columnwidth]{Plots/MMatch_3_hc.pdf}
    \caption{Mismatches for the $h_+$(left) and $h_{\times}$(right) polarizations between hybrids and \phXHM for three different inclinations.
    For edge-on systems we only show results for the $h_+$ polarization since $h_{\times}$ vanishes.
    The minimum, maximum and median are taken over the range of total masses and azimuthal angle $\phi_S$ for the hybrids. The mismatch is numerically optimized over the azimuthal angle $\phi_T$ of the model (see main text for the details).}
    \label{fig:multimode_matches}
\end{figure}

When having a multi-mode waveform, not only it is important to model accurately each individual mode but also the relative phases and time shifts between them. To test this we compute the mismatch for the $h_+$ and $h_{\times}$ polarizations between our hybrids and the model for three inclination values: 0, $\pi/3$ and $\pi/2$ (rad). The polarizations for the hybrids and the model are built with the same inclination, however the azimuthal angle entering the spherical harmonics can be different. We denote by $\phi_S$ and $\phi_T$ the azimuthal angle of the hybrids (source) and model (template) respectively. $\phi_S$ takes the values of an equally spaced grid of five points between 0 and $2\pi$. Then for each value of $\phi_S$ we numerically optimize the value of $\phi_T$ that gives the best mismatch. For each configuration of inclination and $\phi_S$ the mismatch is computed for an array of 8 total masses from 20$M_\odot$ to 300$M_\odot$ logarithmically spaced and then we take the minimum, median and maximum values over all these configurations. Similarly to the single-mode matches we used the Advanced-LIGO design sensitivity Zero-Detuned-High-Power noise curve and a lower cutoff of 20 Hz. The results are shown in Fig.~\ref{fig:multimode_matches}. It can be observed that the mismatches degrade for higher inclinations due to the weaker contribution of the dominant $(2,2)$ mode which is the best modelled mode, although events that are seen close to edge-on are much less likely to be detected due to the reduced signal-to-noise-ratio. For edge-on systems we only show results for the $h_+$ polarization, since the $h_{\times}$ vanishes. Alternative ways of quantifying multi-mode mismatches have been used in the literature, see e.g.~ \cite{seobnrv4hm} or \cite{blackman2017}, with different advantages and drawbacks. The quantities we show here are chosen for simplicity.
%we opted for showing in this way because it is the easiest and clearest since it does not use neither a detector network nor a particular configuration of the detectors. 

\subsection{Recoil}
\label{subsec:recoil}
 
Asymmetric black-hole binaries will radiate gravitational waves anisotropically. This will result in a net emission of linear momentum, at a rate (in geometric units)
\begin{equation}
\frac{dP^{k}}{dt}=\frac{r^2}{16\pi}\int{d\Omega \left( \dot{h}_+^2+\dot{h}_\times^2 \right) n^k},
\end{equation}
where $n^k$ is the radial unit-vector pointing away from the source, leading the final remnant to recoil in the opposite direction. The precise value of the final recoil velocity will depend on the interactions among different GW multipoles. This quantity is extremely sensitive to the relative time and phase shifts among different modes and thus provides an excellent and physically meaningful test-bed for our model. 

\begin{figure}[htpb]
    \centering
    \includegraphics[width=\columnwidth]{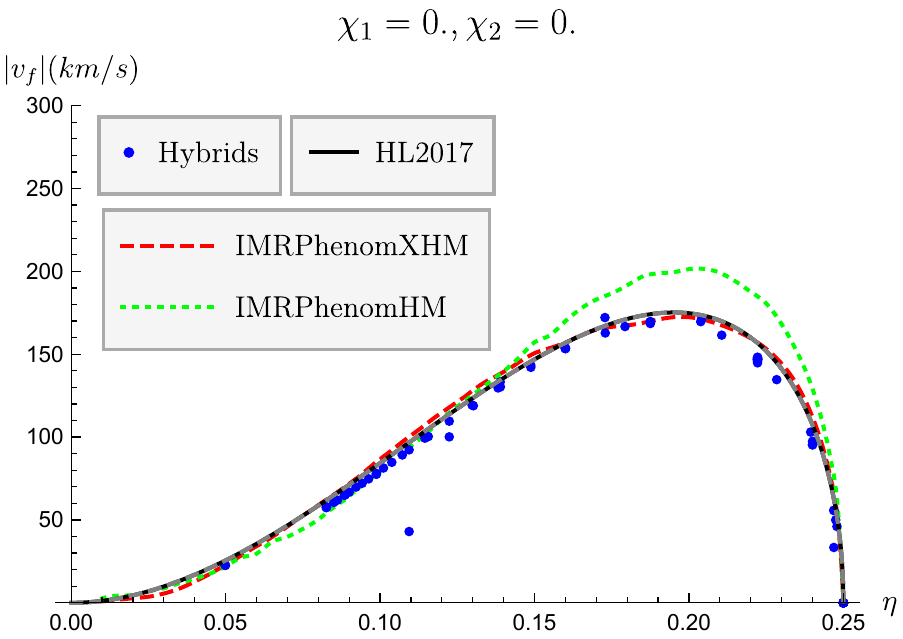}
    \caption{In this plot we show the absolute value of the final recoil velocity for a non-spinning black-hole binary, as computed with \phXHM (in red) \phHM (green), and with the fit of Ref. \cite{HL2017} (black). In blue we show the recoil velocity for all the non-spinning configurations in our calibration dataset. We can see that, when a good number of NR waveform is available, our calibrated model can reproduce with great accuracy the final velocity of the remnant.}
    \label{fig:nonspinning_kicks}
\end{figure}

\begin{figure}[htpb]
    \centering
    \includegraphics[width=\columnwidth]{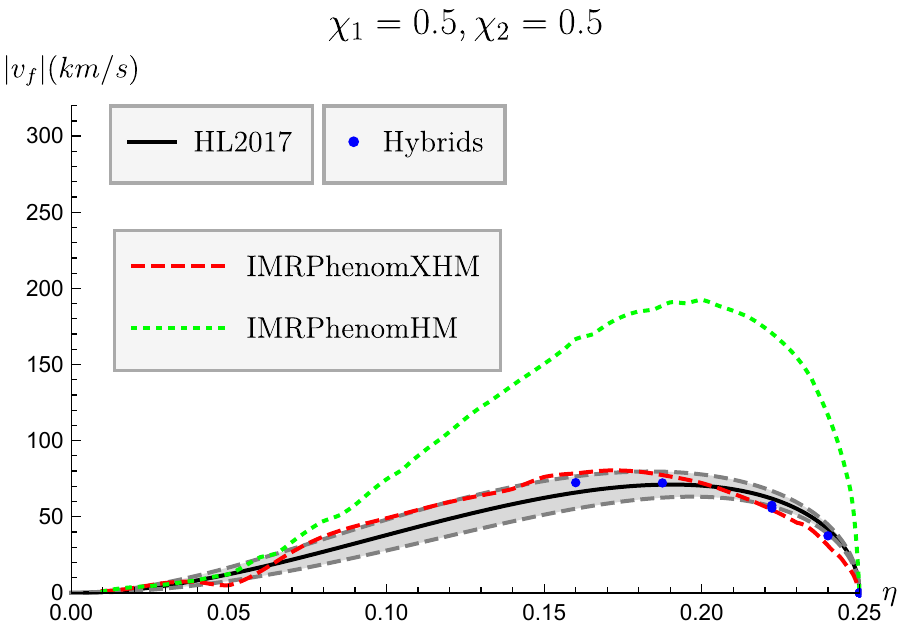}
    \caption{In this plot we show the absolute value of the final recoil velocity $|v_f|$ for an equally-spinning black-hole binary with adimensional spins $\chi_1=\chi_2=0.5$, as computed with \phXHM (in red) \phHM (green), and with the fit of Ref. \cite{HL2017} (black, note that here we also shade in gray the fit's error margins, using the error estimates provided by the authors in Tab.~IV of the aforementioned reference). In blue we show the recoil velocity for all the corresponding configurations in our calibration dataset. Note that, despite the loss of accuracy due to having fewer waveforms than in the non-spinning case, our model returns a value of $|v_f|$ much closer to NR than the uncalibrated version.}
    \label{fig:spinning_kicks}
\end{figure}

We computed the final recoil velocity predicted by \phXHM for different binary configurations and compared the results with those obtained directly from our hybrid waveforms. As new numerical simulations became available, several works presented increasingly improved NR-based fits for the final recoil velocity (see for instance \cite{HLZpaper,HL2017,HealyLousto2018,Varma:2018aht} and the latter reference for further works and comparisons). Below we compare our results to the fit of Ref. \cite{HL2017}, for two test configurations: a black-hole binary where both bodies are non-spinning (Fig. \ref{fig:nonspinning_kicks}), and one where both are spinning with Kerr parameters $\chi_1=\chi_2=0.5$ (Fig. \ref{fig:spinning_kicks}). For comparison, we also show the recoil velocities obtained with \phHM.  

\subsection{Time-domain behaviour}
\begin{figure*}[htpb]
    \centering
    \includegraphics[width=0.8\columnwidth]{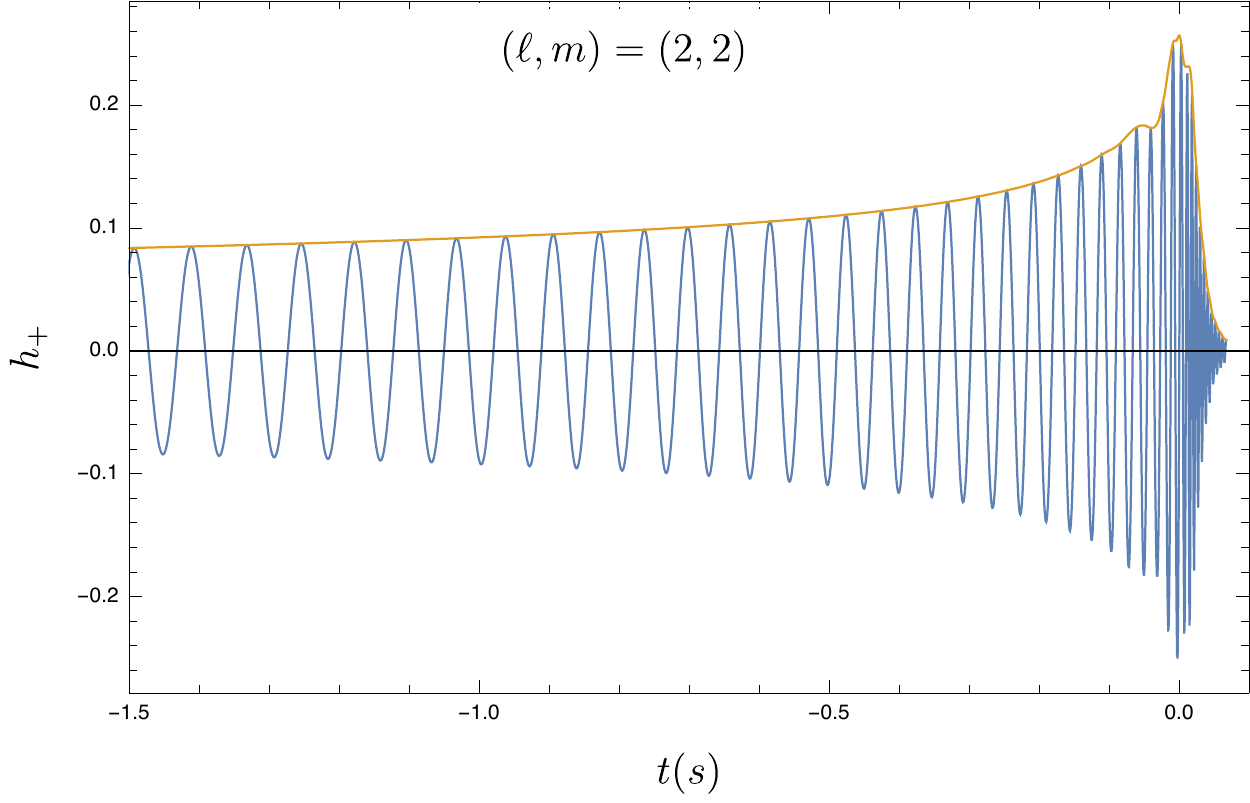}\\
    \includegraphics[width=0.8\columnwidth]{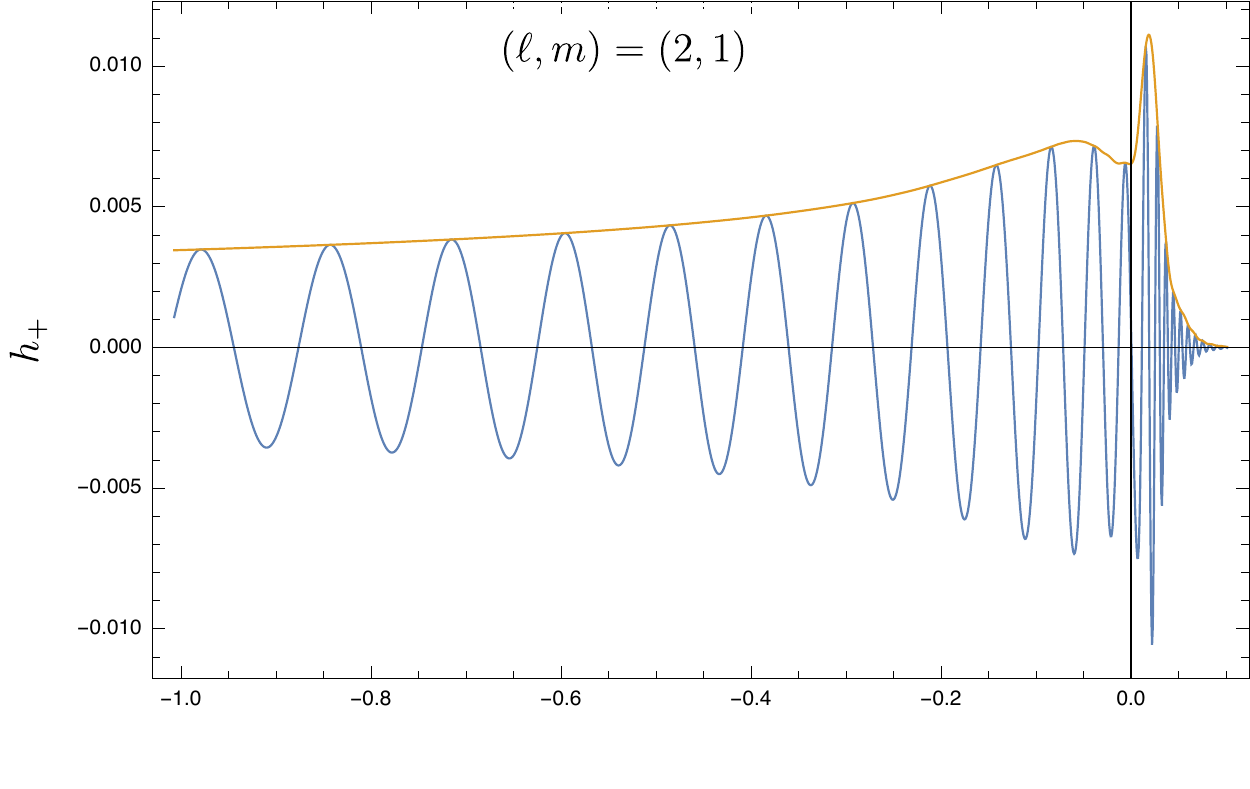}\includegraphics[width=0.8\columnwidth]{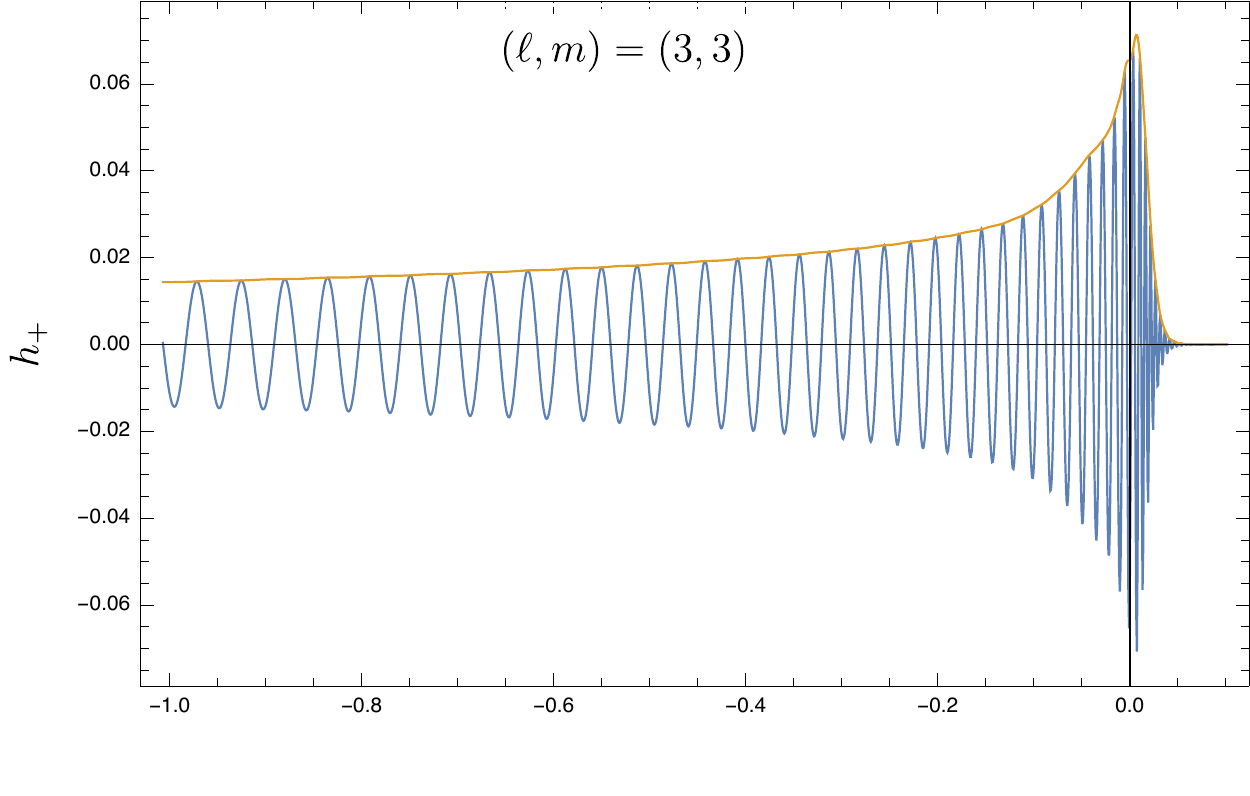}\\
    \includegraphics[width=0.8\columnwidth]{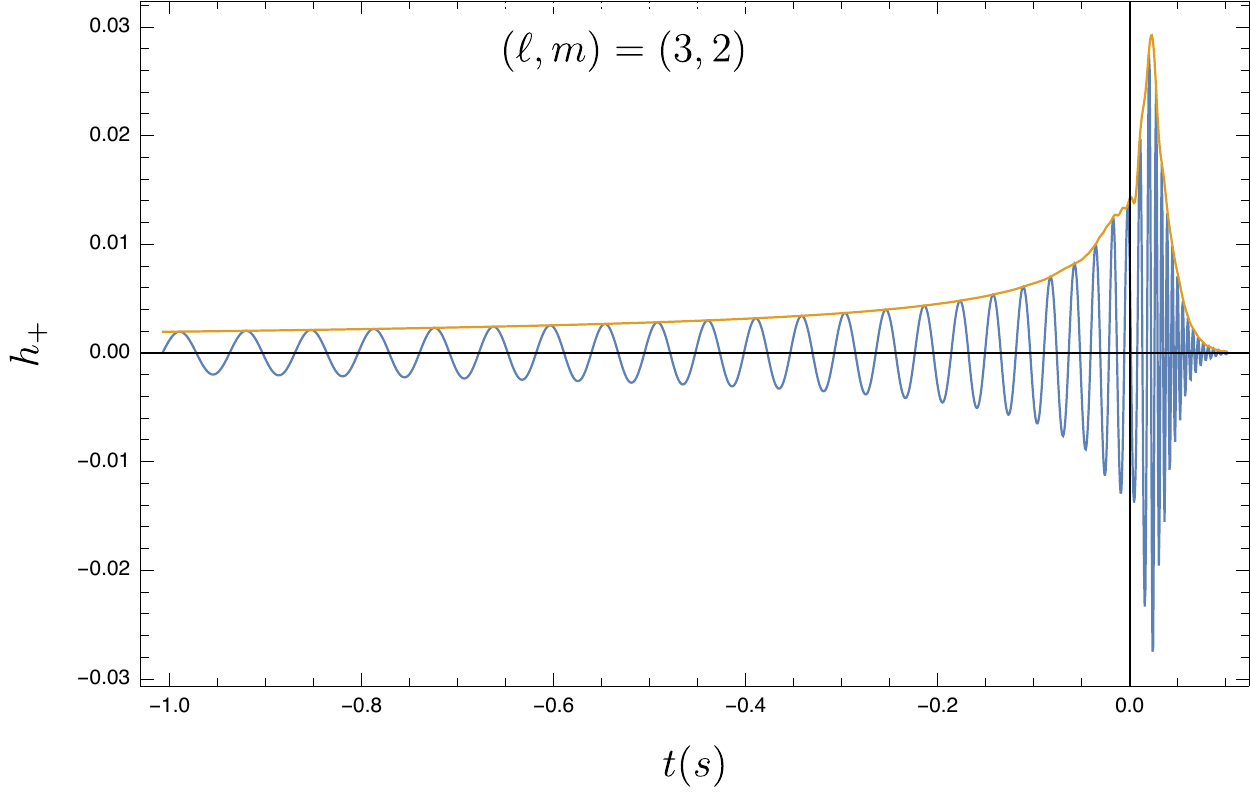}\includegraphics[width=0.8\columnwidth]{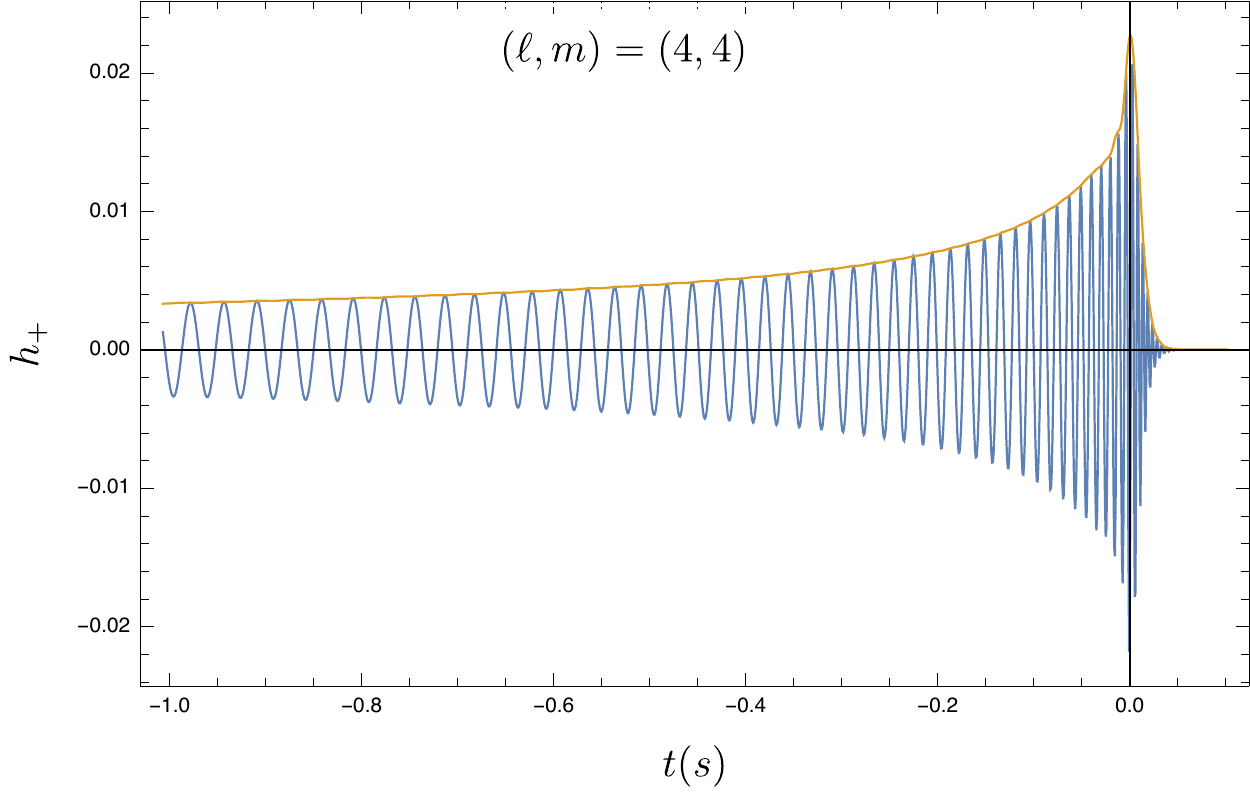}
\caption{Single-mode waveforms for a binary with $q=4$, with maximally spinning black holes. The model appears to extrapolate well beyond its calibration region ($|\chi_{1,2}|\leq0.99$).}
    \label{fig:TD_maximally_spinning}
\end{figure*}

\begin{figure*}[htpb]
    \centering
    \includegraphics[width=0.8\columnwidth]{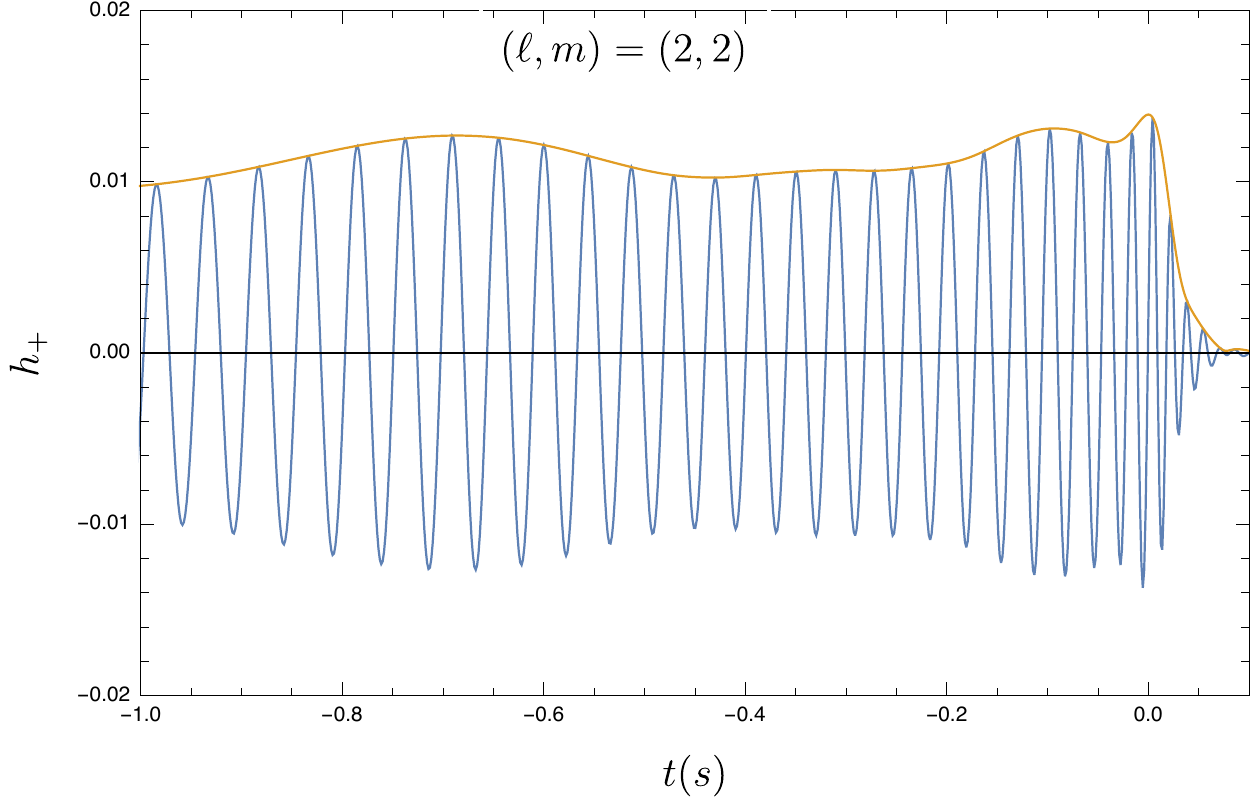}\\
    \includegraphics[width=0.8\columnwidth]{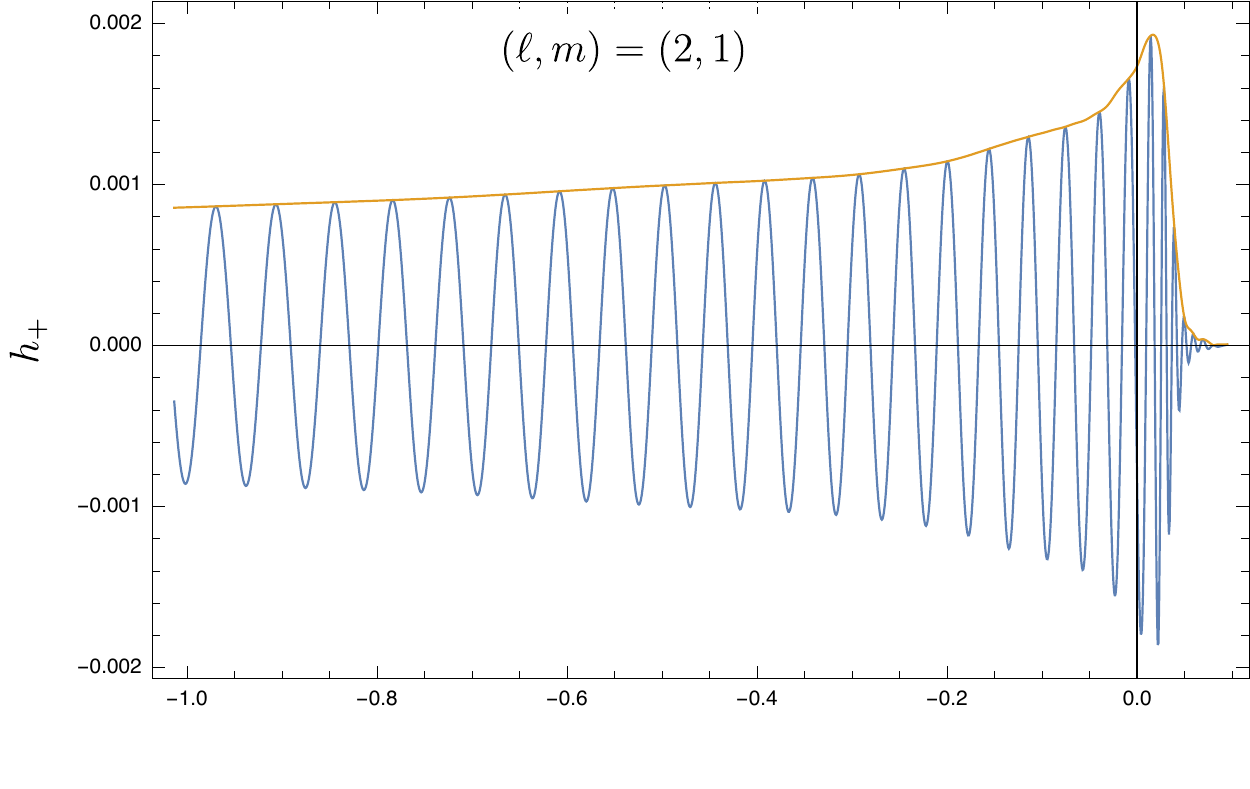}\includegraphics[width=0.8\columnwidth]{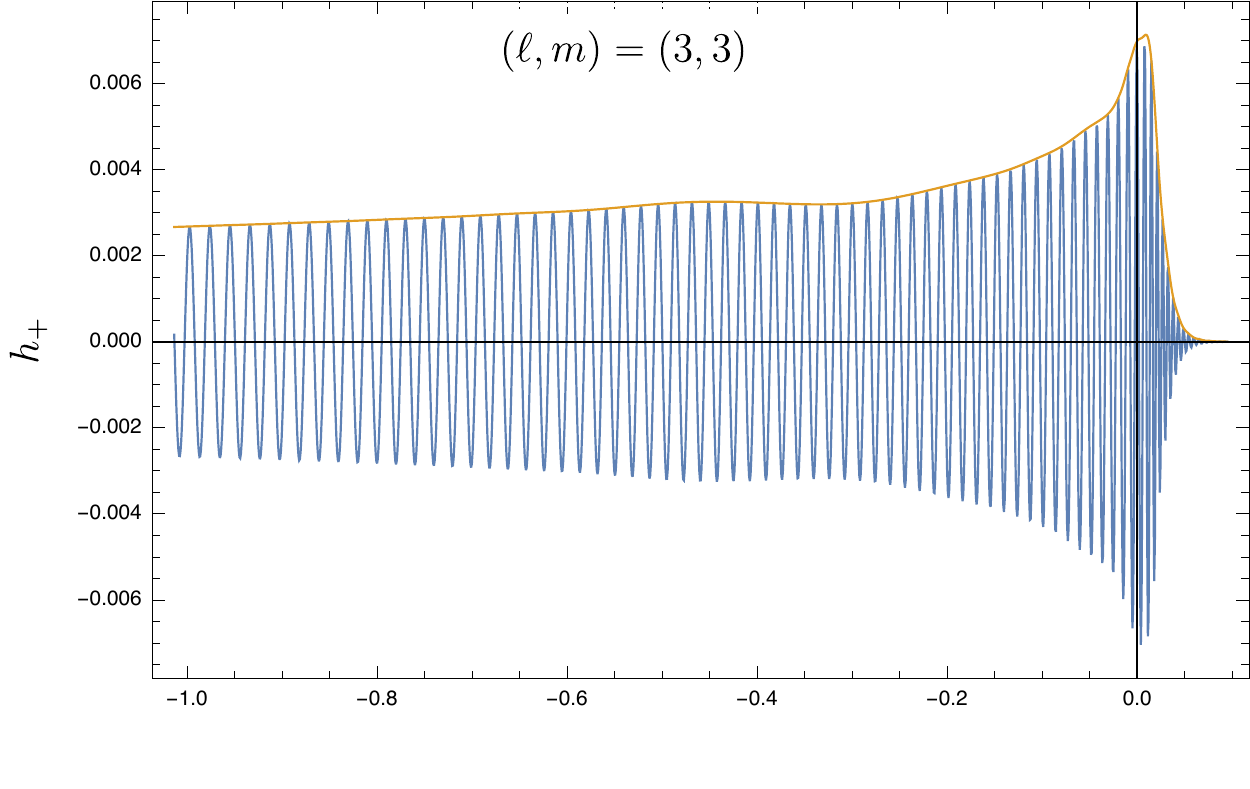}\\
    \includegraphics[width=0.8\columnwidth]{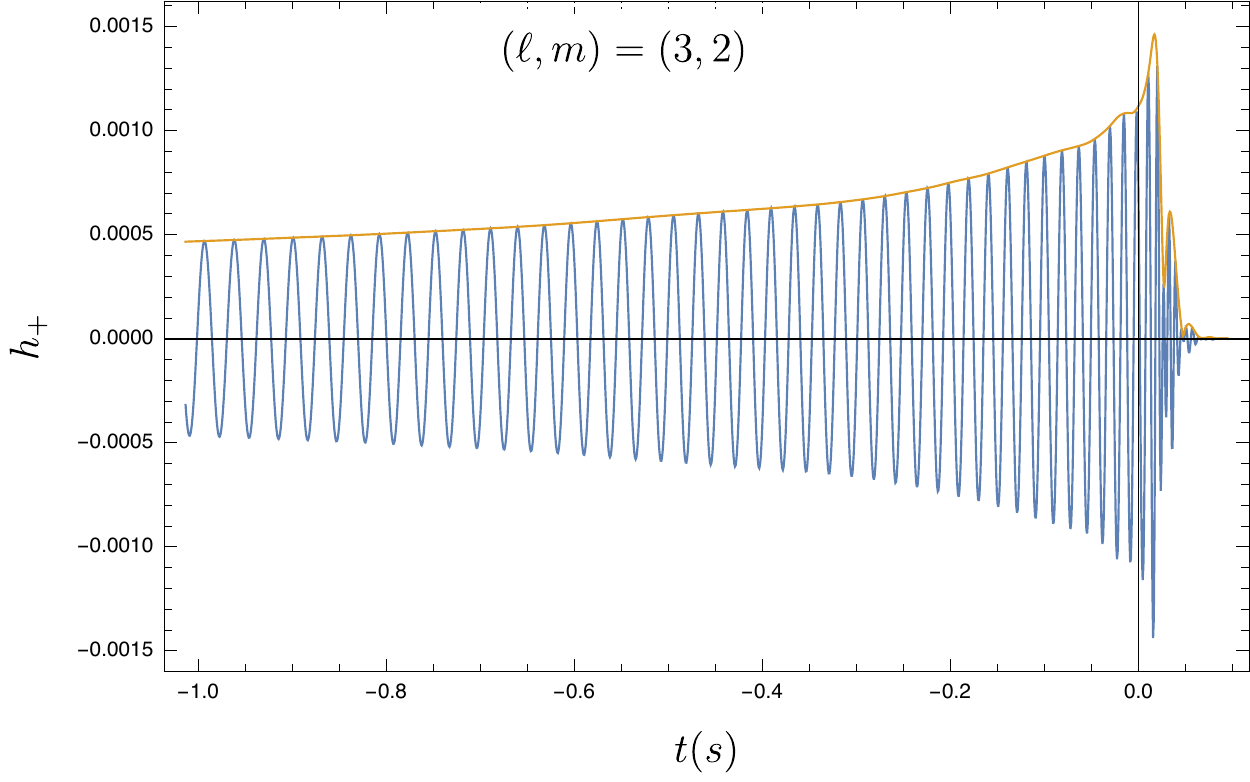}\includegraphics[width=0.8\columnwidth]{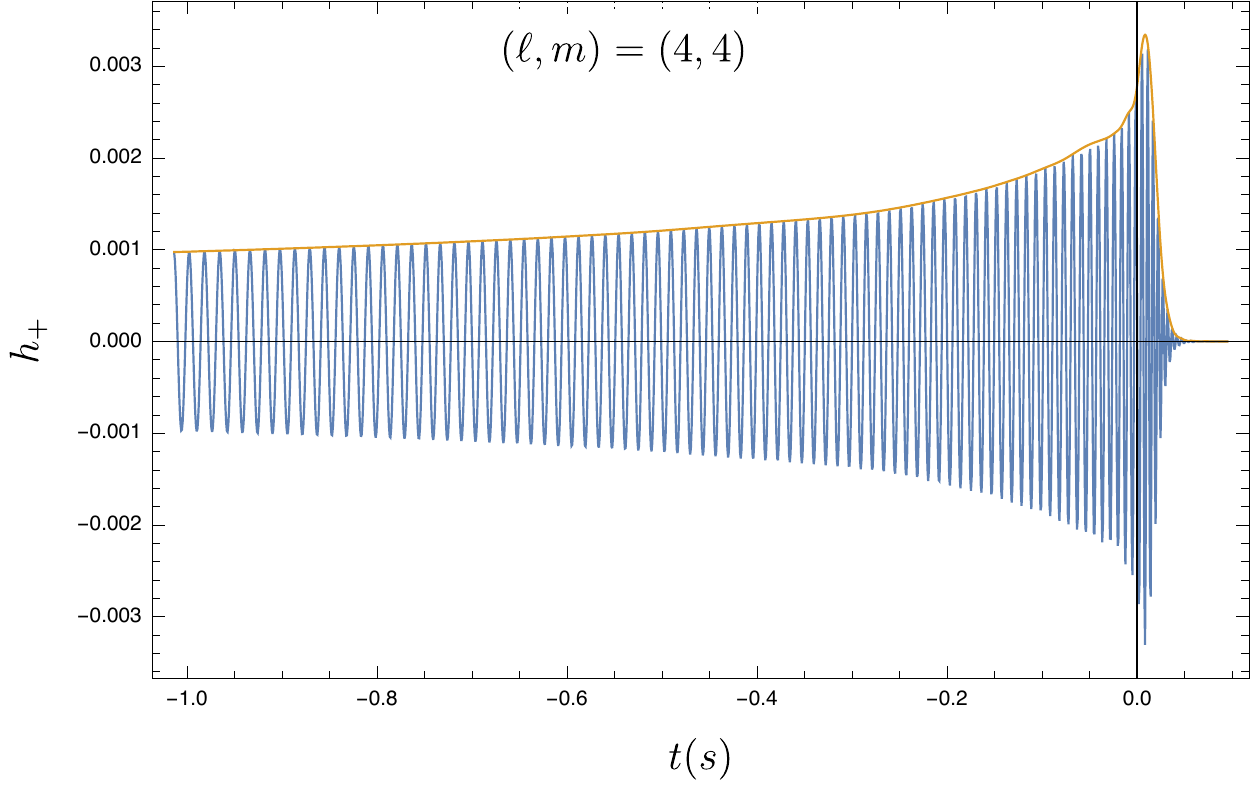}
\caption{Single-mode waveforms for a binary with $q=100,\chi_1=\chi_2=0.7$. There are no NR simulations in our calibration dataset with $q>18$, and the extrapolation to high mass-ratios is done by placing Teukolsky waveforms at the large-q boundary of the parameter space, as explained in Subsec. \ref{subsec:emr_waveforms}. Here we can see that the model achieves a smooth transition between NR and point-particle physics.}
    \label{fig:TD_high_q}
\end{figure*}

We have checked that the model has a reasonable behaviour in regions of the parameter-space where no simulations are available (e.g. $18\leq q \leq 200$) and for extreme spins. We show here example-waveforms to test both these regimes. Figs. \ref{fig:TD_maximally_spinning} and \ref{fig:TD_high_q} show single-mode waveforms for binaries with parameters $(q,\chi_1,\chi_2)=(4,1.,1.)$ and $(q,\chi_1,\chi_2)=(100,0.7,0.7)$, respectively. We can see that in both cases the model returns well-behaved waveforms.

\subsection{Parameter estimation: GW170729}
\label{sec:GW170729}

In our companion paper to present \phX for the $(2,2)$
mode we re-analyzed the data for the first gravitational wave event, GW150914, as an example for an application to parameter estimation. Here we present a re-analysis of GW170729, where the effect of models with subdominant higher harmonics has been discussed in the literature \cite{Chatziioannou:2019dsz}, and we will demonstrate 
broad agreement between \phXHM, \phHM and SEOBNRv4HM for this event. Again we use coherent Bayesian inference methods to determine the posterior distribution $p(\vec{\theta} | \vec{d})$ to derive expected values and error estimates for the parameters of the binary. Following 
\cite{Chatziioannou:2019dsz}, we use the public data for this event from the Gravitational Wave Open Science Center (GWOSC) \cite{Vallisneri:2014vxa,GWOSC,o2data} calibrated by a cubic spline and the PSDs used in \cite{LIGOScientific:2018mvr}. We analyze four seconds of the strain data set with a lower cutoff frequency of 20Hz. For our analysis we use the \texttt{LALInference}
\cite{Veitch:2014wba}
implementation of the nested sampling algorithm. We perform the runs using 2048 'live points' for five different seeds, then merge into a single posterior result. We choose the same priors used in \cite{Chatziioannou:2019dsz}, taking into account that \phXHM is a non-precessing model and we have to use aligned spin priors.

\begin{figure*}[htpb]
    \centering
\includegraphics[width=0.8\columnwidth]{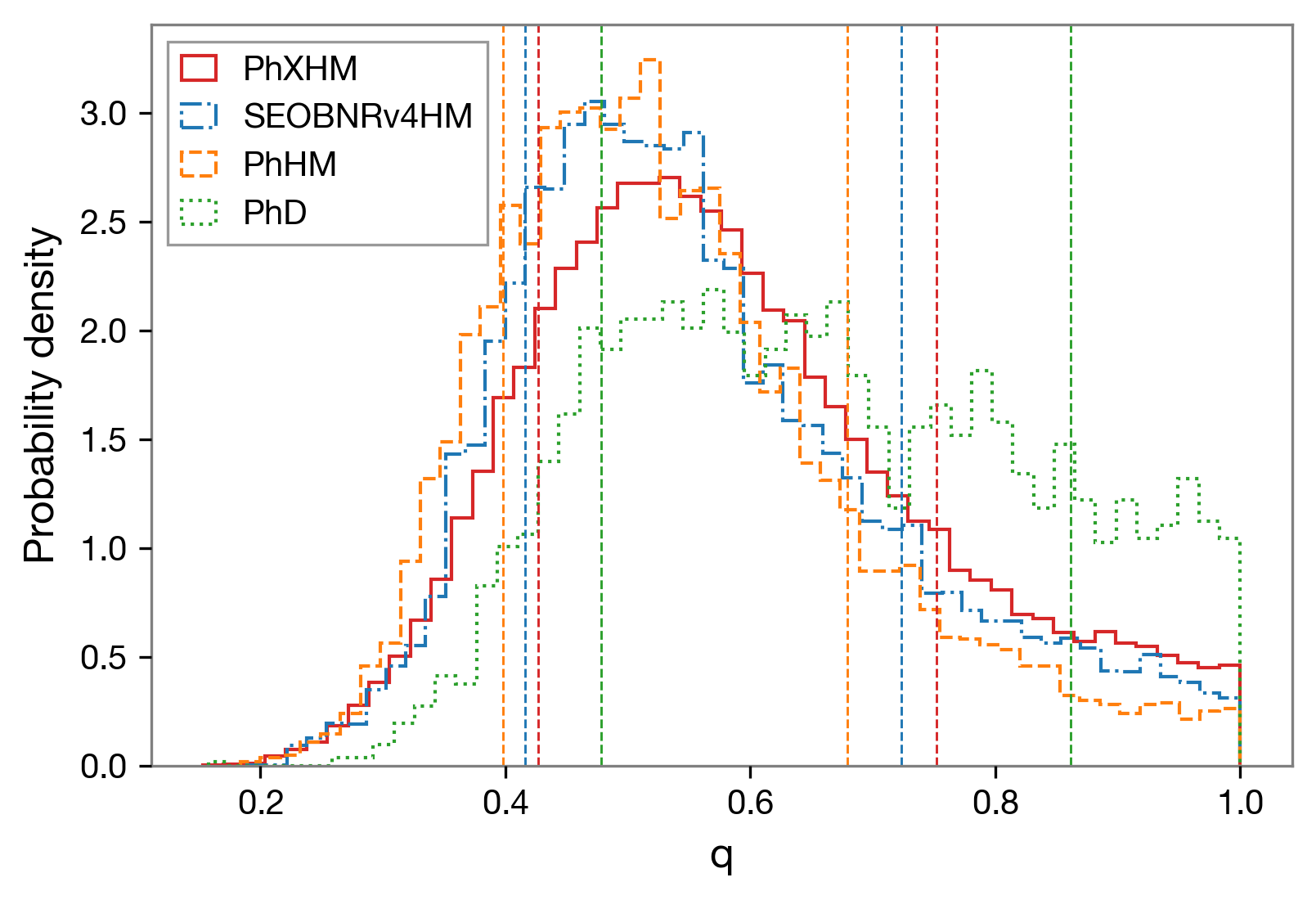}\includegraphics[width=0.8\columnwidth]{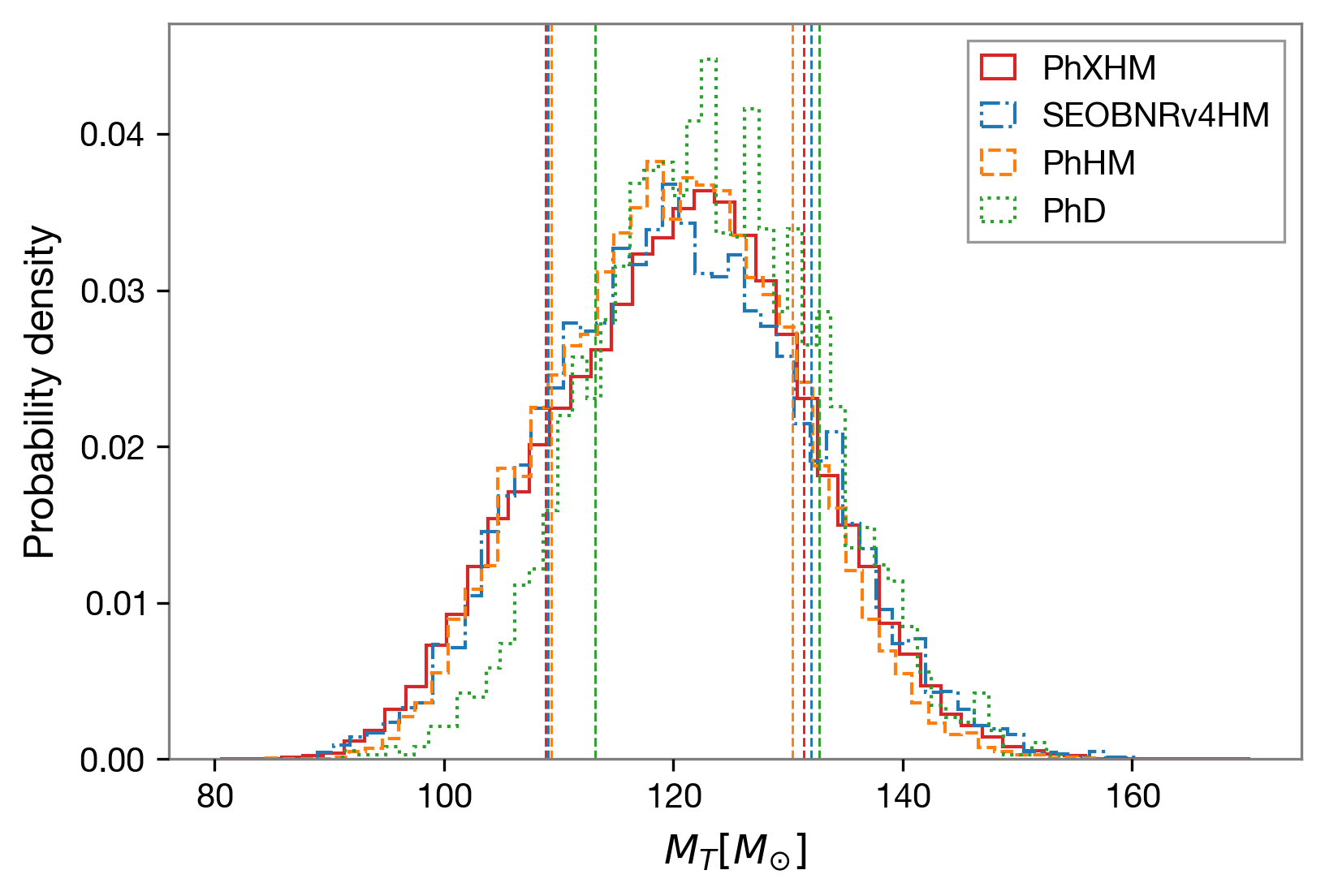}
\includegraphics[width=0.8\columnwidth]{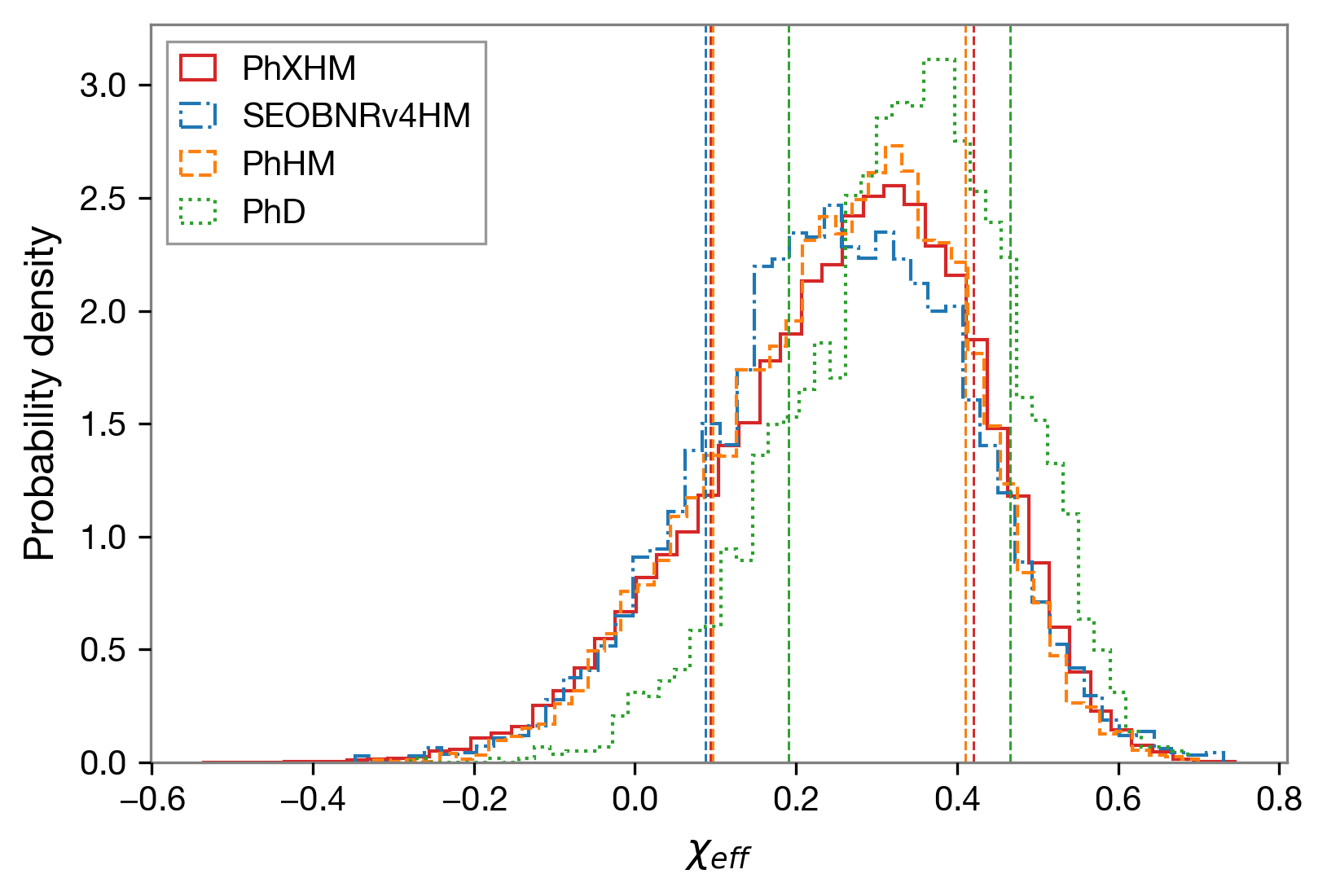}\includegraphics[width=0.8\columnwidth]{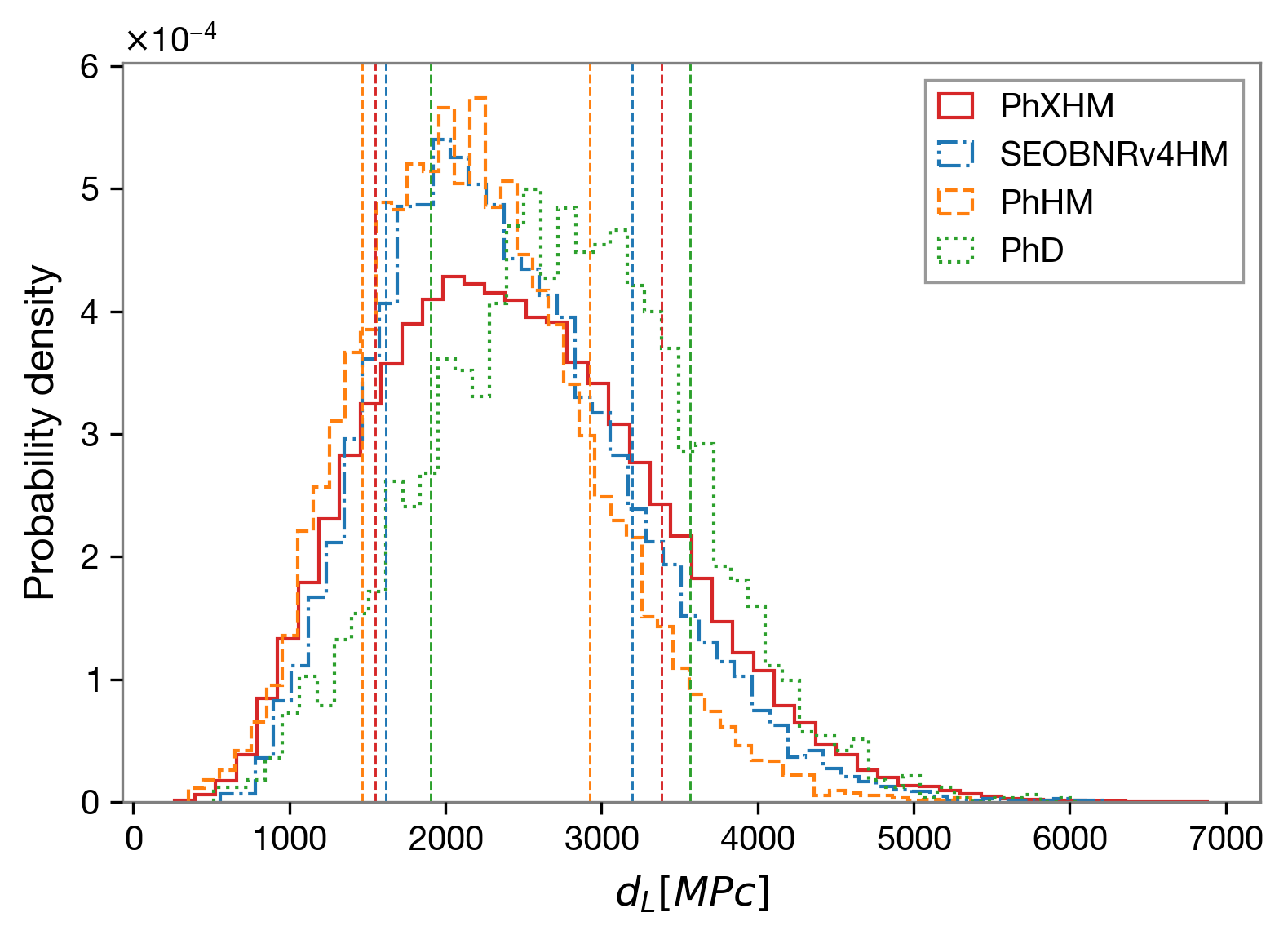}
% results from the literature
    \caption{Comparison between \phXHM, %  and   for
    \phHM, SEOBNRv4HM and \phD (results for the latter three models are taken from \cite{Chatziioannou:2019dsz}) for the
    event GW170729 as discussed in Sec.~\ref{sec:GW170729}. We show posterior distributions of mass ratio, total mass, effective aligned spin and luminosity distance. The dashed vertical lines mark the $90\%$ confidence limits.}
    \label{fig:PE}
\end{figure*}

In Fig.~\ref{fig:PE} we compare our results with the higher mode models (\phHM and SEOBNRv4HM) and the (2,2) mode model results (\phD) published in \cite{Chatziioannou:2019dsz}. 
We find that the posteriors derived from \phXHM are consistent with the two other models that include higher harmonics, which can be distinguished from the results obtained for models that only include the (2,2) mode.

\subsection{NR injection study}
\begin{figure}[htpb]
    \centering
    \includegraphics[width=\columnwidth]{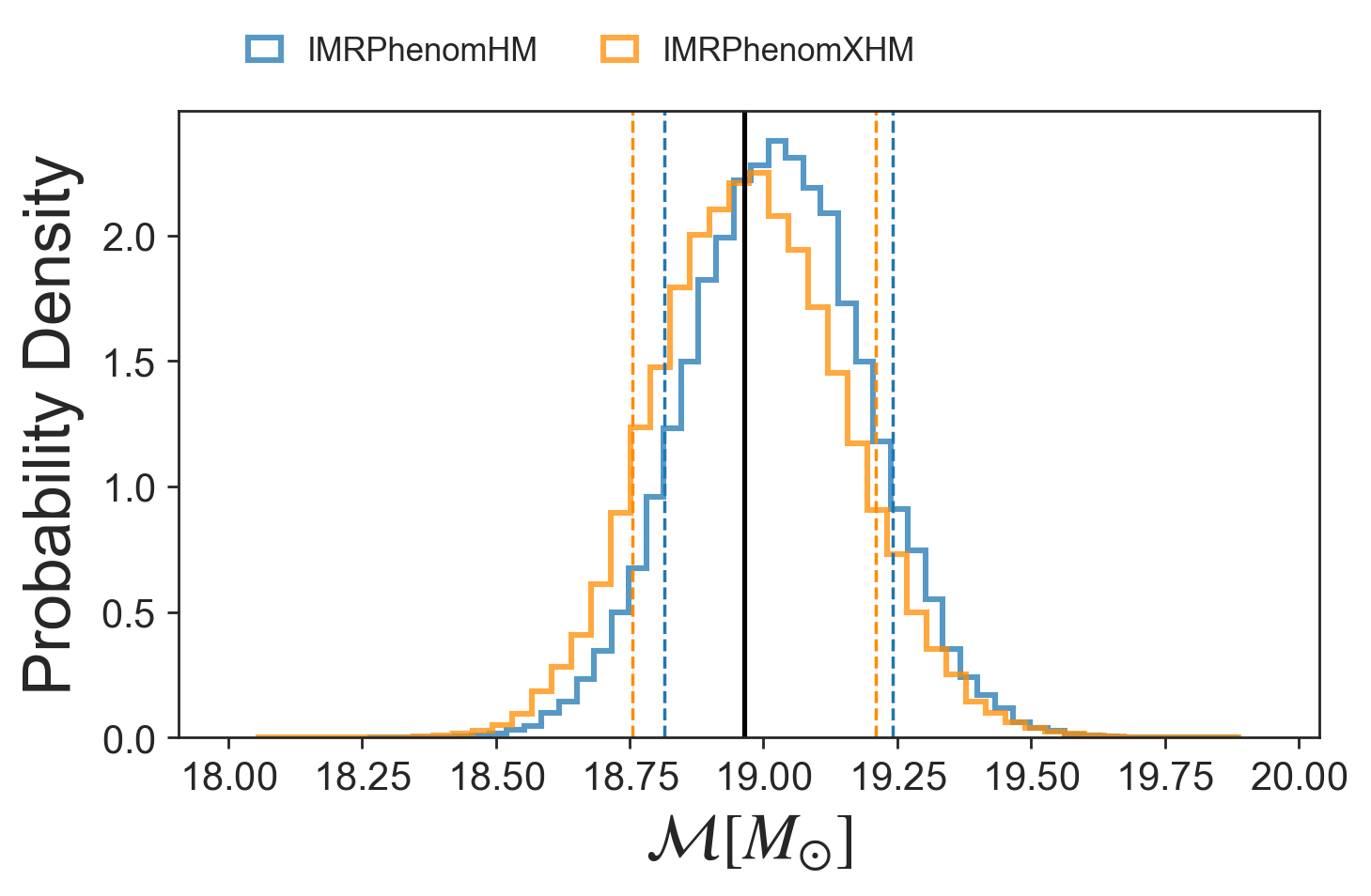}\\
    \includegraphics[width=\columnwidth]{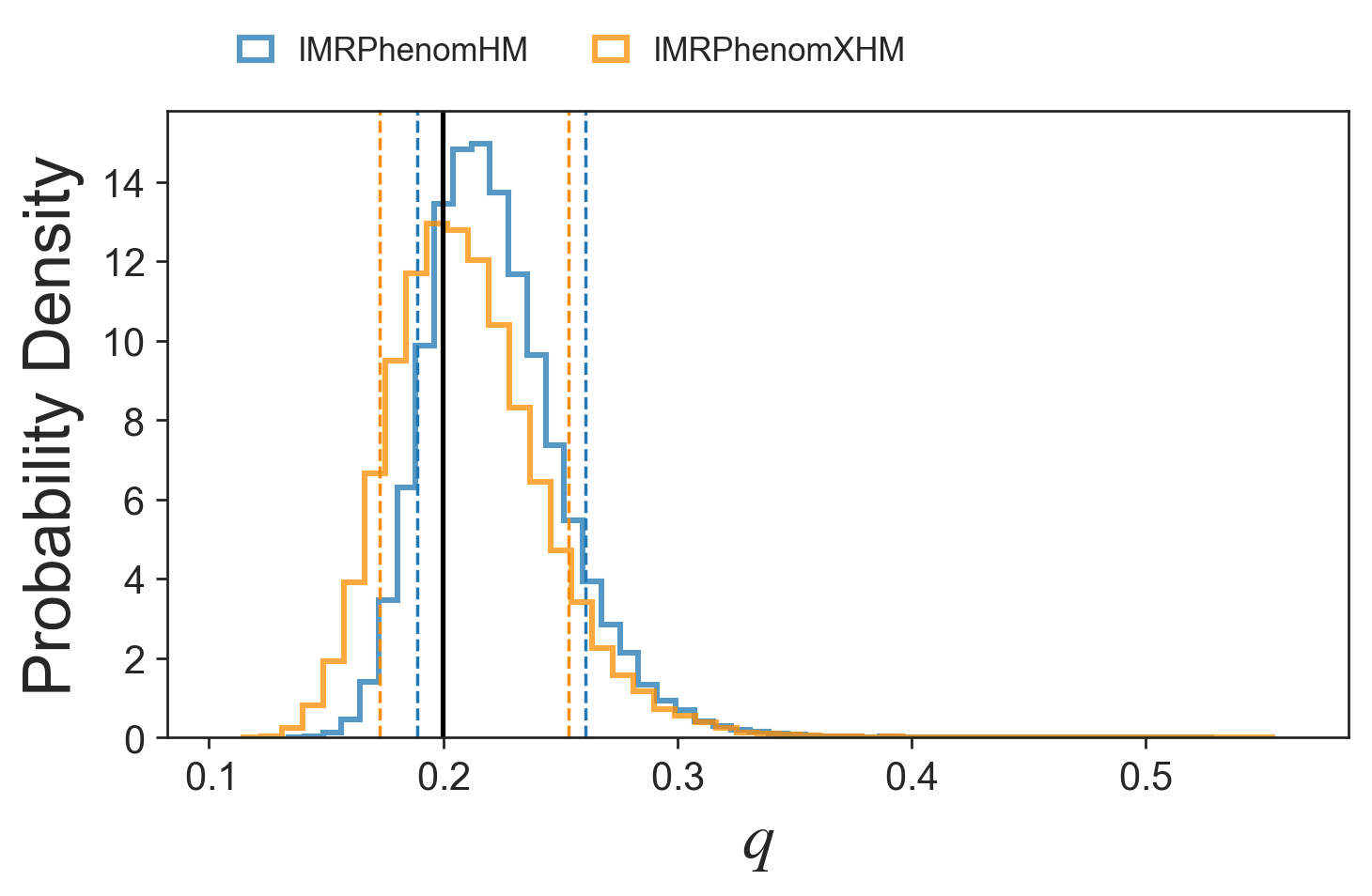}\\
    \includegraphics[width=\columnwidth]{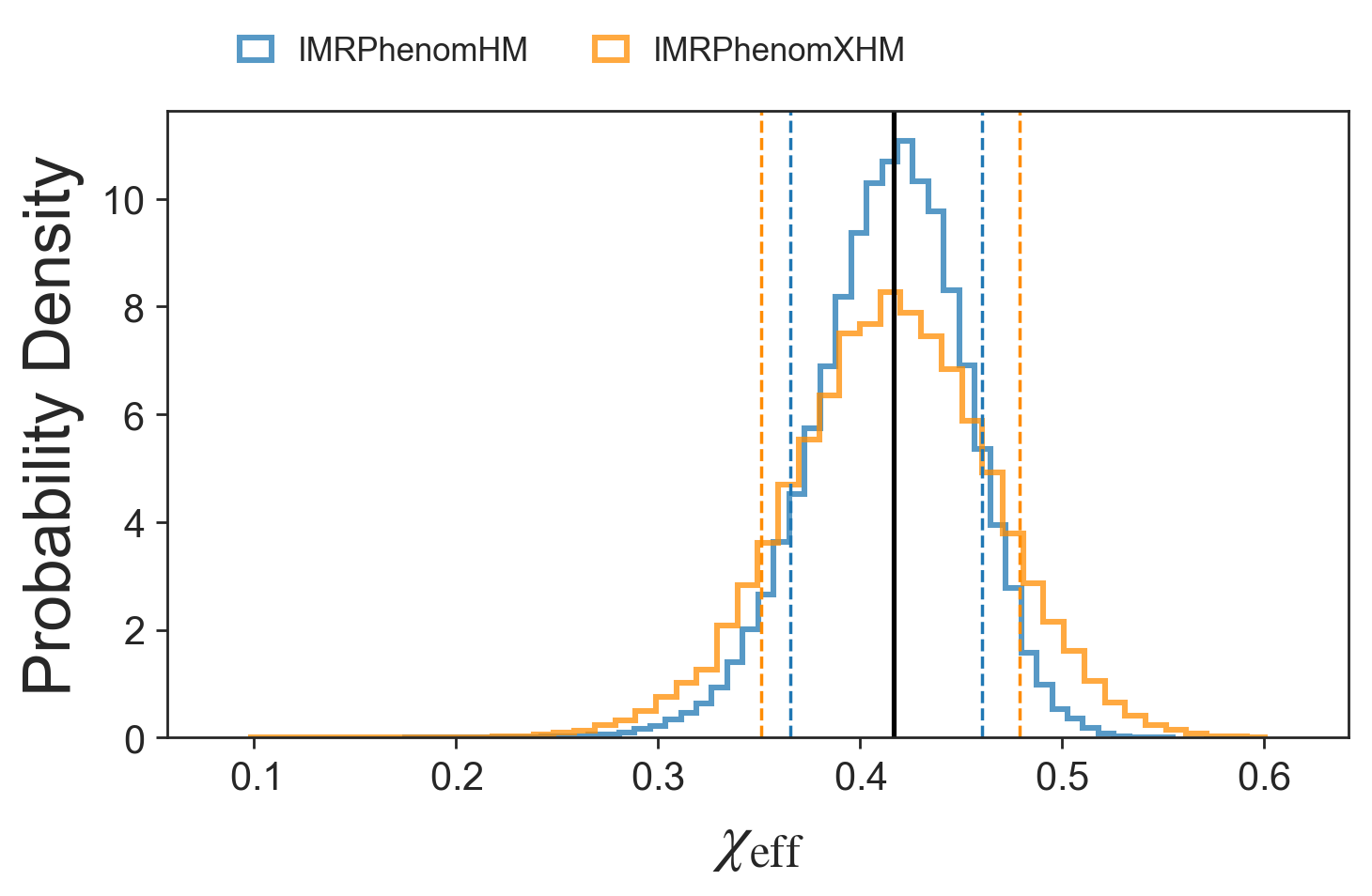}
    \caption{1D posterior distributions for some of the mass and spin parameters characterizing the mock signal generated using the public SXS simulation \textsc{SXS:BBH:0110}. Results obtained with \phXHM and \phHM are plotted in orange and blue respectively. A vertical black line marks the injected value. Dashed lines indicate 90\% credible intervals.}
    \label{fig:posteriors_inj}
\end{figure}

\begin{figure}[htpb]
    \centering
    \includegraphics[width=\columnwidth]{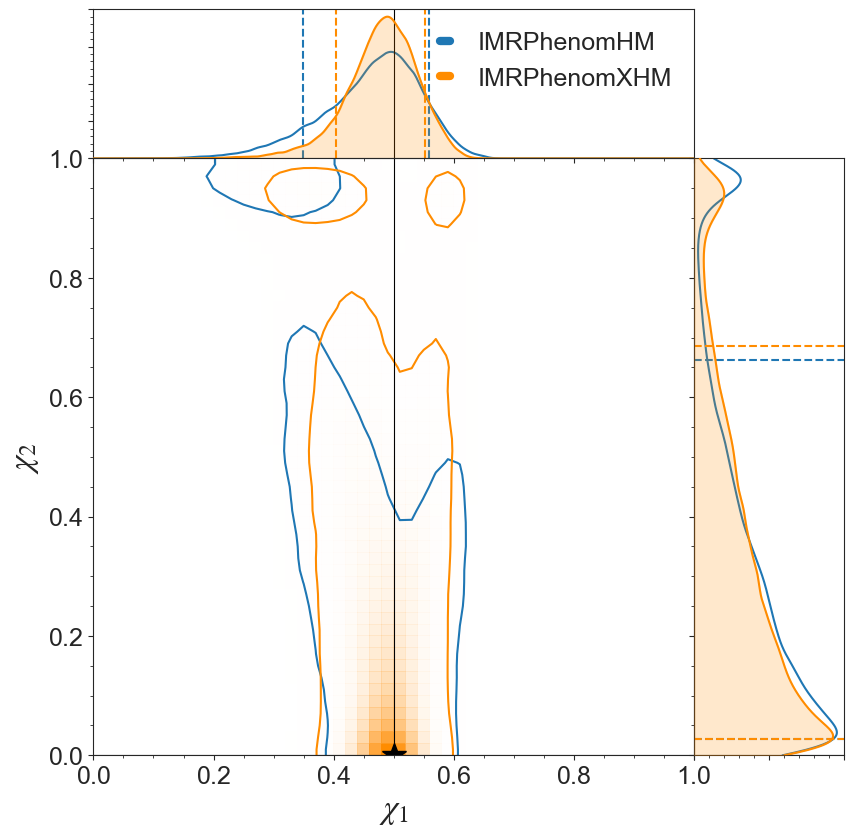}
    \caption{Joint posterior distributions for the individual spins $\chi_1,\chi_2$. Dashed lines indicate 90\% credible intervals. Solid black lines mark the injected values for each parameter (note that the secondary has zero spin, and therefore the line in this case coincides with the x-axis). }
    \label{fig:2D_posteriors_inj}
\end{figure}

As a further test of the improvements brought by \phXHM, we have performed parameter estimation of a synthetic signal generated using the public SXS waveform \textsc{SXS:BBH:0110}. This corresponds to a binary with strongly asymmetric masses ($q=5$), and adimensional spin magnitudes $\chi_1=0.500,\chi_2=0.$ at a reference frequency of 20 Hz. 
We injected the signal into a Hanford-Ligo-Virgo detector network in zero-noise and used Advanced-LIGO design sensitivity PSDs.  The mock signal had chirp mass $\mathcal{M}=18.96 M_\odot$ and total mass $62 M_\odot$ in detector frame, right ascension $\mathrm{ra}=1.4$ rad, declination $\mathrm{dec}=-0.6$ rad and geocentric time $t_c=1126259600.0$ s. The source was placed at a luminosity distance $d_{L}=$ 600 Mpc at an inclination $\iota=\pi/3$. The network SNR of this configuration was $17.9$. 
For this analysis we used the gravitational-wave inference library \textsc{pBilby} \cite{Ashton:2018jfp,smith2019massively} with dynamic nested sampling \cite{10.1093/mnras/staa278} and 2048 live points. 
We present our results in Fig.\,\ref{fig:posteriors_inj} and \ref{fig:2D_posteriors_inj}. \phXHM delivers a better recovery of the mass ratio (centre panel of Fig.\,\ref{fig:posteriors_inj}) and, although the measurement of $\chi_\mathrm{eff}$ is very consistent with \phHM, the spin of the primary appears to be more tightly constrained by the upgraded model. 

\subsection{Computational cost}
We will now compare the computational cost of the evaluation of different models available in \texttt{LALSimulation} compared to the IMRPhenomX family. Since the different models include a different number of modes we also show the evaluation time per mode. Using the \texttt{GenerateSimulation} executable within \texttt{LALSimulation} we compute the average evaluation time over 100 repetitions for a non-spinning case $(q,\chi_1,\chi_2)=(1.5,0,0)$ for a frequency range of 10 to 2048 Hz. We vary the total mass of the system from 3 to 300 solar masses and the frequency spacing $df$ is automatically chosen by the function \texttt{SimInspiralFD} to take into account the length of the waveform in the time domain for the given parameters. All the timing calculations were carried out in the LIGO cluster CIT to allow comparison with the benchmarks we have shown in \cite{our_mb} to compare different accuracy thresholds of multi-banding which is a technique that accelerate the evaluation of the model by evaluating it in a coarser non-uniform frequency grid and using interpolation to get the waveform in the final fine uniform grid, reducing considerably the computational cost. 

\begin{figure}[htpb]
    \centering
    \includegraphics[width=\columnwidth]{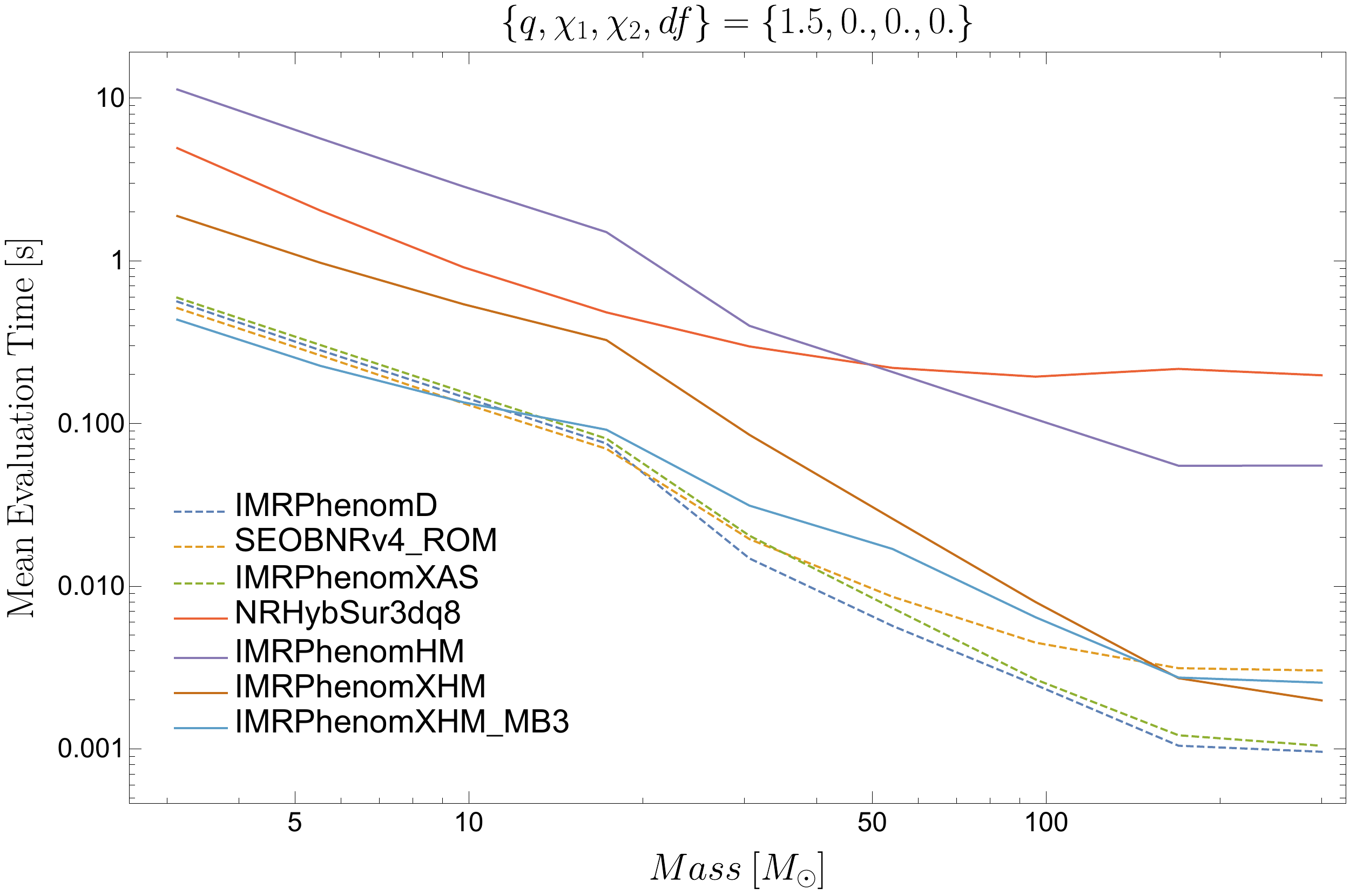}
    \includegraphics[width=\columnwidth]{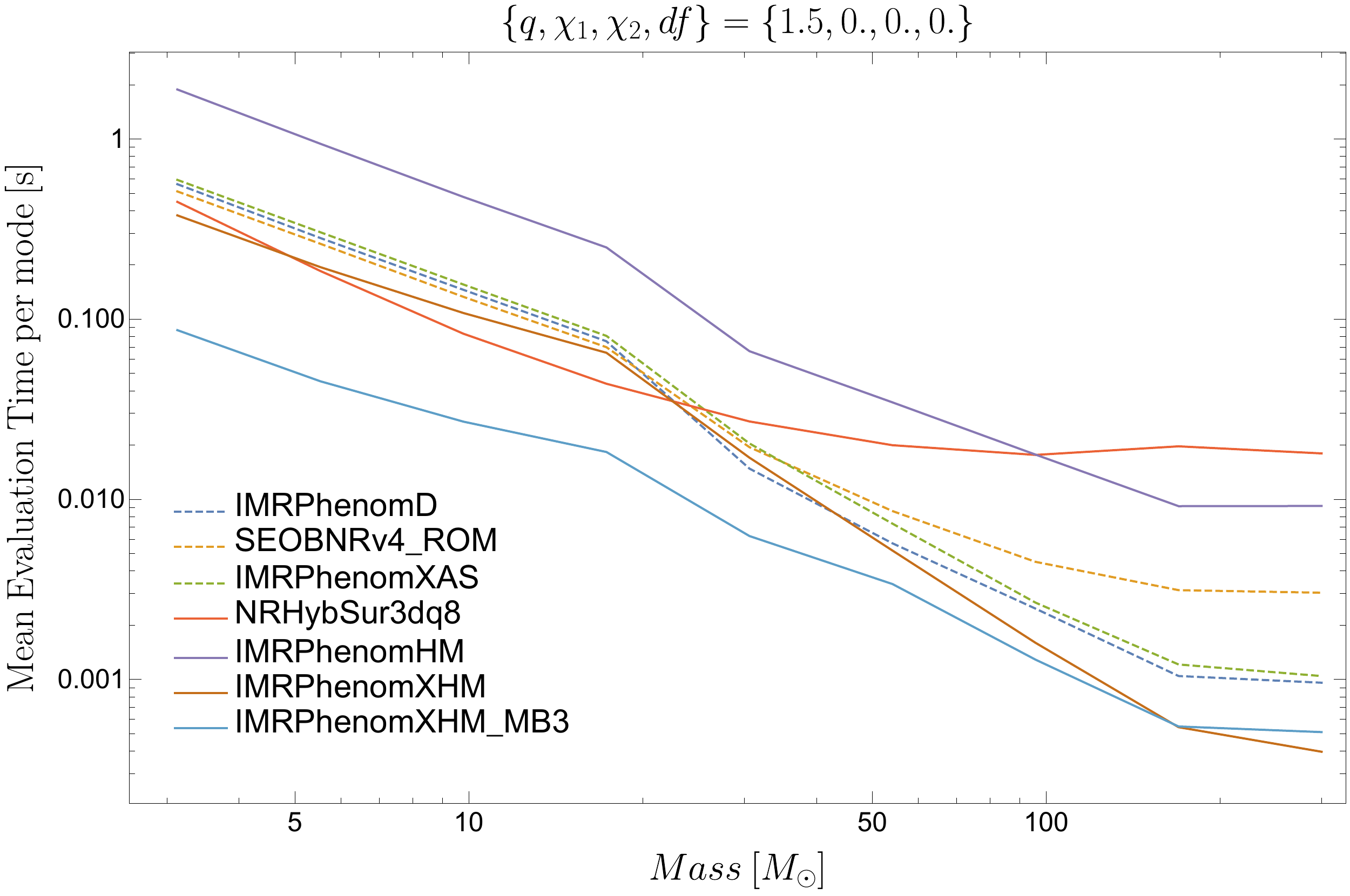}
    \caption{Evaluation time of different waveform models in \texttt{LALSimulation}. Top panel: we show the total contribution of the evaluation of the quadrupolar/multimode waveform. Bottom panel: evaluation time per mode, the models include different numbers of modes, so we average over this number for a fair comparison.}
    \label{fig:benchmarks}
\end{figure}

In Fig.~\ref{fig:benchmarks} the dashed lines represent models for only the 22 mode (\phD, \vfourom and \phX). We see that the three models show very similar performance for low masses, while for higher masses the IMRPhenom models are faster.  
Models that include higher modes are shown with solid lines: \surro (11 modes), \phHM (6 modes)  and  \phXHM (5 modes), the latter is shown with and without the acceleration technique of multibanding \cite{our_mb}. Since \surro is a time domain model, for many applications the actual evaluation time would also include the time for the Fourier transformation to the frequency domain, which can also lead to requirements for a lower start frequency and windowing to avoid artefacts from the Fourier transforms, which again would increase evaluation time. We also see that the new model \phXHM without multibanding is already significantly faster than the previous \phHM model. Comparing the new model to the surrogate, \phXHM without multibanding is significantly faster when considering all the modes, but the evaluation cost per mode is only lower for high masses. However, using the multibanding technique \cite{our_mb} with the threshold value of $10^{-3}$, which is the default setting when calling the model in \texttt{LALSuite} (IMRPhenomXHM\_MB3 in the plot),
\phXHM is significantly faster also for the evaluation time per mode. The threshold value can be adjusted to control the speed and accuracy of the algorithm as explained in Appendix~\ref{appendix:LAL} and in \cite{our_mb}, where we have shown that for an example injection of a relatively high signal-to-noise ratio 28, even at a threshold of $10^{-1}$, which evaluates significantly faster than the conservative default setting, differences in posteriors are hardly visible.

%%%%%%%%%%%%%%%%%%%%%%%%%%%%%%%%%%%%%%%%%%%%%%%%%%%%%%%%
\section{Conclusions} \label{sec:conclusions}

Phenomenological waveform models in the frequency domain have become a standard tool for gravitational wave parameter estimation \cite{LIGOScientific:2018mvr}
due to their computational efficiency, accuracy \cite{Kumar:2016dhh}, and simplicity. The current generation of such models has been built on the \phD model for the (2,2) mode of the gravitational wave signal of non-precessing and non-eccentric coalescing black holes, which has been extended to precession by the \phP \cite{Hannam:2013oca,PhenomPv2} and \pvthree \cite{Khan:2018fmp} models, to sub-dominant harmonics by the \phHM model, and to tidal deformations by the \pvtwotidal model \cite{Dietrich:2018uni,Dietrich:2019kaq}. 

The present paper is the second in a series to provide a thorough update of the family of phenomenological frequency domain models: In a parallel paper \cite{phenX} we have presented \phX, which extends \phD to a genuine double spin model, includes a calibration to extreme mass ratios, and improves the general accuracy of the model. In the present work we extend \phX to subdominant modes. Contrary to \phHM, the \phXHM model we present here is calibrated to numerical hybrid waveforms, and we have tested in Sec.~\ref{sec:quality} that the new model is indeed significantly more accurate than \phHM.

Calibration of the model to comparable mass numerical data has proceeded in two steps: we have started with a data set based on numerical relativity simulations we have performed with the BAM and Einstein Toolkit codes, and the data set corresponding to the 2013 edition of the SXS waveform catalog \cite{SXS:catalog} (including updates up to 2018). The quality and number of waveforms available at the time has determined the number of modes we model in this paper, i.e.~the $(2,1), (3,2), (3,3)$ and $(4,4)$ spherical harmonics.
During the implementation of the model in \texttt{LALSuite} \cite{lalsuite}, the 2019 edition of the SXS waveform catalog \cite{Boyle:2019kee} became available, and we have subsequently upgraded the calibration of \phX and of the subdominant mode phases to the 2019 SXS catalog.
We have not updated the amplitude calibrations, which are more involved but contribute less to the accuracy of the model. 
Instead, we plan to update the amplitude model in future work, where we will include further harmonics, in particular the $(4,3)$ and $(5,5)$ modes, which we can now calibrate to numerical date thanks to the increased number of waveforms and improved waveform quality of the latest SXS catalog.

The computational performance and a method to accelerate the waveform evaluation by means of evaluation on appropriately chosen unequally spaced grids and interpolation is presented in a companion paper \cite{our_mb}.

While \phXHM resolves various shortcomings of \phHM, further improvements are called for by the continuous improvement of gravitational wave detectors: 
We will need to address the complex phenomenology of the (2,1) harmonic for close to equal masses, add further modes as indicated above, and include non-oscillatory $m=0$ modes.

\phXHM can be extended to precession following \cite{Hannam:2013oca,PhenomPv2,Khan:2018fmp}. Regarding mode mixing in the context of precession one will however take into account that mixing then occurs between precessing modes \cite{ToniPrecession}, while current phenomenological precession extensions handle mode-mixing at the level of the co-precessing modes. 

%\foreach \n in {Insp1, Insp2, Insp3, Int1, Int2, ALambda, Lambda, Sigma}{%
%\begin{table}[H]
%\begin{center}
%\input{FitsStats/\n_21}%
%\end{center}
%\caption{\n\_21}%
%\end{table}
%}
%%%%%%%%%%%%%%%%%%%%%%%%%%%%%%%%%%%%%%%%%%%%%%%%%%%%%%%%

\section*{Acknowledgements} 

We thank Maria Haney for carefully reading the manuscript and valuable feedback, and the internal code reviewers of the LIGO and Virgo collaboration
for their work on checking our \texttt{LALSuite} code implementation.
We thank Alessandro Nagar, Sebastiano Bernuzzi and Enno Harms for giving us access to $\it{Teukode}$ \cite{Harms_2014}, which was used to generate our extreme-mass-ratio waveforms. 

This work was supported by European Union FEDER funds, the Ministry of Science, 
Innovation and Universities and the Spanish Agencia Estatal de Investigaci{\'o}n grants FPA2016-76821-P,        % FPA2017-90687-REDC, FPA2017-90566-REDC. 
RED2018-102661-T,    % RENATA
RED2018-102573-E,    % REDES ESTRATÉGICAS: Participación Española en Estructuras Euro... 
FPA2017-90687-REDC,  % CPAN
Vicepresidència i Conselleria d’Innovació, Recerca i Turisme, Conselleria d’Educació, i Universitats del Govern de les Illes Balears i Fons Social Europeu, 
Generalitat Valenciana (PROMETEO/2019/071),  
EU COST Actions CA18108, CA17137, CA16214, and CA16104, and
the Spanish Ministry of Education, Culture and Sport grants FPU15/03344 and FPU15/01319.
MC acknowledges funding from the European Union's Horizon 2020 research and innovation programme, under the Marie Skłodowska-Curie grant agreement No. 751492.
The authors thankfully acknowledge the computer resources at MareNostrum and the technical support provided by Barcelona Supercomputing Center (BSC) through Grants No. AECT-2019-2-0010, AECT-2019-1-0022, AECT-2019-1-0014, AECT-2018-3-0017, AECT-2018-2-0022, AECT-2018-1-0009, AECT-2017-3-0013, AECT-2017-2-0017, AECT-2017-1-0017, AECT-2016-3-0014, AECT2016-2-0009,  from the Red Española de Supercomputación (RES) and PRACE (Grant No. 2015133131). BAM and ET simulations were carried out on the BSC MareNostrum computer under PRACE and RES (Red Española de Supercomputación) allocations and on the FONER computer at the University of the Balearic Islands. 
Benchmarks calculations were carried out on the cluster CIT provided by LIGO Laboratory and supported by National Science Foundation Grants PHY-0757058 and PHY-0823459. Parameter Estimation results used the public data obtained from the Gravitational Wave Open Science Center (https://www.gw-openscience.org), a service of LIGO Laboratory, the LIGO Scientific Collaboration and the Virgo Collaboration. LIGO is funded by the U.S. National Science Foundation. Virgo is funded by the French Centre National de Recherche Scientifique (CNRS), the Italian Istituto Nazionale della Fisica Nucleare (INFN) and the Dutch Nikhef, with contributions by Polish and Hungarian institutes.

%%%% a few appendices, for good measure %%%%%%%%

\appendix

\section{Conversion from spheroidal to spherical-harmonic modes}
\label{appendix:spheroidal_spherical}
%%%%%%%%%%%%%%%%%%%%%%%%%%%%%%%%%%%%%%%
Spin-weighted spheroidal harmonics can be written as a linear combination of spherical ones:
\begin{align}
\,_{s}S(a,\theta,\phi)^{m}_{l}=\sum_{l'=2}^{\infty}{\alpha_{l\,l'm}(a) \,_{s}\,Y(\theta,\phi)^{m}_{l'}}.
\end{align}
In the sum above, each spherical harmonic is weighted by a mixing coefficient $\alpha_{l\,l'm}$ measuring its overlap with the corresponding spherical harmonic: 
\begin{align}
\alpha_{l\,l'm}=\int{d\Omega\, _{s}S(a,\theta,\phi)^{m}_{l}  \,_{s}\,Y^{*}(\theta,\phi)^{m}_{l'} }.
\end{align} 
Note that the mixing coefficients are functions of the final spin only. 
Although in theory the $(3,2)$ couples to all the modes with $m=2$, in practice we find that the strongest source of mode-mixing comes from the mixing with the $(2,2)$. Therefore, we choose to neglect the coupling to modes with $l>3$.

Under this assumption 
% long explanation...
%one can re-expand the spheroidal-harmonic modes of the strain as
%\begin{align}
%h^{S}_{22} \ \ _{-2}S^{2}_{2}&=h^{S}_{22}(\alpha_{222} \ \ _{-2}\,Y^{2}_{2}+\alpha_{232} \ \ _{-2}Y^{2}_{3})\\
%h^{S}_{32} \ \ _{-2}S^{3}_{2}&=h^{S}_{32} (\alpha_{322}\ \ _{-2}Y^{2}_{2}+\alpha_{332} \ \  _{-2}Y^{2}_{3}),
%\end{align}
%where we fixed $s=-2$, as we are working with quantities derived from $\psi_4$. Furthermore, one has:
%\begin{align}
%h=\sum_{l,m}h^{S}_{\ell m} \, _{-2}S^{l}_{m}=\sum_{l,m}h_{l m}\, _{-2}Y^{l}_{m}.
%\end{align}
%Combining the two equations above, 
the coefficients of the strain in the two bases of harmonics are related via the following simple linear transformation:
\begin{gather}
 \begin{pmatrix} h_{22} \\ h_{32}  \end{pmatrix}
 =
  \begin{pmatrix}
   \alpha_{222} &
   \alpha_{232} \\
    \alpha_{322}  &
    \alpha_{332}  
   \end{pmatrix}
    \begin{pmatrix} h_{22}^{S} \\ h_{32}^{S}  \end{pmatrix}.
\end{gather}

The mixing coefficients have been computed in \cite{BertiMixingCoeffs} for black holes spinning up to $\chi_f=0.9999$. To improve accuracy for extreme spins, we perform a quadratic-in-spin fit of all the data-points with $\chi_f\in\left[0.999,0.9999\right]$ and use it to obtain the values of the mixing coefficients extrapolated at $|\chi_f|=1$.

\section{Testing tetrad conventions}\label{appendix:HMconventions}

\begin{figure*}[htbp]
    \centering
    \includegraphics[width=\columnwidth]{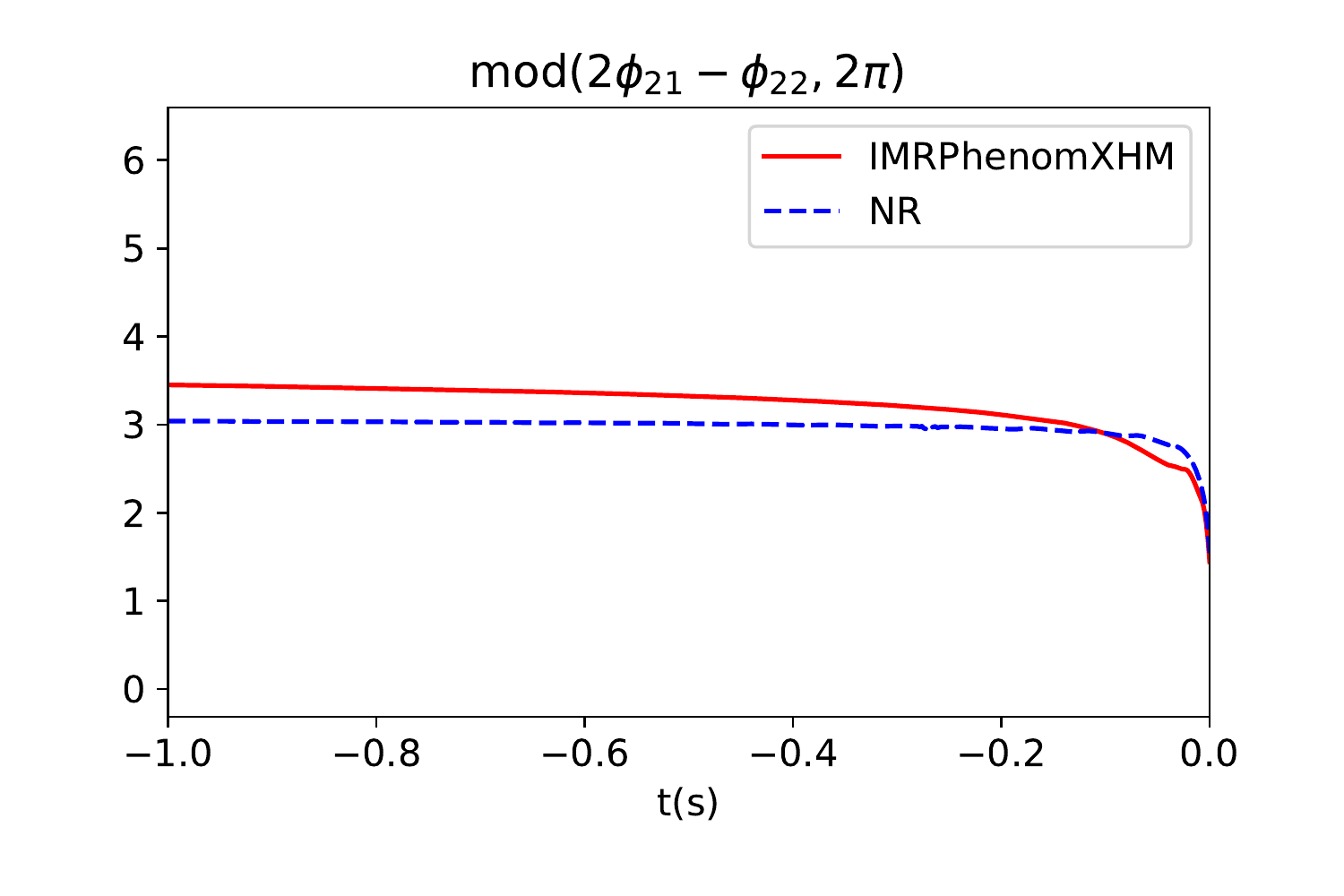}\includegraphics[width=\columnwidth]{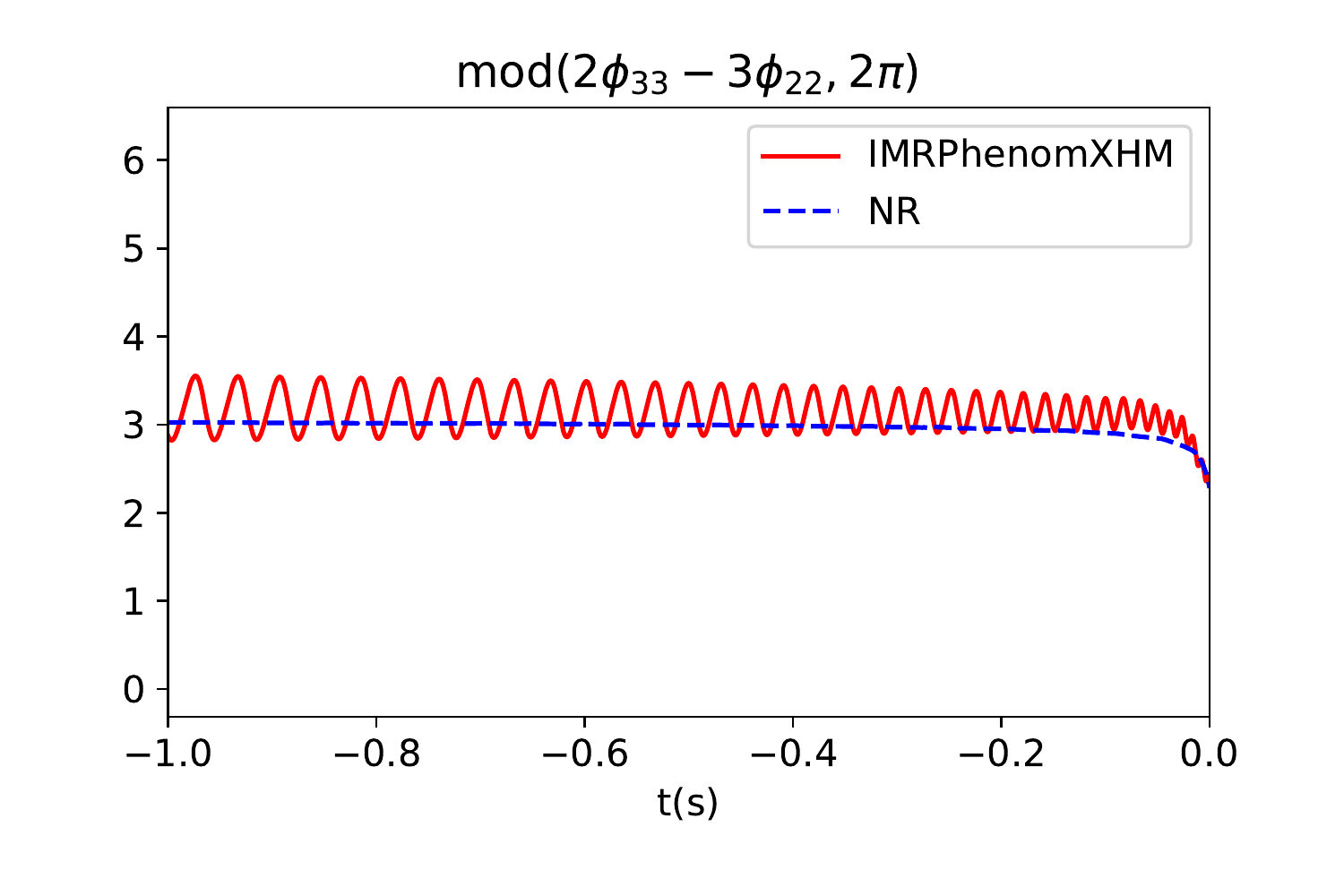}\\
    \includegraphics[width=\columnwidth]{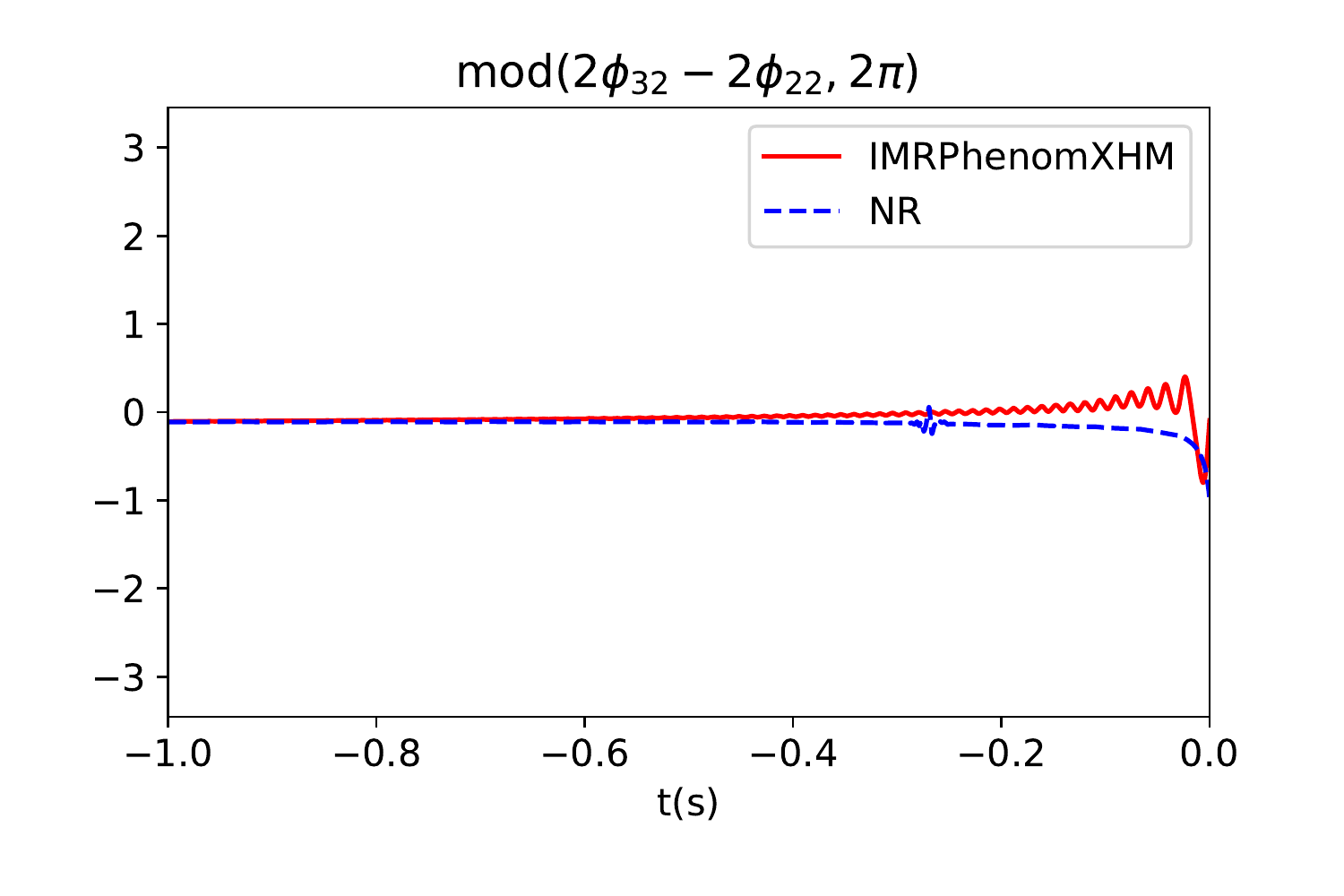}\includegraphics[width=\columnwidth]{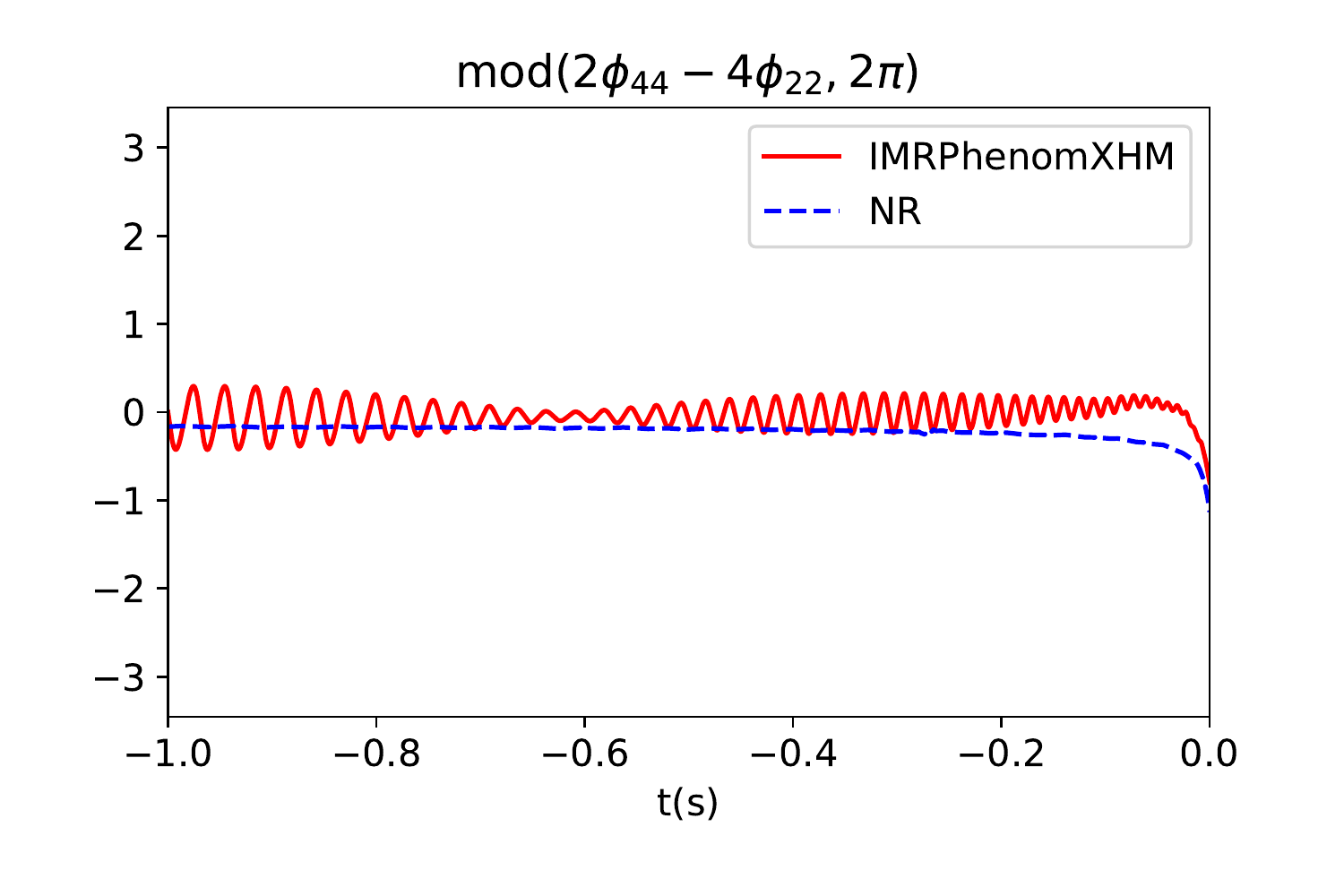}
    \caption{The quantity $\mod(2\Phi_{\ell m}-m \Phi_{22},2\pi)$ can be used to discriminate between different tetrad conventions. Here we show that the time-domain conversion of \phXHM follows the same tetrad choice implemented in the LVCNR catalog.}
    \label{fig:check_tetrad}
\end{figure*}

The relative phase-alignment of the different modes is established trough Eq.\,(\ref{eq:phase_alignment}), which implies a specific choice of tetrad convention. One can check that, when calling the model in time-domain, \phXHM returns modes that follow the same convention adopted for the LVCNR catalog \cite{Schmidt:2017btt}. One has that
\begin{equation*}
    \mod(2\Phi_{\ell m}-m \Phi_{22},2\pi)=\begin{cases}
    \pi \ \ \text{m odd}
    \\
    0 \ \ \text{m even}
    \end{cases}
\end{equation*}
holds for both the LVCNR catalog and the \phXHM higher-multipoles, as shown in Fig.\,\ref{fig:check_tetrad}. 

%%%%%%%%%%%%%%%%%%%%%%%%%%%%%%%%%%%%%%%%%%%%%%%%%%%%%%%%%%%%%%%%%%%%%%

\section{Notes on the implementation of the \phXHM model in the LIGO Algorithms Library}\label{appendix:LAL}
%%%%%%%%%%%%%%%%%%%%%%%%%%%%%%%%%%%%%%%%%%%%%%%%%%%%%%%%%%%%%%%%%%%%%%

The \phXHM model is implemented in the C language as part of the
{\tt LALSimIMR} package of inspiral-merger-ringdown waveform models, which is 
part of the {\tt LALSimulation} collection of code for gravitational waveform and noise generation within \texttt{LALSuite} \cite{lalsuite}.
Online Doxygen documentation is available at {\tt https://lscsoft.docs.ligo.org/lalsuite}, with top level information 
for the {\tt LALSimIMR} package
provided through the {\tt LALSimIMR.h} header file.
%
%For a LALSuite coding style guide see {\tt https://lscsoft.docs.ligo.org/lalsuite/lal/group\_\_lal\_\_spec.html}.
Externally callable functions of the 
\phXHM model follow the \texttt{XLAL} coding standard of \texttt{LALSuite}.

Following our implementation of the \phX model, our \phXHM implementation is highly modularised, such that the inspiral, intermediate and ringdown parts can be updated independently, they are also tracked with independent version numbers, and are implemented in different files of the source code.
Note that the \texttt{XLAL} standard implies that all the source code files are included via the C preprocessor into the main driver file, 
{\tt LALSimIMRPhenomXHM.c}.

The model can be called both in the native Fourier domain, and in the time domain, where an inverse fast Fourier transformation is applied by the \texttt{LALSuite} code. The SWIG \cite{SWIG:code} software development tool is used to automatically create Python interfaces to all \texttt{XLAL} functions of our code, which can be used alternatively to the C interfaces.

Special attention is due for the time and phase alignment of our \texttt{LALSuite} implementation. As mentioned in Sec.~\ref{sec:prelim_phenomenology}, our hybrid waveforms are aligned in time such that the Newman-Penrose scalar for the $\ell=\vert m \vert = 2$ modes peaks $500 M$ before the end of the waveform. In the \texttt{LALSuite} implementation, we first apply a global time-shift of $500M$ to our reconstructed waveforms, and then a parametric fit that accounts for the time-difference between the peak-time of $\psi_4$ and that of strain. An inverse Fourier transformation of the Fourier domain waveform, as produced by \texttt{LALSuite}, will then return a strain peaking around the end of the waveform.

When calling the model in the time domain through \texttt{LALSimulation}'s {\tt ChooseTDWaveform} interface, the time coordinate is chosen such that
$t=0$ for the peak of the sum of the square of the polarizations: 
\begin{align}
  &A(t)   = h^2_{+}(t) + h^2_{\times}(t),  \\
  &A(t=0) = A_{peak}.
\end{align}
These polarizations include all the modes used to generate the model and also depend on the line of sight from the detector to the source through the inclination and azimuthal angle. This choice is consistent with the choice made for the \phHM model.

In \texttt{LALSimulation} the model is called through the function \texttt{ChooseFDWaveform}, whose input parameters \texttt{f\_ref} and \texttt{phiRef} are used to define the phase of the 22 mode at some particular reference frequency. The rest of the modes are built with the correct relative phases with respect to the 22 mode. The argument  \texttt{phiRef} is defined as the orbital phase at the frequency \texttt{f\_ref}. 
See our discussion in the context of \phX \cite{phenX} for further details.
We relate this with the frequency domain 22 phase by means of the SPA (see also our discussion in \cite{phenX}).
\begin{equation}
\label{eq:phiRef}
\Psi_{2-2}(\texttt{f\_ref}) = 2 \texttt{phiRef} - 2\pi \: \texttt{f\_ref} \:t_{\texttt{f\_ref}} + \frac{\pi}{4}.
\end{equation}
Note that when talking about positive frequencies we have to refer to the negative mode, although we usually skip the minus sign for economy of the language.

Since our model is built in the Fourier domain we can not compute the quantity $t_{\texttt{f\_ref}}$ without a Fourier transformation to the time domain plus a numerical root finding, and we currently set it to  $t_{\texttt{f\_ref}}=0$. 
Furthermore, the expression \ref{eq:phiRef} would not be valid if \texttt{f\_ref} is situated in the merger-ringdown part of the waveform, because the SPA approximation is only reliable for the inspiral. This means that when comparing a time-domain model with our model with the exact same parameters we    can only expect them to agree up to rotations,
and would thus have to optimize over \texttt{phiRef} to achieve agreement.
%The EOB-family models that are in time-domain can not neither return their waveforms with the correct phase at \texttt{f\_ref}. However, what we can do is to take the phase of the 22 mode at \texttt{f\_ref} of the Fourier transform of any time-domain model and call our model with a new \texttt{phiRef} that reproduces this Fourier domain phase. In this way the two models should agree at least in the Fourier domain.
%The new \texttt{phiRef\_new} is given in terms of the Fourier 22 phase by
% \begin{equation}
%     \texttt{phiRef\_new} = \frac{1}{2} \left( \Phi_{2-2}(\texttt{f\_ref}) -\frac{\pi}{4} \right)
% \end{equation}

In \texttt{LALSimulation} the azimuthal angle that enters in the spin-weighted spherical harmonics is defined as $\beta = \frac{\pi}{2} - \texttt{phiRef}$, this means that changing the parameter \texttt{phiRef} is equivalent to rotating the system: For example, by increasing the \texttt{phiRef}  by a quantity $\delta \phi$, we would rotate the system an angle $-\delta \phi$. When the 22 mode only is considered, \texttt{phiRef} acts just as a global phase factor for the waveform ($e^{i \: 2 \texttt{phiRef}}$) and the match is not affected since it maximizes over phase and time shifts. However, when higher modes are included this is not satisfied anymore since the term $e^{i \: m \texttt{phiRef}}$ is different for every mode and can not be factored out.
Note that in our \texttt{LALSuite} code, this rotation is applied to every individual mode, such that individual modes and the mode sum are consistent with respect to rotations.
%We have built the model such that the individual modes already take into account the rotation prescribed by \texttt{phiRef}.
%In consequence, when summing all the modes to get the multimode waveform we do not include again this rotation in the spherical harmonics, and these are called using 
%$Y_{\ell m}(\iota = \iota, \beta = \pi/2)$.

The user is free to specify the spherical harmonic modes that should be used to construct the waveform. The default behaviour is to use
all the modes available: $\{$22, 2-2, 21, 2-1, 33, 3-3, 32, 3-2, 44, 4-4$\}$,
below we describe how the modes can be chosen through the different interfaces available for \texttt{LALSuite} waveforms.

Furthermore, the model implemented in \texttt{LALSuite} supports acceleration of waveform evaluation by interpolation of an unequispaced frequency grid broadly following the ``multibanding'' of \cite{Vinciguerra:2017ngf}. Our version of the algorithm is described in 
\cite{our_mb} to do the evaluation faster and can also use a custom list of modes specified by the user.
The multibanding algorithm is parameterized by a threshold, which describes the permitted local interpolation error for the phase in radians, lower values 
thus correspond to higher accuracy.
The default value is set to a value of $10^{-3}$.  

Extensive debugging information can be enabled at compile time with the C preprocessor flag {\tt -D PHENOMXHMDEBUG}.
%This value also is a measure of the absolute error that the multibanding algorithm introduced. In that way, if it is set to $10^{-3}$ the phase difference of the waveforms with and without multibanding should be lower than $10^{-3}$ radians.

%If we do not specify any option the default behaviour is to use the multibanding with a threshold parameter of $10^{-4}$ (this may be relaxed later) and

\paragraph{\texttt{Python} Interface.}

To call the model with the default behaviour we use the function \texttt{SimInspiralChooseFDWaveform} from \texttt{lalsimulation} with the argument \texttt{lalparams} being an empty \texttt{LALSuite} dictionary \texttt{lalparams=lal.CreateDict()}.
The threshold of the multibanding and the mode array can be changed by adding their values to the \texttt{LALSuite} dictionary in the following way: 

\begin{widetext}
\begin{verbatim}
lalsimulation.SimInspiralWaveformParamsInsertPhenomXHMThresholdMband(lalparams, threshold)
ModeArray = lalsimulation.SimInspiralCreateModeArray()
for mode in [[2,2],[2,-2],[2,1],[2,-1]]:
    lalsimulation.SimInspiralModeArrayActivateMode(ModeArray, mode[0], mode[1])
lalsimulation.SimInspiralWaveformParamsInsertModeArray(lalparams, ModeArray).
\end{verbatim}
\end{widetext}

If threshold=0 then multibanding is switched off. By calling \texttt{ChooseFDWaveform} with this \texttt{LALSuite} dictionary we would get the hp and hc polarizations from the contribution of the 22, 2-2, 21, and 2-1 modes without using multibanding.

\paragraph{\texttt{GenerateSimulation} Interface.}

This is an executable in \texttt{LALSimulation} called through command line. The parameters to evaluate the model are passed by options like {\tt --m1, --spin1z}, etc. The multibanding threshold and the mode array are specified as follows
\begin{verbatim}
./GenerateSimulation --approximant IMRPhenomXHM
...waveform params... 
--phenomXHMMband 0. 
--modesList "2,2, 2,-2, 2,1, 2,-1".    
\end{verbatim}
%This calls the model for the same example than in the python interface..

%\textbf{\texttt{LALInference} and \texttt{Bilby}}

\paragraph{\texttt{LALInference} and \texttt{Bilby}.}

We also included the options in the two standard codes to perform Bayesian inference in gravitational wave data analysis: \texttt{LALInference}\cite{LALInference} and \texttt{Bilby} \cite{Ashton:2018jfp}. \texttt{LALInference} uses the same syntax than \texttt{GenerateSimulation} when called through the command line. You can also add these options to the config file and the example we have employed so far can be called as:
\begin{verbatim}
    [engine]
    ...
    approx = IMRPhenomXHMpseudoFourPN
    modesList = '2,2, 2,-2, 2,1, 2,-1'
    phenomXHMMband = 0
    ...
\end{verbatim}
Note that in the current version of \texttt{LALInference} the string \texttt{pseudoFourPN} has to be added to the name of the approximant.
For \texttt{Bilby} these options are specified in the \texttt{waveform\_argument dictionary} defined in the configuration file. The equivalent example would be called as:
\begin{verbatim}
waveform_arguments = 
dict(waveform_approximant='IMRPhenomXHM',
     reference_frequency=50.,
     minimum_frequency=20.,
     mode_array=[[2,2],[2,-2],[2,1],[2,-1]],
     phenomXHMMband=0.).
\end{verbatim}

The released version of \texttt{Bilby} does not support the multibanding option yet, however a private branch that support this option can be downloaded with \texttt{git clone -b imrphenomx \url{https://git.ligo.org/cecilio.garcia-quiros/bilby.git}}.  Equivalently we provide a branch for the \texttt{PyCBC} software \cite{pycbc} which can be obtained with the command \texttt{git clone -b imrphenomx \url{https://github.com/Ceciliogq/pycbc.git}}.

\section{Inspiral phase: higher-modes extension of \phX}
\label{app:inspiral_hm}

The inspiral orbital phase calibrated in \phX can be written as a pseudo-PN expansion:
\begin{equation}
\phi_{22}(f)=\mathcal{N} (Mf)^{-5/3}\sum_{i=0}^{9}(Mf)^{i/3}\left(c^{i}_{22}+d_{22}^{i}\log{f}\right),
\end{equation}
where $\mathcal{N}$ is a certain normalization constant. In the reconstruction of the higher-mode inspiral phase we need $\frac{m}{2} \phi_{22}(\frac{2}{m} f)$ (see Eq.\,(\ref{inspiral_scaling}) ). To avoid recomputing the $(2,2)$-phase on a new frequency array for each mode, we wish to rewrite this quantity as
\begin{equation}
\frac{m}{2} \phi_{22}\left( \frac{2}{m} f \right)=\mathcal{N} (Mf)^{-5/3}\sum_{i=0}^{9}(Mf)^{i/3}\left(c^{i}_{m}+d_{m}^{i}\log{f}\right),
\end{equation}
where all the rescaling factors have been reabsorbed in frequency-independent coefficients. It is straightforward to verify that the coefficients of the two expansions above are related as follows:
\begin{align}
c_{m}^{i}&=\left(\frac{m}{2}\right)^{(8-i)/3}\left(c_{22}^{i}-d_{22}^{i}\log{\frac{m}{2}}\right),\nonumber\\
d_{m}^{i}&=\left(\frac{m}{2}\right)^{(8-i)/3}d_{22}^{i}.
\end{align}

\section{Fourier Domain Post-Newtonian amplitudes}\label{appendix:FPN}

When comparing the Fourier domain  expressions for the 
spherical harmonic mode amplitudes given in 
equations (11-12) of \cite{PNamps} we found significant discrepancies with our numerical data. We have thus recomputed the mode amplitudes as outlined below, and include the explicit expressions we have used (which deviate from \cite{PNamps} at 2PN order), and which resolve the observed discrepancies with the numerical data, at the end of this appendix.

%In this section we briefly comment on how we derive the Fourier domain post-Newtonian amplitudes starting from the time domain PN expressions that are found in the literature \cite{Arun:2008kb} and write the result explicitly for several modes.

The time domain PN spherical harmonic modes are typically written in the form
\begin{equation}\label{eq:global}
h_{\ell m} = A_{\ell m} e^{-i m \phi}, \quad A_{\ell m}(x) = 2 \:\eta \: x \sqrt{\frac{16 \pi}{5}} \hat{h}_{\ell m}.
\end{equation}
Expressions for the $\hat{h}_{\ell m}$ can be found in \cite{Blanchet2014}, \cite{Arun:2008kb} and \cite{Buonanno},
including non-spinning terms up to 3PN order, and spinning
terms up to 2PN order.
The quantities $\hat{h}_{\ell m}$, and with them the time domain amplitudes  $A_{\ell m}$ are complex functions. 
According to the SPA, the modes in the frequency domain can then be approximated as
\begin{equation}
\tilde{h}_{\ell m}(f) \approx A_{\ell m}(x) \sqrt{\frac{2 \pi}{m \ddot{\phi}(x)}} e^{i \:\Psi_{\ell m}(f)},
\end{equation}
and we define $A^{\mathrm{SPA}}_{\ell m}(f) = A_{\ell m}(x) \sqrt{\frac{2 \pi}{m \ddot{\phi}(x)}}$
(compare also with the expressions in \cite{Santamaria:2010yb}).
The orbital phase $\phi$ is related to the freqency and the PN expansion parameter $x$ by $\ddot{\phi} = \dot{\omega} = (3/2)\sqrt{x} \dot{x}$. The frequency $f$, which acts as the independent variable in the Fourier domain is related to $x$ by $x= \left( \frac{2\pi \:f}{m} \right)^{2/3}$. We then obtain
\begin{equation}\label{eq:spaamp}
A^{\mathrm{SPA}}_{\ell m}(f) = A_{\ell m}(x) \sqrt{\frac{2 \pi}{m (3/2)\sqrt{x} \dot{x}}}.
\end{equation}
We now need to compute $\dot{x}$. Using the TaylorT4 expression \cite{Taylors} we finally obtain:
\begin{widetext}
\begin{align}\label{eq:xdot}
\dot{x} = &\frac{-m_1 m_2}{32744250 (m_1 + m_2)^2} \bigg[ \:x^5 \left( -419126400 m_1^2-838252800 m_1 m_2-419126400 m_2^2\right) 
\\\nonumber
+& x^6 
\begin{aligned}[t]
	\big( & 1152597600 m_1^3 m_2+1247400 m_1^2 \big(1848 			m_2^2+743\big)+ \\ \nonumber
 	&2494800 m_1 m_2 \big(462 m_2^2+743\big)+926818200 m_2^2\big)
 \end{aligned}
\\\nonumber
+& x^7 
\begin{aligned}[t]
\big(&-1373803200 m_1^4 m_2^2-2747606400 m_1^3 m_2^3 
\\ &+207900 m_1^3 m_2 \big(10206 \chi_1^2-19908 \chi_1 \chi_2+10206 \chi_2^2-13661\big) 
   \\ &-23100 m_1^2 \big(59472 m_2^4 +91854 \chi_1^2+34103
   \\&-18 m_2^2 \big(10206 \chi_1^2-19908 \chi_1 \chi_2+10206 \chi_2^2-13661\big)\big) 
   \\&+23100 m_1 m_2 \big(9 m_2^2 \big(10206 \chi_1^2-19908 \chi_1 \chi_2+10206 \chi_2^2-13661\big)
   \\&-2 \big(45927 \chi_1^2+45927 \chi_2^2+34103\big)\big)
  -23100 m_2^2 \big(91854 \chi_2^2+34103\big)\big)
\end{aligned}
\\
+&x^{13/2} 
\begin{aligned}[t]
	\big(&6566313600 m_1^4 \chi_1+13132627200 m_1^3 m_2 \chi_1-2619540000 m_1^3 \chi_1
	\\&-34927200 m_1^2 (48 \pi -m_2 (188 m_2
   (\chi_1+\chi_2)+75 \chi_2))
   \\&-34927200 m_1 m_2 (96 \pi -m_2 (376 m_2 \chi_2+75 \chi_1))
   \\&-34927200 m_2^2 ((75-188 m_2) m_2
   \chi_2+48 \pi )\big)
\end{aligned}
\\
+& x^{15/2}
\begin{aligned}[t]
	 \big(&-34962127200 m_1^5 m_2 \chi_1-69924254400 m_1^4 m_2^2 \chi_1+14721814800 m_1^4 m_2 \chi_1
	 \\&+17059026600 m_1^4 \chi_1-34962127200 m_1^3 m_2^3 (\chi_1+\chi_2)-14721814800 m_1^3 m_2^2 \chi_2
   \\&+5821200 m_1^3 m_2 (5861 \chi_1+1701 \pi )-4036586400
   m_1^3 \chi_1
   \\&+207900 m_1^2 \big(3 \pi  \big(31752 m_2^2+4159\big)+2 m_2 \big(-168168 m_2^3 \chi_2-35406 m_2^2 \chi_1
   \\&+41027 m_2
   (\chi_1+\chi_2)+9708 \chi_2\big)\big)+415800 m_1 m_2 \big(-84084 m_2^4 \chi_2+35406 m_2^3 \chi_2
   \\&+82054 m_2^2 \chi_2+3
   \pi  \big(7938 m_2^2+4159\big)+9708 m_2 \chi_1\big)
   \\&+207900 m_2^2 (2 m_2 (41027 m_2-9708) \chi_2+12477 \pi )\big)
\end{aligned}
\\
+&x^{17/2}
\begin{aligned}[t]
 \big(&84184254000 m_1^6 m_2^2 \chi_1+170726094000 m_1^5 m_2^3 \chi_1+2357586000 m_1^5 m_2^3 \chi_2
 \\&-35665037100 m_1^5 m_2^2
   \chi_1-198816225300 m_1^5 m_2 \chi_1+88899426000 m_1^4 m_2^4 (\chi_1+\chi_2)
   \\&+35665037100 m_1^4 m_2^3 \chi_2-138600
   m_1^4 m_2^2 (3399633 \chi_1+530712 \chi_2+182990 \pi )
   \\&+33313480200 m_1^4 m_2 \chi_1+87143248500 m_1^4 \chi_1+9702000 m_1^3 m_2^5
   (243 \chi_1+17597 \chi_2)
   \\&+35665037100 m_1^3 m_2^4 \chi_1-69300 m_1^3 m_2^3 (4991769 (\chi_1+\chi_2)+731960 \pi )
   \\&-33313480200 m_1^3
   m_2^2 \chi_2+1925 m_1^3 m_2 (97151928 \chi_1+6613488 \chi_2-12912300 \pi )
   \\&-14891068500 m_1^3 \chi_1-11550 m_1^2 \big(15 \pi  \big(146392
   m_2^4+286940 m_2^2-2649\big)
   \\&+2 m_2 \big(-3644340 m_2^5 \chi_2+1543941 m_2^4 \chi_2+54 m_2^3 (58968 \chi_1+377737 \chi_2)
   \\&+1442142
   m_2^2 \chi_1-4874683 m_2 (\chi_1+\chi_2)-644635 \chi_2\big)\big)
   \\&-23100 m_1 m_2 \big(m_2 \big(m_2 \big(8606763
   m_2^2-1442142 m_2-8095994\big) \chi_2
   \\&+(-551124 m_2-644635) \chi_1\big)+15 \pi  \big(71735 m_2^2-2649\big)\big)
   \\&+57750 m_2^2 (2 m_2 (754487
   m_2-128927) \chi_2+7947 \pi )\bigg].
\end{aligned}
\end{align}
\end{widetext}

We can now compute the complex non-polynomial Fourier Domain PN amplitudes
$A_{\ell m}$, that we re-expand up to 3PN order, as we have done for the 
$\ell=\vert m\vert = 2$ modes in \phX. 
We list the resulting complex Fourier Domain PN amplitudes following this method. We write them as a function of $v= \sqrt{x}= \left( \frac{2\pi \:f}{m} \right)^{1/3}$ and factor out a common term to simplify the comparison with \cite{PNamps} (note that there  $\nu=\eta$, $V_m=v$ and $m\Psi_{\mathrm{SPA}} + \pi/4 = \Psi_{\ell m}$), finally obtaining
\begin{equation}\label{eq:spaamp_2}
A^{\mathrm{SPA}}_{\ell m}(f) = \pi \sqrt{\frac{2 \eta}{3}}v^{-7/2} \hat{H}_{\ell m}
\end{equation}

The Post-Newtonian expressions that we use to calibrate the inspiral part of the amplitude are given by the expanded expressions below, as used in
eq.~(\ref{eq:spaamp}),
except for the $(2,1)$ mode. We observed that for some cases with $q<40$ the re-expansion of the $(2,1)$ mode breaks down before reaching the cutting frequency of the inspiral. For those cases the $(2,1)$ amplitude is very small (see Section \ref{sec:prelim_phenomenology}) and the spinning contribution of higher PN terms is more important due to the different competing effects which can lead to  cancellations in the waveform which are not captured by our 3PN accurate qusi-circular expressions. For the $(2,1)$ mode we therefore do not re-expand in a power series but keep the form of expression (\ref{eq:spaamp}) when $q<40$. The expression (\ref{eq:spaamp}) is not used for $q>40$ because it shows a divergence that appears before the inspiral cutting frequency for high spins.

\begin{widetext}
\begin{align}
\hat{H}_{22} =& 
\begin{aligned}[t]
1 &+ \left(\frac{451 \eta }{168}-\frac{323}{224}\right) v^2  +v^3 \left(\frac{27 \delta  \chi_a^z}{8}-\frac{11 \eta  \chi_s^z}{6}+\frac{27 \chi_s^z}{8}\right) +
   \\ 
   &+ v^4 \left(-\frac{49 \delta  \chi_a^z \chi_s^z}{16}+\frac{105271 \eta^2}{24192}+6 \eta  (\chi_a^z)^2+\frac{\eta  \left(\chi_s^z\right)^2}{8}
   -\frac{1975055 \eta }{338688}-\frac{49 (\chi_a^z)^2}{32}-\frac{49 \left(\chi_s^z\right)^2}{32}-\frac{27312085}{8128512}\right) +
  \\
  &+ v^6 \left(\frac{107291 \delta  \eta  \chi_a \chi_s}{2688}-\frac{875047 \delta 
 \chi_a \chi_s}{32256}+\frac{31 \pi  \delta  \chi_a}{12}+\frac{34473079 \eta ^3}{6386688}+\frac{491 \eta^2 {\chi_a}^2}{84}-\frac{51329 \eta^2 \chi_s^2}{4032}- \right.
  \\
  &  -\frac{3248849057 \eta^2}{178827264}+\frac{129367 \eta  \chi_a^2}{2304}+\frac{8517 \eta  {\chi_s}^2}{224}-\frac{7 \pi  \eta  \chi_s}{3}-\frac{205 \pi^2 \eta
  }{48}+\frac{545384828789 \eta }{5007163392}-\frac{875047 {\chi_a}^2}{64512} -
  \\
  & \left. -\frac{875047 \chi_s^2}{64512}+\frac{31 \pi  {\chi_s}}{12}+\frac{428 i \pi }{105}-\frac{177520268561}{8583708672} \right)  
\end{aligned}
\\
\hat{H}_{21} =& 
\begin{aligned}[t]
\label{eq:21amp}
\frac{1}{3} i \sqrt{2} \Bigg[& v \delta + v^2
  \left(-\frac{3 \delta  \chi_s^z}{2}-\frac{3 \chi_a^z}{2}\right)
 +v^3 \left(\frac{117 \delta  \eta }{56}+\frac{335 \delta }{672}\right) +
 \\+&v^4 \left( -\frac{965}{336} \delta  \eta  \chi_s^z+\frac{3427 \delta  \chi_s^z}{1344}-\pi  \delta -\frac{i \delta }{2}-\frac{1}{2}   i \delta  \log (16)-\frac{2101 \eta  \chi_a^z}{336}+\frac{3427 \chi_a^z}{1344}\right)+
 \\
 +& v^5 \Bigg(\frac{21365 \delta  \eta^2}{8064}+10 \delta  \eta  \chi_a^2+\frac{39}{8}
   \delta  \eta  \chi_s^2-\frac{36529 \delta  \eta }{12544}-\frac{307 \delta 
   \chi_a^2}{32}-\frac{307 \delta  \chi_s^2}{32}+3 \pi  \delta  \chi_s -
   \\-&\frac{964357 \delta }{8128512}+\frac{213 \eta  \chi_a {\chi_s}}{4}-\frac{307 \chi_a \chi_s}{16}+3 \pi  \chi_a\Bigg) +
\\+&  v^6 \Bigg(-\frac{547}{768} \delta  \eta^2 \chi_s-15 \delta  \eta  \chi_a^2
  \chi_s-\frac{3}{16} \delta  \eta  \chi_s^3-\frac{7049629 \delta  \eta 
  \chi_s}{225792}+\frac{417 \pi  \delta  \eta }{112}-\frac{1489 i \delta  \eta
  }{112}-\frac{89}{28} i \delta  \eta  \log (2)+\frac{729}{64} \delta  \chi_a^2
  \chi_s
  \\+& \frac{243 \delta  \chi_s^3}{64}+\frac{143063173 \delta 
  \chi_s}{5419008}-\frac{2455 \pi  \delta }{1344}-\frac{335 i \delta
  }{1344}-\frac{335}{336} i \delta  \log (2)+\frac{42617 \eta^2 \chi_a}{1792}-15
  \eta  \chi_a^3-\frac{489}{16} \eta  \chi_a {\chi_s}^2
  \\-&\frac{22758317 \eta  \chi_a}{225792}+\frac{243 {\chi_a}^3}{64}+\frac{729 \chi_a \chi_s^2}{64}+\frac{143063173 {\chi_a}}{5419008}\Bigg)
\Bigg]
\end{aligned}
\\
\hat{H}_{33} =& 
\begin{aligned}[t]
\label{eq:33amp}
&
-\frac{3}{4} i \sqrt{\frac{5}{7}} \Bigg[ 
v \delta  +v^3 \delta \left(\frac{27}{8}\eta-\frac{1945  }{672}\right) 
    + v^4 \left(-\frac{2}{3} \delta  \eta  \chi_s^z+\frac{65 \delta  \chi_s^z}{24}+\pi  \delta -\frac{21 i \delta }{5}+6 i \delta 
  \log \left(\frac{3}{2}\right)-\frac{28 \eta  \chi_a^z}{3}+\frac{65 \chi_a^z}{24}\right)+
  \\&+v^5 \left(\frac{420389 \delta  \eta^2}{63360}+10 \delta  \eta  \chi_a^2+\frac{1}{8}
   \delta  \eta  \chi_s^2-\frac{11758073 \delta  \eta }{887040}-\frac{81 \delta 
   \chi_a^2}{32}-\frac{81 \delta  \chi_s^2}{32}-\frac{1077664867 \delta
   }{447068160}+\frac{81 \eta  \chi_a \chi_s}{4}-\frac{81 \chi_a
   \chi_s}{16}\right)+
   \\&+v^6 \Bigg(-\frac{67}{24} \delta  \eta^2 \chi_s-\frac{58745 \delta  \eta  {\chi_s}}{4032}+\frac{131 \pi  \delta  \eta }{16}-\frac{440957 i \delta  \eta
   }{9720}+\frac{69}{4} i \delta  \eta  \log \left(\frac{3}{2}\right)+\frac{163021 \delta 
   \chi_s}{16128}-\frac{5675 \pi  \delta }{1344}+\frac{389 i \delta
   }{32}
   \\& -\frac{1945}{112} i \delta  \log \left(\frac{3}{2}\right)
    - \frac{137 \eta^2
   \chi_a}{24}-\frac{148501 \eta  \chi_a}{4032}+\frac{163021 {\chi_a}}{16128}\Bigg)
  \Bigg]
\end{aligned}
\\
\hat{H}_{32} =& \frac{1}{3} \sqrt{\frac{5}{7}}
\label{eq:32amp}
\begin{aligned}[t]
\Bigg[&v^2(1-3 \eta )  + v^3 4 \eta   \chi_s^z + v^4\left(-\frac{589 \eta^2}{72}+\frac{12325 \eta
  }{2016}-\frac{10471}{10080}\right)  +
  \\ +& v^5 \left(\eta  \left(\frac{113 \delta  \chi_a}{8}+\frac{1081 {\chi_s}}{84}-\frac{66 i}{5}\right)+\frac{1}{24} (-113 \delta  \chi_a-113 {\chi_s}+72 i)-15 \eta^2 \chi_s\right)
   \\+& v^6 \Bigg(\eta  \left(-\frac{1633 \delta  \chi_a \chi_s}{48}-\frac{563
   \chi_a^2}{32}-\frac{2549 \chi_s^2}{96}+8 \pi  {\chi_s}-\frac{8689883}{149022720}\right)+\frac{81 \delta  \chi_a {\chi_s}}{16}
   \\+&\frac{837223 \eta ^3}{63360}+\eta^2 \left(30 \chi_a^2+\frac{313
   \chi_s^2}{24}-\frac{78584047}{2661120}\right)+\frac{81 {\chi_a}^2}{32}+\frac{81 \chi_s^2}{32}+\frac{824173699}{447068160}\Bigg)
   \Bigg]
\end{aligned}
 \\
\hat{H}_{31} =& 
\begin{aligned}[t]
\frac{i}{12
  \sqrt{7}} \Bigg[& v \delta + v^3
  \left(\frac{17 \delta  \eta }{24}-\frac{1049 \delta }{672}\right) 
  \\+& v^4 \left(\frac{10 \delta  \eta  \chi_s^z}{3}+\frac{65 \delta 
  \chi_s^z}{24}-\pi  \delta -\frac{7 i \delta }{5}-\frac{1}{5} i \delta  \log
  (1024)-\frac{40 \eta  \chi_a^z}{3}+\frac{65 \chi_a^z}{24}\right) +  
  \\+& v^5 \left(-\frac{4085 \delta  \eta^2}{4224}+10 \delta  \eta  \chi_a^2+\frac{1}{8}
   \delta  \eta  \chi_s^2-\frac{272311 \delta  \eta }{59136}-\frac{81 \delta 
   \chi_a^2}{32}-\frac{81 \delta  \chi_s^2}{32}+\frac{90411961 \delta
   }{89413632}+\frac{81 \eta  \chi_a \chi_s}{4}-\frac{81 \chi_a
   \chi_s}{16} \right) +
   \\ + &
   v^6 \left(\frac{803}{72} \delta  \eta^2 \chi_s-\frac{36187 \delta  \eta  \chi_s}{1344}+\frac{245 \pi  \delta  \eta }{48}-\frac{239 i \delta  \eta }{120}-\frac{5}{12} i
   \delta  \eta  \log (2)+\frac{264269 \delta  \chi_s}{16128}+\frac{313 \pi  \delta
   }{1344}+\frac{1049 i \delta }{480}+ \right.
   \\+& \left. 
   \frac{1049}{336} i \delta  \log (2)+\frac{2809 \eta^2
   \chi_a}{72}-\frac{318205 \eta  \chi_a}{4032}+\frac{264269 \chi_a}{16128}\right)
   \Bigg]
  \end{aligned}
\\
\hat{H}_{44} =& 
\begin{aligned}[t]
\label{eq:44amp}
\frac{4}{9} \sqrt{\frac{10}{7}}  \Bigg[& v^2 (3 \eta -1)+ v^4\left(\frac{1063 \eta^2}{88}-\frac{128221 \eta }{7392 }+\frac{158383}{36960}\right) +
\\+& v^5 \Bigg(\pi  (2-6 \eta )+\frac{1}{120} \Bigg(-\eta  (1695 \delta  \chi_a+2075
   \chi_s-3579 i+2880 i \log (2)) +
   \\+& 565 \delta  \chi_a+1140 \eta^2
   \chi_s+565 \chi_s-1008 i+960 i \log (2)\Bigg)\Bigg) +
   \\+& v^6 \Bigg(\eta  \left(\frac{243 \delta  \chi_a \chi_s}{16}+\frac{563
   \chi_a^2}{32}+\frac{247 {\chi_s}^2}{32}-\frac{22580029007}{880588800}\right)-\frac{81 \delta  \chi_a
   \chi_s}{16}-\frac{7606537 \eta ^3}{274560}
   \\+&\eta^2 \left(-30 {\chi_a}^2-\frac{3 \chi_s^2}{8}+\frac{901461137}{11531520}\right)-\frac{81 {\chi_a}^2}{32}-\frac{81 \chi_s^2}{32}+\frac{7888301437}{29059430400}\Bigg)
\Bigg]
\end{aligned}
  \\
\hat{H}_{43} =&  
\begin{aligned}[t]
\frac{3}{4} i \sqrt{\frac{3}{35}} \Bigg[& v^3 \left(2 \delta  \eta -\delta\right)
   +v^4 \left(\frac{5 \eta\chi_a}{2}-\frac{5 \delta  \eta  \chi_s}{2}\right)
   +v^5 \left(\frac{887 \delta  \eta^2}{132}-\frac{10795 \delta \eta}{1232}+\frac{18035 \delta }{7392}\right)
  \\+& v^6 \Bigg(-\frac{469}{48} \delta  \eta^2 \chi_s+\frac{4399 \delta  \eta  \chi_s}{448}+2 \pi  \delta  \eta -\frac{16301 i \delta  \eta }{810}+12 i \delta  \eta % \log
   \left(\frac{3}{2}\right)-\frac{113 \delta \chi_s}{24}-\pi  \delta +\frac{32 i
   \delta }{5}+
   \\&-6 i \delta  \log \left(\frac{3}{2}\right)-\frac{1643 \eta^2 \chi_a}{48}+\frac{41683 \eta \chi_a}{1344}-\frac{113 \chi_a}{24}\Bigg)
   \Bigg]
\end{aligned}
\end{align}
\end{widetext}

\bibliography{uib,phenomx,eob_refs,nr_refs}

\end{document}